\documentclass[fleqn,usenatbib]{mnras}
\usepackage{newtxtext,newtxmath}

\usepackage[T1]{fontenc}
\usepackage{longtable}

\usepackage{graphicx}
\usepackage{bm}
\usepackage{natbib}
\usepackage{booktabs}
\usepackage{color}
\usepackage{units}
\usepackage{orcidlink}

\usepackage[normalem]{ulem}

\newcommand{\program}[1]{\textsc{#1}}
\newcommand{\citeg}[1]{\citep[e.g.,][]{#1}}
\newcommand{\be}{\begin{equation}}
\newcommand{\ee}{\end{equation}}

\title[SN~Ia physics from ZTF DR2]{Lightcurve Modelling of 2,205 ZTF DR2 Type~Ia Supernovae: Implications for SN Ia Physics and Cosmology}

\author[N.~Sarin et al.]{Nikhil Sarin$^{1, 2}$\thanks{E-mail:nsarin.astro@gmail.com}\orcidlink{0000-0003-2700-1030}, Ellen Lindsj\"o$^{3}$, 
Lisa Kelsey$^{1,2}$\orcidlink{0000-0003-0313-0487}, 
Matthew Grayling$^{1,2}$, 
Jesper Sollerman$^{4}$\orcidlink{0000-0003-1546-6615}
\newauthor
Steve Schulze$^{5}$,
Adam A.~Miller$^{6,5,7}$\orcidlink{0000-0001-9515-478X}
Madeleine Ginolin$^{1,2}$\orcidlink{0009-0004-5311-9301}, 
Erin E. Hayes$^{1,2}$\orcidlink{0000-0003-3847-0780}, 
Conor M. B. Omand$^{8}$\orcidlink{0000-0002-9646-8710}, 
\newauthor
Kaisey~S.~Mandel$^{1,2}$\orcidlink{0000-0001-9846-4417},
Aaron Do$^{1, 2}$\orcidlink{0000-0003-3429-7845},
Suhail.~Dhawan$^{9}$\orcidlink{0000-0002-2376-6979},
Joel Johansson$^{10}$\orcidlink{0000-0001-5975-290X}
\\
$^{1}$Kavli Institute for Cosmology, University of Cambridge, Madingley Road, CB3 0HA, UK\\
$^{2}$Institute of Astronomy, University of Cambridge, Madingley Road, CB3 0HA, UK\\
$^{3}$Department of Physics, KTH Royal Institute of Technology, The Oskar Klein Centre, AlbaNova, SE-106 91 Stockholm, Sweden\\
$^{4}$Oskar Klein Centre, Department of Astronomy, Stockholm University, SE-106 91, Stockholm, Sweden\\
$^5$Center for Interdisciplinary Exploration and Research in Astrophysics (CIERA), Northwestern University, 1800 Sherman Ave., Evanston, IL 60201, USA\\
$^{6}$Department of Physics and Astronomy, Northwestern University, 2145 Sheridan Road, Evanston, IL 60208, USA\\
$^{7}$NSF-Simons AI Institute for the Sky (SkAI), 172 E. Chestnut St., Chicago, IL 60611, USA\\
$^{8}$Astrophysics Research Institute, Liverpool John Moores University, Liverpool Science Park IC2, 146 Brownlow Hill, Liverpool, UK, L3 5R\\
$^{9}$School of Physics and Astronomy, University of Birmingham, Birmingham B15 2TT, UK \\
$^{10}$Oskar Klein Centre, Department of Physics, Stockholm University, SE-106 91, Stockholm, Sweden\\
}
\date{Accepted XXX. Received YYY; in original form ZZZ}

\pubyear{\the\year{}}

\begin{document}
\label{firstpage}
\pagerange{\pageref{firstpage}--\pageref{lastpage}}
\maketitle

\begin{abstract}
We fit the multi-band light curves of 2,205 Type Ia supernovae (SNe~Ia) from the Zwicky Transient Facility DR2 with a one-zone radioactive decay model with a phenomenological addition to include Fe recombination physics. We find a strong correlation between inferred nickel mass and SALT2 stretch, which our simplified modelling links to longer diffusion times in more massive ejecta, offering a physical basis for the brighter-slower relation. SNe~Ia in low-mass hosts ($\log_{10}(M_*/M_\odot) < 10$) produce $\approx 12\%$ more $^{56}$Ni than those in high-mass hosts, linking the host-galaxy mass step to ejecta properties and hinting at metallicity or age-dependent burning efficiencies. A pseudo-bolometric comparison provides lower limits on the nickel masses, highlighting their sensitivity to SED-level assumptions. Injection-and-recovery tests with realistic ZTF sampling and the same model recover the nickel scale but show significant sensitivity to distance and opacity assumptions; individual-event point estimates are therefore model-dependent. Accounting for selection biases and broad individual-event posteriors, hierarchical modelling of 902 SNe ($z \leq 0.06$) gives Gaussian population distributions with $\mu_{\rm ej} = 1.26 \pm 0.01~M_\odot$ ($\sigma_{\rm ej} = 0.33 \pm 0.01~M_\odot$) and $\mu_{\rm Ni} = 0.64 \pm 0.06~M_\odot$ ($\sigma_{\rm Ni} = 0.42 \pm 0.02~M_\odot$). 
This work provides a step towards physical characterization of the local SN~Ia population while quantifying diversity and environmental dependencies relevant to progenitor physics and precision cosmology.

\end{abstract}
\begin{keywords}
supernovae: general -- cosmology: observations -- stars: white dwarfs -- transients: supernovae
\end{keywords}
\section{Introduction}
\label{sec:intro}

Type Ia supernovae (SNe~Ia) are fundamental to astrophysics and cosmology as the standardizable candles that revealed the accelerating expansion of the Universe~\citep{Riess1998,Schmidt1998, Perlmutter1999} and as major producers of iron-group elements~\citeg{Ruiter2025}. Their utility in cosmology relies on the empirical \lq{brighter-slower}\rq\ relation~\citep{Rust1974,Pskovskii1977,Phillips1993}, which standardizes peak luminosities based on light-curve decline rate, and the \lq{brighter-bluer}\rq\ relation~\citep{Riess1996,Tripp1998} which standardizes based on optical colour. However, understanding of the physical processes underlying these relations and their connections to explosion mechanisms remains incomplete, motivating detailed studies of SN~Ia explosion physics.

Despite their importance, a complete physical picture of SNe~Ia remains elusive. The most popular model involves the thermonuclear runaway of a carbon-oxygen white dwarf (WD) in a binary system, but the specific progenitor channel---single-degenerate (SD) accretion or double-degenerate (DD) merger---is highly debated~\citep[e.g.,][]{Hillebrandt2013, Maoz2014, Ruiter2025}. These channels are linked to distinct explosion physics. In Chandrasekhar-mass ($M_{\rm Ch} \approx 1.4$ M$_\odot$) models, the explosion proceeds via delayed detonation or deflagration-to-detonation transition~\citep{Khokhlov1991,Hoeflich1996}. In sub-$M_{\rm Ch}$ models, double-detonation is triggered on a less massive WD \citep{Fink2010,Sim2010}. Double detonations may also occur in dynamically driven merger scenarios, such as D6-like channels~\citep{Shen2018}. More exotic scenarios, such as violent WD mergers~\citep{Pakmor2010,Pakmor2012}, can produce normal or under-luminous SNe~Ia depending on the progenitor masses and densities. Differentiating these scenarios requires linking observable properties to the fundamental physical parameters of the explosion.

A key parameter distinguishing these models is the total ejected mass, $M_{\rm ej}$. Measuring the $M_{\rm ej}$ distribution for a large SN~Ia sample provides a direct test of the prevalence of competing explosion channels. However, previous efforts have been limited by small sample sizes; to our knowledge, the largest systematic study to date derived physical parameters for 337 SNe~Ia~\citep{Scalzo2014}. However, this was based on empirical relations for ejecta properties, which have known limitations~\citeg{Khatami2019}.

SN~Ia light curves are powered by radioactive decay of $^{56}$Ni synthesized in the explosion \citep{Colgate1969}. The synthesized nickel mass ($M_{\rm Ni}$) is a key contributor to peak luminosity among normal SNe~Ia, with larger $M_{\rm Ni}$ producing higher peak luminosity and longer photon diffusion timescales~\citeg{Hoeflich1996, Khatami2019}. This provides a plausible physical explanation for the brighter-slower relation: brighter SNe decline more slowly because they synthesize more nickel driving larger diffusion times due to higher temperatures or due to larger ejecta masses~\citeg{Wygoda2019}. However, $M_{\rm Ni}$ alone cannot explain the full width-luminosity relation; radiative-transfer calculations show that decline rate also depends on opacity, ejecta structure, ionisation evolution, composition, and viewing angle~\citep[e.g.,][]{Blondin2017}.

Beyond simple radioactively powered diffusion, colour evolution reveals additional physics through secondary maxima in red and near-infrared bands ($i$, $z$, $J$, $H$) occurring $\sim$30--40 days post-explosion~\citep{Kasen2006}. The physical origin is attributed to Fe~III $\rightarrow$ Fe~II recombination at $T \sim 5000$--7000~K, which changes the already wavelength-dependent opacity and emissivity structure: $\sim 10^5$ Fe~II absorption lines increase the opacity at blue wavelengths while red/near-IR wavelengths remain transparent \citep{Pinto2000,Hoflich2002}. As the ejecta cool, a recombination wave moves inward and, once it reaches iron-rich material, the change from doubly to singly ionised iron increases the emissivity at red and near-infrared wavelengths~\citep{Kasen2006}. This process requires stratification in the ejecta and therefore cannot be reproduced exactly by a one-zone model. Standard gray-opacity (i.e., wavelength and temperature independent) one-zone models cannot reproduce this behaviour, motivating wavelength-dependent extensions and multi-zone models, which can also capture other effects such as the distribution of $M_{\rm Ni}$~\citep{Sarin2026_mixing}.

The so-called Arnett's rule ~\citep{Arnett1982, Nadyozhin1994, Pinto2000} provides a relationship between peak luminosity and $M_{\rm Ni}$, which has been used to provide a quick estimate of the $M_{\rm Ni}$ synthesized in different SNe. However, these models' accuracy is limited by simplifying assumptions~\citep{Khatami2019}. Moreover, the rule does not allow for robust extraction of properties beyond $M_{\rm Ni}$ due to inherent degeneracies. Although other methods to estimating $M_{\rm Ni}$ also exist, each with their own limitations~\citep{Katz2013, Gaba2025}. A more robust approach towards extracting multiple ejecta parameters is to fit light curves with models incorporating wavelength-dependent opacity, simultaneously constraining $M_{\rm Ni}$, $M_{\rm ej}$, kinetic energy ($E_{\rm kin}$), and marginalising over uncertain parameters such as the optical and gamma-ray opacities. Ideally, this should also be performed on bolometric rather than multi-band light curves~\citep[e.g.,][]{Scalzo2014}, so that wavelength-dependent physics, such as the effect of recombination can be ignored. However, we often do not have sufficiently extensive multi-wavelength observations to construct the bolometric light curve, instead relying on some bolometric corrections.

The application of physical models has evolved with survey scale. Early studies on small CfA samples established the methodology \citep{Hicken2009}, while large surveys like SDSS-II ($N \sim 500$) enabled statistical studies of light-curve diversity \citep{Frieman2008}. The Nearby Supernova Factory provided precise $M_{\rm Ni}$ and $M_{\rm ej}$ measurements for smaller samples, revealing evidence for sub-$M_{\rm Ch}$ populations \citep{Scalzo2014}. However, detailed physical modelling of large samples ($N > 1000$) has remained challenging, with most analyses limited to empirical standardization (e.g., SALT2; \citealt{Guy2007}) and analyses of luminosity functions and rates~\citep{Desai2026}.

An additional detail revealed by empirical modelling is the host galaxy mass step: SNe~Ia in low-mass hosts ($\log_{10}(M_*/M_\odot) < 10$) exhibit $\sim$0.06 mag offset in Hubble residuals (the difference between observed and model brightness) after SALT2 standardization~\citep{Kelly2010,Sullivan2010, Lampeitl2010}. The physical origin behind this step remains unclear, with proposed explanations including progenitor age~\citep{Rigault2020}, metallicity effects~\citep{Childress2013}, dust variations~\citep{BroutScolnic2021}, or standardization errors~\citep{Howell2009, Ginolin2025}, or some combination of them all~\citep{Rose2021}. Measuring physical parameters ($M_{\rm Ni}$, $M_{\rm ej}$) as a function of host properties provides direct insight into whether this reflects intrinsic explosion physics or observational systematics.

The Zwicky Transient Facility (ZTF) Data Release 2 (DR2) presents the largest homogeneous sample of well-observed SNe~Ia ever assembled \citep{Rigault2025}, containing 3,628 spectroscopically confirmed SNe~Ia, of which 2,667 pass quality cuts for cosmological analysis. This represents an order-of-magnitude increase over previous surveys, enabling a transition from discovery of SN~Ia diversity to precise quantification of physical parameter distributions and detection of rare sub-populations~\citep{Dimitriadis2025}.

In this paper, we leverage the ZTF DR2 sample to perform the largest systematic analysis of SN~Ia physical properties to date. We develop a wavelength-dependent one-zone radioactive decay model incorporating, phenomenologically, Fe~II recombination physics to simultaneously model primary and secondary maxima in multi-band ($gri$) light curves. Using Bayesian inference, we fit 2,205 cosmology-quality SN~Ia light curves, deriving posterior distributions for physical properties for all events, including $M_{\rm Ni}$, $M_{\rm ej}$, $E_{\rm kin}$, ejecta velocity ($v_{\rm ej}$), and wavelength-dependent opacity parameters.

Our primary goals are: (1) construct statistically robust distributions of $M_{\rm Ni}$, $M_{\rm ej}$, and $E_{\rm kin}$ for the SN~Ia population, achieving $\sim 100\times$ improvement in number of events analysed over previous studies; (2) use the $M_{\rm ej}$ distribution to constrain relative contributions of $M_{\rm Ch}$ and sub-$M_{\rm Ch}$ explosion channels; (3) investigating the physical origin of the brighter-slower relation and host galaxy mass step by correlating physical parameters with SALT2 observables and host properties; (4) characterize wavelength-dependent opacity effects across the population; and (5) provide a characterization of the local SN~Ia population as a benchmark for theoretical models and future hierarchical Bayesian inference of population properties.

This paper is structured as follows. In Sec.~\ref{sec:sample}, we describe the ZTF DR2 dataset and the sample selection. In Sec.~\ref{sec:methods}, we present our wavelength-dependent one-zone model and Bayesian inference framework. In Sec.~\ref{sec:results}, we present physical parameter distributions, correlations with SALT2 parameters, and host galaxy dependencies. In Sec.~\ref{sec:discussion}, we discuss implications for progenitors, explosion models, and cosmology. We conclude in Sec.~\ref{sec:conclusions}.

\section{Sample Selection and Properties}
\label{sec:sample}

\begin{figure}
\centering
\includegraphics[width=0.48\textwidth]{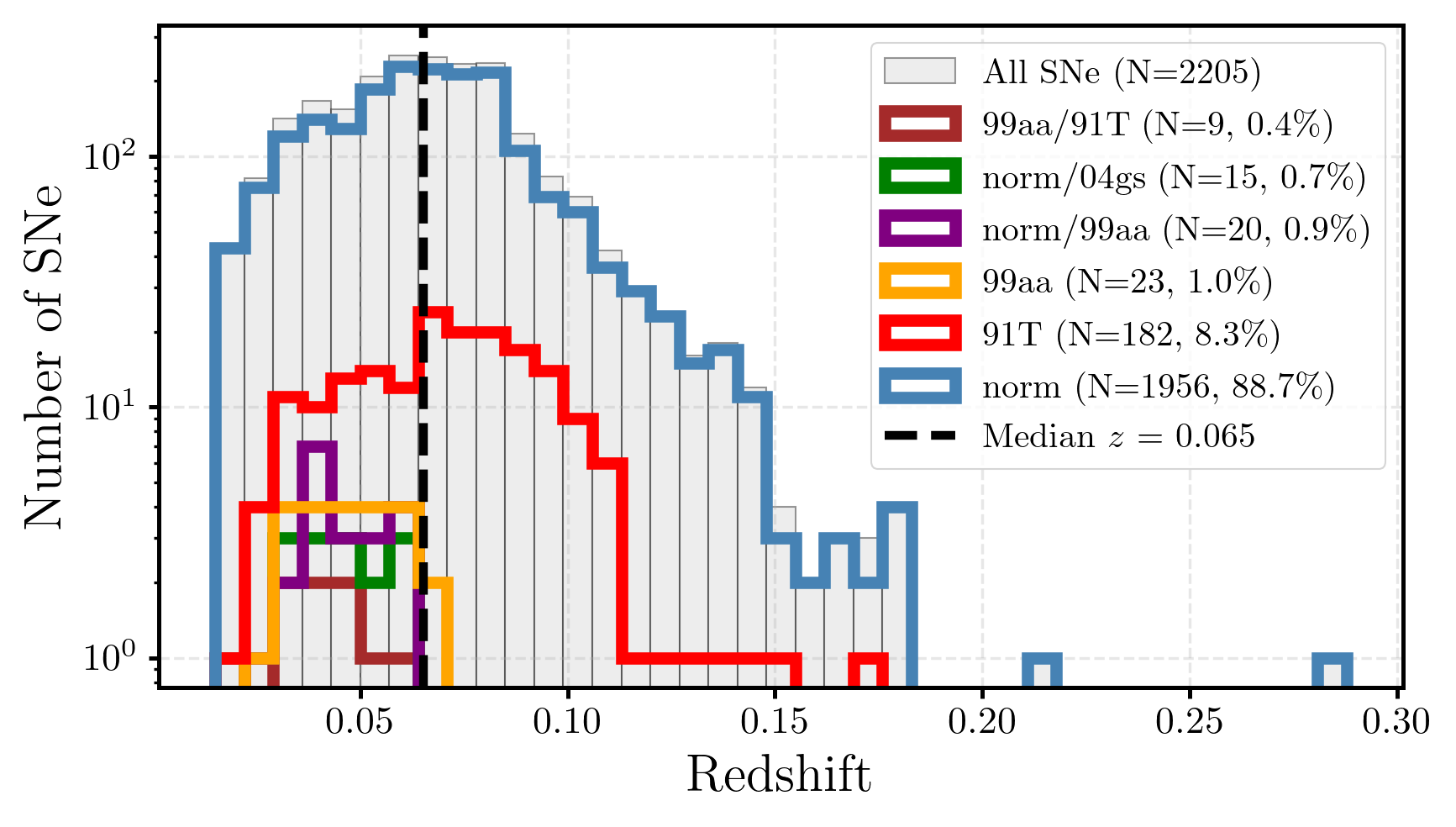}
\caption{Summary of the sample analysed in this work, their redshift distribution and subtypes. All classifications are from~\citet{Rigault2025}.}
\label{fig:sample}
\end{figure}

Our analysis draws on the Zwicky Transient Facility (ZTF) Data Release~2 (DR2) sample of spectroscopically confirmed Type~Ia supernovae~\citep{Rigault2025}. ZTF monitors the northern sky in custom $g$, $r$, and $i$ filters, delivering roughly three-day cadence photometry that tracks light curves from discovery to up to a $100$d post maximum. DR2 provides 3,628 SN~Ia light curves with homogeneous reductions, SALT2 fits \citep{Guy2007}, subtype classifications, as well as additional quantities, such as host and local galaxy mass measurements~\citep{Senzel2025, Burgaz2025, Ginolin2025}. Throughout this paper, we use properties extracted from this release, aside from the parameters estimated in the present work.

To obtain a uniform, well-observed set for light curve modelling we apply three quality cuts to the parent catalogue. We first retain objects flagged as \texttt{snia-cosmo}, leaving 2,800 of 3,628 SNe ($77\%$) that meet the ZTF cosmology standards and exclude clearly peculiar, unclassified, or subtypes not included in cosmological studies such as 91bg-like or Ia-csm events. We additionally require that the light curve is well sampled, which is already flagged in the release with a coverage flag \texttt{lccoverage\_flag = 1}. The events removed due to this constraint are typically high-redshift discoveries or those that are faint, leaving 2,354 SNe that satisfy the above constraints. We further require that data are well described by the empirical SALT2 model, also provided as a flag in the DR2 release, \texttt{fitquality\_flag = 1}. This selection removes an additional 149 objects ensuring we are left with a homogeneous sample of light curves with good coverage and belonging to the cosmological spectroscopic class~\citep{Rigault2025}. This implicitly also removes spectroscopic subtypes such as 91bg-like, SN~Iax, CSM-interacting, SN~2003fg-like, and other peculiar classes, leaving us with spectroscopically normal SNe~Ia together with the overluminous SN~1991T- and SN~1999aa-like events.

Our final working sample contains 2,205 SNe~Ia ($61\%$ of the parent catalogue), the largest homogeneous set of multi-band SN~Ia light curves analysed with a single physical model to date. In Figure~\ref{fig:sample} we show histograms of the redshift distribution of our full sample and for the different subtypes. We refer to these 2205 events as our selected sample. 

The redshift distribution spans $0.015 < z \leq 0.29$ with a median of $z=0.065$. Normal SNe~Ia dominate (1,956 objects, $88.7\%$), accompanied by 182 SN~1991T-like ($8.3\%$) and 23 SN~1999aa-like ($1.0\%$) events; the remaining 44 light curves show mixed or uncertain spectroscopic classifications. Our selection procedure does induce some bias that will be reflected in our final distributions. In particular, our sample does not include any subluminous SN~1991bg-like and SN~Iax explosions, which likely have a lower nickel mass, therefore biasing us towards larger ejecta and nickel mass estimates. Given that this sample is dominated by spectroscopically normal, cosmology-quality SNe~Ia, any inferred nickel fractions that appear inconsistent with normal SN~Ia spectra should be interpreted as possible modelling systematics rather than evidence that the cosmological sample is intrinsically non-normal.

The SALT2 parameters of the selected sample behave as expected for cosmology-quality light curves: the stretch distribution centres on $x_1 = 0.07$ with $44\%$ of events having $x_1 < 0$, while the colour distribution peaks at $c = -0.01$ with $-0.3 < c < 0.4$. Median peak magnitude is $m_r^{\rm peak} = 17.9$~mag, and $90\%$ of the events are brighter than 19.5~mag at maximum light. Host-galaxy stellar masses are available for 2,151 SNe ($97.6\%$) of the sample considered in this work~\citep{Rigault2025} and span $6.0 < \log_{10}(M_*/M_\odot) < 11.5$ with a median of 9.8; the sample is evenly divided across the canonical $\log_{10}(M_*/M_\odot) = 10$ mass step threshold \citep[e.g.,][]{Sullivan2010}, providing leverage on environmental trends.
These selection biases induced in our sample are mostly intrinsic to cosmology samples and precisely match the population used for Hubble diagram analyses. Throughout the bulk of the paper, we therefore interpret the derived physical parameters as representative of the ``cosmologically useful'' SN~Ia population rather than the full diversity of thermonuclear explosions. In Sec.~\ref{sec:hierarchicalinference}, we perform an analysis to constrain the true distributions by correcting for selection biases through hierarchical inference.

The 2,205-object sample analysed here is roughly two orders of magnitude larger than the largest earlier study that fitted physical models to SN~Ia light curves (e.g. \citealt{Bora2024}, which modelled 28 SNe) while retaining the advantages of a single systematic, non-targeted survey, with uniform photometry, and consistent modelling assumptions. The combination of size, homogeneity, and $gri$ coverage makes the ZTF DR2 cosmology sample an ideal laboratory for measuring population-level explosion properties and for revisiting long-standing questions about progenitor channels and environmental dependencies. We note that~\citet{Scalzo2014} analysed 334 supernovae to estimate ejecta properties of the population. However, their analysis relied on empirical relations rather than fits to the lightcurve.

\section{Methodology}\label{sec:methods}
Our aim is to extract physical model parameters from multi-band SN~Ia light curves from ZTF. We approach this by fitting the light curves with a one-zone radioactive decay model. Below, we describe the general features and limitations of the model, followed by the general procedure for inference that we apply to our full sample.
\subsection{Model}\label{sec:model_physics}
The cosmological subsample of SNe~Ia is powered by the radioactive decay of $^{56}$Ni (and subsequent species) as compared to other subtypes such as SNe~Ia-CSM, which have an additional power-source from the interaction of the SN ejecta with CSM.

The standard one-zone radioactive decay model sometimes also called the ``Arnett'' model~\citep{Arnett1982, Nadyozhin1994, Pinto2000}, describes the bolometric
luminosity of a spherical SN with ejecta, $M_{\rm ej}$, expanding homologously with characteristic velocity, $v_{\rm ej}$.

The instantaneous radioactive heating
rate from $^{56}$Ni and $^{56}$Co decay is given by, \begin{equation}
\dot{Q}(t) \approx M_{\rm Ni} \left[ \epsilon_{1} e^{-t/\tau_{\rm Ni}} + \epsilon_{\rm 2}
e^{-t/\tau_{\rm Co}} \right],
\end{equation}
where $M_{\rm Ni} = f_{\rm Ni} M_{\rm ej}$ is the $^{56}$Ni mass, with $f_{\rm Ni}$ the $^{56}$Ni fraction of the ejecta mass (i.e., a burning efficiency).
The specific heating rates are $\epsilon_{\rm 1} = 6.45 \times 10^{43}$ erg s$^{-1}$ M$_\odot^{-1}$ and
$\epsilon_{\rm 2} = 1.45 \times 10^{43}$ erg s$^{-1}$ M$_\odot$$^{-1}$, and the decay timescales are $\tau_{\rm Ni} = 8.8$ days and $\tau_{\rm Co} = 111.3$ days.

The observed bolometric luminosity is determined by photon diffusion through the ejecta and gamma-ray trapping efficiency, with
\begin{equation} L_{\rm bol}(t) =
-\frac{2}{t_{\rm diff}^2} \left[1 - e^{-A_\gamma(t)} \right] \int_0^t
\dot{Q}(t') t' e^{(t'^2 - t^2)/t_{\rm diff}^2} dt', \end{equation} where the
photon diffusion timescale is
\begin{equation}
t_{\rm diff} = \sqrt{\frac{2
\kappa M_{\rm ej}}{13.7 c v_{\rm ej}}},
\end{equation} where $\kappa$ is the optical/UV gray opacity, i.e., no dependence on temperature or wavelength.

We additionally account for gamma-ray leakage via, \begin{equation}\label{eq:gamma-rayleakage}
A_\gamma(t) = \frac{3 \kappa_\gamma M_{\rm ej}}{4\pi v_{\rm ej}^2 t^2},
\end{equation}
where $\kappa_\gamma$ is the effective gamma-ray opacity. This is the common parameterisation of gamma-ray leakage across a range of radioactive-decay powered transient models~\citep[e.g.,][]{Wang2015, Nicholl2017, Sarin2022, Omand2024}.
This term $[1 - e^{-A_\gamma(t)}]$ represents the fraction of gamma-rays that are trapped and
thermalised in the ejecta at time $t$. Early in the evolution, when the ejecta are optically thick to gamma-rays ($A_\gamma \gg 1$), nearly all gamma-ray energy is
deposited, while at late times ($A_\gamma \ll 1$), gamma-rays freely escape and the luminosity declines more steeply than a pure diffusion model would predict.

Standard implementations of one-zone radioactive decay models (such as the model described above) assume gray opacity and convert $L_{\rm bol}$ to photometry using a single-temperature blackbody emission (with the temperature estimated from Stefan-Boltzmann law). However, SNe~Ia exhibit secondary maxima in the red/near-IR bands ($i$, $z$, $J$, $H$) at $\sim$30--40 days~\citep[e.g.,][]{Elias81,Ford93, Krisciunas2001,Mandel09,Folatelli10,Mandel2022,Deckers2025}, which manifests as a re-brightening. The physical origin is theorized to be the Fe~III $\rightarrow$ Fe~II recombination at $T \sim 7000$~K, which causes a dramatic increase in near-infrared line emissivity and an effective wavelength-dependent opacity change that redistributes flux from UV/optical to longer wavelengths~\citep{Kasen2006}. This is intrinsically a stratified radiative-transfer effect and the one-zone prescription below should therefore be viewed as a phenomenological approximation to the observed colour evolution rather than a reliable model of the recombination wave.

We implement the effect of this recombination by modifying our opacity to become a function of wavelength $\lambda$ and time $t$. In particular, instead of a constant, gray opacity, we assume,
\begin{equation}
\kappa(\lambda, t) = \kappa_{\rm opt} \times \left[1 + \Delta\kappa(\lambda, t)\right]
\end{equation}
where $\kappa_{\rm opt}$ is the base optical opacity at $\lambda = 5000$~\AA, and
\begin{equation}
\Delta\kappa(\lambda, t) = 0.9 \times \tanh\left[s \cdot w(\lambda) \cdot g(t)\right]
\end{equation}
The wavelength factor is
\begin{equation}
w(\lambda) = -\frac{\lambda - \lambda_{\rm cen}}{\lambda_{\rm width}}
\end{equation}
with $\lambda_{\rm cen} = 6500$~\AA\ and $\lambda_{\rm width} = 2000$~\AA, such that blue wavelengths have $w > 0$ (increased opacity) and red wavelengths have $w < 0$ (decreased opacity). As this is also a time-dependent effect, we introduce the time dependence by folding in a Gaussian term centered on the recombination time, $t_{\rm recomb}$ that lasts for $\sigma_{\rm recomb}$, with an extra floor term that controls how fast the opacity returns to the level before recombination began. In particular,
\begin{equation}
g(t) = \exp\left[-\frac{(t - t_{\rm recomb})^2}{2\sigma_{\rm recomb}^2}\right] + f_{\rm floor} \cdot R(t) \cdot D(t)
\end{equation}
where $t_{\rm recomb}$ is the recombination time, $\sigma_{\rm recomb}$ is the width, $f_{\rm floor}$ controls late-time effects, $R(t)$ is a rise function that activates after $t_{\rm recomb}$ with timescale $2\sigma_{\rm recomb}$, and $D(t)$ is an exponential decay beginning at $t_{\rm decay} = t_{\rm recomb} + 10$ days with decay timescale $\tau_{\rm decay} = 3\sigma_{\rm recomb}$. The parameter $s$ (``opacity strength'') controls the amplitude of the changing opacity, with $\tanh$ ensuring saturation at $\pm 90\%$ even for large $s$. We stress that these modifications are all phenomenological, and motivated by the real physics seen in radiative-transfer simulations~\citep[e.g.,][]{Kasen2006, Blondin2022}, which is otherwise too computationally expensive to implement in inference-capable models.

Our photospheric properties are now derived similarly to the standard blackbody assumption, but both the temperature and radius of the photosphere are now also wavelength dependent. In particular,
\begin{equation}
R_{\rm ph}(\lambda, t) = v_{\rm ej} t \times \sqrt{\frac{\kappa(\lambda, t)}{\kappa_{\rm opt}}},
\end{equation}
while the photospheric temperature is
\begin{equation}
T_{\rm ph}(\lambda, t) = \left[\frac{L_{\rm bol}(t)}{4\pi \sigma_{\rm SB} R_{\rm ph}(\lambda, t)^2}\right]^{1/4}.
\end{equation}

For each wavelength and time, we can compute blackbody flux density via,
\begin{equation}
F_\lambda(\lambda, t) = \frac{2\pi h \nu^3}{c^2} \frac{1}{e^{h\nu/k_B T_{\rm ph}(\lambda,t)} - 1} \times \frac{R_{\rm ph}(\lambda, t)^2}{d_L^2} \times (1 + z)
\end{equation}
where $d_L$ is the luminosity distance, assuming Planck18 cosmology~\citep{Planck2020} with redshift, $z$. We note a cosmology consistent to one inferred from SN Ia will produce lower nickel mass estimates than our analysis, given the smaller distances for the same redshift.
The SED is constructed on a grid spanning 100--60,000~\AA\ (500 points), then integrated through ZTF $g$, $r$, $i$ filter bandpasses to produce synthetic magnitudes. Critically, our phenomenological modification enables us to capture the effects of re-brightening at longer wavelengths, as the lower opacity in red bands yields smaller $R_{\rm ph}$ and higher $T_{\rm ph}$, producing a brighter flux. This prescription is one-zone and phenomenological, so the absolute nickel fractions should be interpreted with caution and checked against bolometric estimates where possible; parameters such as $M_{\rm ej}$ and $v_{\rm ej}$ retain their usual diffusion-model meaning, but can remain correlated with assumptions about the SED and opacity prescription.

In total, our model has 10 free physical parameters. Four of these are explosion parameters, $M_{\rm ej}$ (ejecta mass, M$_\odot$), $f_{\rm Ni} = M_{\rm Ni}/M_{\rm ej}$ (nickel fraction), $v_{\rm ej}$ (ejecta velocity, km s$^{-1}$), $t_0$ (explosion time, MJD). Five parameters describe the opacity and recombination effects, $\kappa_{\rm opt}$ (base opacity at 5000~\AA, cm$^2$ g$^{-1}$), $s$ (wavelength-dependent strength), $t_{\rm recomb}$ (recombination epoch, days), $\sigma_{\rm recomb}$ (width, days), $f_{\rm floor}$ (late-time amplitude). We also include an additional parameter to capture the effect of host-extinction from dust, $A_V$ (host extinction, mag) which we model with an $R_V = 3.1$ following~\citet{Fitzpatrick1999}, we fix the milky-way extinction to the same value as derived in~\citet{Rigault2025}. Throughout, we fix the gamma-ray opacity, $\kappa_\gamma = 0.03$ cm$^2$ g$^{-1}$ ($\gamma$-ray opacity), a standard value in analyses of SNe~Ia~\citeg{Scalzo2014}. We note that SN cosmology analyses now routinely use a distribution of $R_V$ instead of a fixed value, but we expect our choice to have little impact on our results given all other uncertainties. 

\subsection{Inference}\label{sec:inference}
The DR2 release contains light curves of all supernovae. However, the data are not homogeneous, and include significant outliers, and flux limits and background detections well before and after a supernova. To reduce the impact of such erroneous data on our inference, we first restrict our ZTF $gri$ photometry to the window from $-20$ to $+40$ days from maximum light. Although the prior on $t_{\rm recomb}$ extends to $+45$ days, events without late-time i-band coverage have weak constraints on this parameter, in such cases the posterior is prior-driven on all recombination-related parameters. 
For several objects, this photometry still contains significant outliers and scattered and inconsistent data points, so we include an additional magnitude error of $0.05$ mag as a systematic noise, which is added in quadrature to the original data errors. To further safeguard against outliers significantly affecting our fit, we adopt a mixture Gaussian likelihood, where our likelihood is defined as,
\begin{align}
\mathcal{L}_i = \frac{\alpha}{\sqrt{2\pi\sigma_{m,i}^2}} \exp\left[-\frac{(m_i - m_{\rm model,i})^2}{2\sigma_{m,i}^2}\right] \\ \nonumber
+  \frac{1-\alpha}{\sqrt{2\pi\sigma_{\rm out}^2}} \exp\left[-\frac{(m_i - m_{\rm model,i})^2}{2\sigma_{\rm out}^2}\right]
\end{align}
where $\alpha$ is the inlier fraction and $\sigma_{\rm out}$ is the outlier noise. The total log-likelihood is,
\begin{equation}
\ln \mathcal{L}_{\rm total} = \sum_{i=1}^{N_{\rm obs}} \ln \mathcal{L}_i.
\end{equation}
This likelihood effectively deals with heavily scattered data points, ensuring that they do not affect the fitting by simply assigning such data points to be part of the outlier distribution.

In Table~\ref{tab:priors} we summarise the full set of parameters and prior distributions, including the $10$ model parameters and the $2$ parameters in the likelihood.
\begin{table}
\centering
\caption{Prior distributions for model parameters.}
\label{tab:priors}
\begin{tabular}{lccc}
\hline\hline
Parameter & Symbol & Prior & Range \\
Ejecta mass & $M_{\rm ej}$ & U & 0.01--2.0 M$_\odot$ \\
Nickel fraction & $f_{\rm Ni}$ & U & 0.001--1.0 \\
Velocity & $v_{\rm ej}$ & U & 1,000--40,000 km s$^{-1}$ \\
Explosion time & $t_0$ & U & $t_{\rm first} - 30$ to $t_{\rm first} - 0.1$ d \\
Optical opacity & $\kappa_{\rm opt}$ & U & 0.05--0.15 cm$^2$ g$^{-1}$ \\
Opacity strength & $s$ & LogU & 0.05--10 \\
Recombination time & $t_{\rm recomb}$ & U & $t_{\rm peak} + 15$ to $t_{\rm peak} + 45$ d \\
Width & $\sigma_{\rm recomb}$ & U & 2--15 days \\
Late-time floor & $f_{\rm floor}$ & U & 0.001--1.0 \\
$A_V$ & $A_V$ & U & 0--3 mag \\
\hline
Inlier fraction & $\alpha$ & U & 0.85--0.99 \\
Outlier noise & $\sigma_{\rm out}$ & U & 5--10 mag \\
\hline
\end{tabular}
\end{table}

To sample the posterior we use the model and likelihoods implemented in {\sc Redback}~\citep{Sarin2024} with the the nested sampler {\tt PyMultiNest}~\citep{Feroz2009, Buchner2016} via the {\sc Bilby}~\citep{Ashton2019} interface. Each supernova analysis takes $\approx 10$ minutes on a single {\tt EPYC 7543} CPU, i.e., the entire sample can be analysed in $\approx 370$ CPU-hours.

For all supernovae, we compute additional posteriors on the nickel mass and kinetic energy. As a relative fit-quality diagnostic, we define $\chi^2_{\rm pseudo}/\nu=-2\ln\mathcal{L}_{\max}/(N_{\rm data}-N_{\rm model})$, where $\mathcal{L}_{\max}$ is the maximum sampled likelihood and $N_{\rm model}$ excludes the two mixture-likelihood nuisance parameters. This is not a conventional reduced $\chi^2$: the fitted Gaussian-mixture likelihood includes normalization and outlier terms, so its absolute value has no standard $\chi^2$ interpretation and may be negative. We use it only to rank fits evaluated with the same likelihood. For all parameters we obtain full posterior samples, however to simplify our analyses in later sections, for the majority of the following results we work exclusively with summary statistics for each event such as the posterior median, standard deviation, and $68\%$ credible intervals. 

\begin{figure*}
\centering
\includegraphics[width=0.98\textwidth]{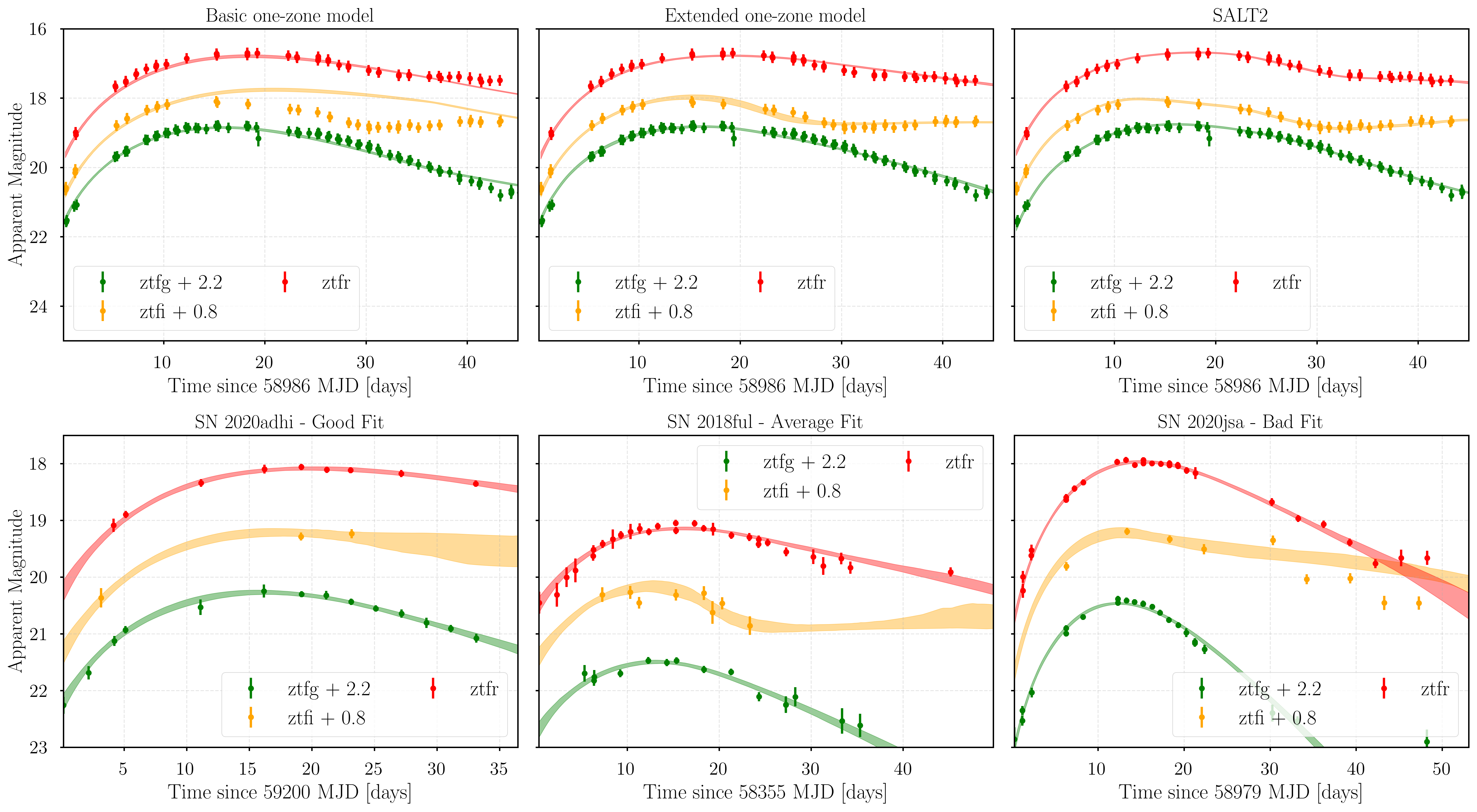}
\caption{Top panels: Light curve fits for the same supernova, SN~2020kej for the basic one-zone, extended model discussed in this paper, and the empirical SALT model, respectively. Bottom panels: Extended-model fits nearest the 10th (left), 50th (middle), and 90th (right) percentiles of $\chi^2_{\rm pseudo}/\nu$ among events with $i$-band coverage.} The model successfully captures primary peaks, decline rates, and secondary maxima across the diversity of ZTF light curves.
\label{fig:fit_quality}
\end{figure*}

\begin{table*}
\centering
\scriptsize
\setlength{\tabcolsep}{2pt}
\caption{Injection-and-recovery validation using synthetic ZTF Ia observations. The evidence offsets are quoted relative to the best recovery model for each synthetic event. Recovered parameters are posterior medians with $68\%$ credible intervals; injected rows show true values. The Arnett rows use a grey bolometric model and therefore test the impact of removing the wavelength-dependent opacity structure.}
\label{tab:ztf_validation}
\begin{tabular}{p{0.24\textwidth}p{0.11\textwidth}p{0.12\textwidth}ccccc}
\hline
Case & Recovery & Distance & $\Delta\ln Z$ & $f_{\rm Ni}$ & $M_{\rm Ni}$ & $M_{\rm ej}$ & $v_{\rm ej}$ \\
 &  &  &  &  & ($M_\odot$) & ($M_\odot$) & (km s$^{-1}$) \\
\hline
Fiducial truth; 65 detections & Injected & $H_0=73.04$ & -- & 0.45 & 0.56 & 1.25 & 11000 \\
Fiducial & Opacity model & $H_0=73.04$ & $-0.27$ & $0.71^{+0.18}_{-0.20}$ & $0.60^{+0.07}_{-0.05}$ & $0.86^{+0.33}_{-0.15}$ & $11045^{+151}_{-146}$ \\
Fiducial & Opacity model & marginalised $H_0$ & 0.00 & $0.72^{+0.17}_{-0.21}$ & $0.64^{+0.08}_{-0.06}$ & $0.90^{+0.35}_{-0.16}$ & $11438^{+405}_{-373}$ \\
Fiducial & Arnett model & marginalised $H_0$ & $-230.12$ & $0.78^{+0.15}_{-0.24}$ & $0.52^{+0.04}_{-0.03}$ & $0.67^{+0.29}_{-0.12}$ & $10823^{+390}_{-341}$ \\
High-$M_{\rm Ni}$ sparse truth; 44 detections & Injected & $H_0=73.04$ & -- & 0.64 & 0.90 & 1.40 & 11000 \\
High-$M_{\rm Ni}$ sparse & Opacity model & $H_0=73.04$ & $-0.25$ & $0.79^{+0.14}_{-0.19}$ & $0.88^{+0.08}_{-0.05}$ & $1.14^{+0.37}_{-0.18}$ & $10992^{+207}_{-200}$ \\
High-$M_{\rm Ni}$ sparse & Opacity model & marginalised $H_0$ & 0.00 & $0.80^{+0.14}_{-0.18}$ & $0.95^{+0.10}_{-0.08}$ & $1.21^{+0.35}_{-0.20}$ & $11353^{+422}_{-373}$ \\
High-$M_{\rm Ni}$ sparse & Arnett model & marginalised $H_0$ & $-15.98$ & $0.79^{+0.14}_{-0.21}$ & $0.85^{+0.06}_{-0.05}$ & $1.08^{+0.38}_{-0.17}$ & $12092^{+468}_{-392}$ \\
\hline
\end{tabular}
\end{table*}

\begin{figure*}
\centering
\includegraphics[width=0.95\textwidth]{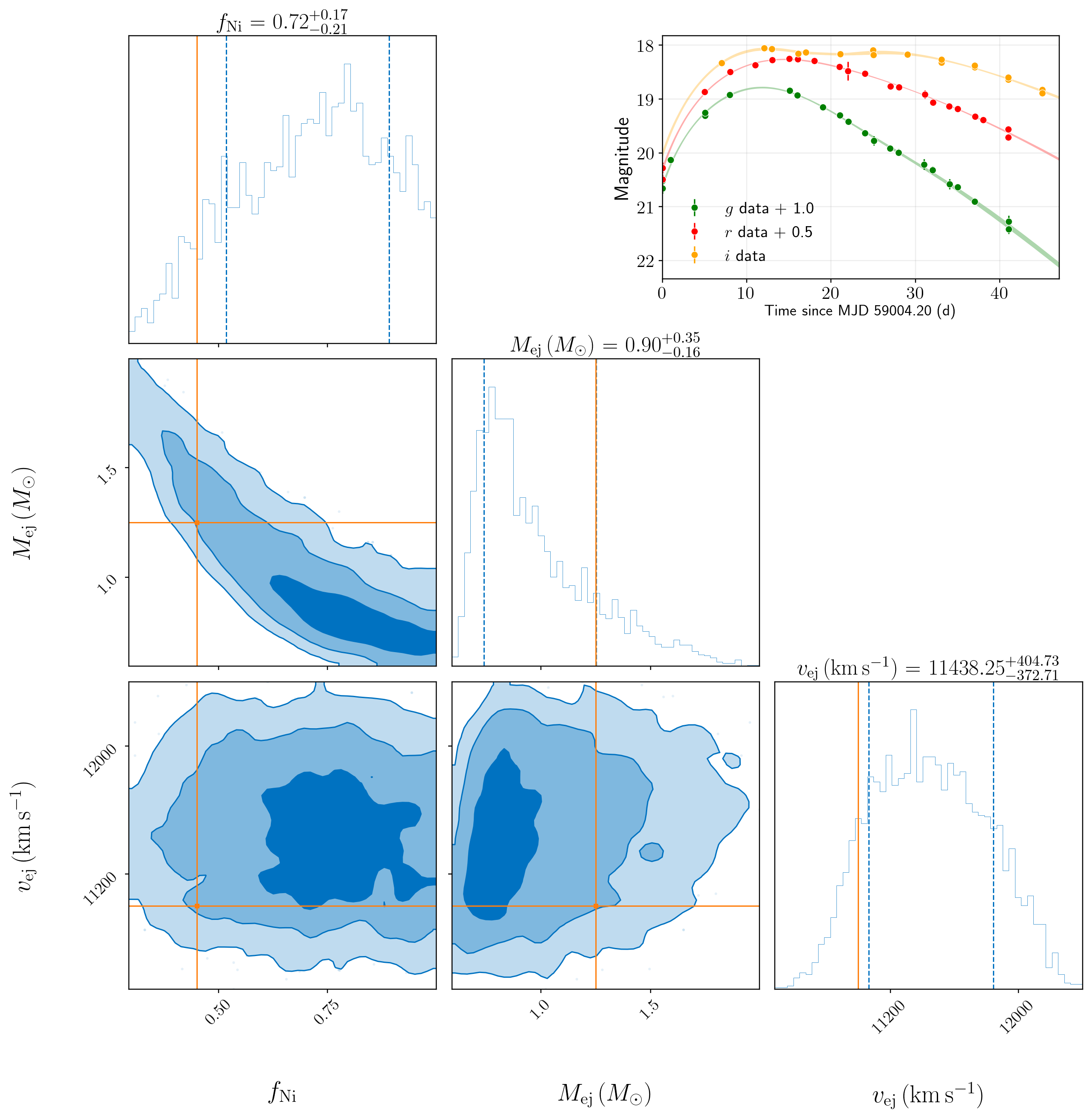}
\caption{Validation diagnostic for the fiducial ZTF-like injection recovered with the wavelength-dependent opacity model while marginalising over $H_0$. Orange lines mark the injected values and blue contours show the recovered posterior for $f_{\rm Ni}$, $M_{\rm ej}$, and $v_{\rm ej}$. The upper-right panel shows the corresponding posterior-predictive light curve as a function of time since the first synthetic observation. Points are the synthetic ZTF observations with additive band offsets and shaded regions are the central $90\%$ credible intervals from posterior draws. The quoted intervals are the $68\%$ credible interval on each parameter. The figure shows that the multi-band light curve is well reproduced while the physical parameters remain broad and degenerate, especially through the anti-correlation between $M_{\rm ej}$ and $f_{\rm Ni}$. This reinforces the need to treat posterior medians as summary statistics rather than complete descriptions of the individual-event physical constraints.}
\label{fig:validation_corner}
\end{figure*}

In Fig.~\ref{fig:fit_quality}, we show some representative fits from our analysis of the sample. In particular, in the top row, we show our analysis of SN~2020kej, where the three panels show fits with the basic one-zone model (without wavelength dependent opacities), the extended model described above, and the empirical SALT2 model, respectively. The bottom row shows events nearest the 10th, 50th, and 90th percentiles of $\chi^2_{\rm pseudo}/\nu$ after restricting to events with $i$-band coverage. These fixed quantiles were chosen to show a good, typical, and relatively poor fit while retaining the red-band information relevant to the opacity prescription. The top panels also demonstrate that the extended model works and performs almost identically to the empirical model, at least for the multi-wavelength data analysed here, providing confidence in our model. 
We note that more detailed multi-wavelength modelling and inference is required to investigate whether our model can correctly characterise the full SED of a SN~Ia at these phases, where we already know the SALT template works well; however, these fits provide good confidence for the analysis of the ZTF sample in this work.

\subsection{Validation with a ZTF-like synthetic event}
\label{sec:ztfvalidation}
To test the inference pipeline in a controlled setting, we perform injection-and-recovery analyses using synthetic observations generated using the observing logs of ZTF via {\sc Redback}. We generate the synthetic events with the same wavelength-dependent model described in Sec.~\ref{sec:model_physics}, include host extinction with $A_V=0.08$ mag, use a SN-calibrated value of $H_0=73~{\rm km~s^{-1}~Mpc^{-1}}$, and apply the ZTF limiting-magnitude and detection cuts in the simulation. This validation is not intended to be a population-level simulation study, but rather a controlled check that the fitting machinery, distance treatment, cadence, and opacity prescription behave sensibly for realistic ZTF-like data.
We consider two representative cases. The first is a fiducial SN~Ia with $M_{\rm ej}=1.25~M_\odot$, $M_{\rm Ni}=0.56~M_\odot$, and $v_{\rm ej}=11000~{\rm km~s^{-1}}$, producing 65 detected points split approximately evenly across $gri$ bands. The second is a more luminous event with $M_{\rm ej}=1.40~M_\odot$, $M_{\rm Ni}=0.90~M_\odot$, and the same ejecta velocity, but with the cadence thinned to resemble the typical ZTF sampling of our real sample: 20 $g$-band, 20 $r$-band, and only 4 early $i$-band detections. This second case tests whether the conclusions depend on having unusually informative red-band coverage.
For each synthetic event we fit the data with the injected wavelength-dependent opacity model and with a grey Arnett model. For each recovery model we repeat the fit with the correct distance scale, with an intentionally inconsistent Planck-like value of $H_0=67.67~{\rm km~s^{-1}~Mpc^{-1}}$, and with $H_0$ marginalised over a broad prior. Table~\ref{tab:ztf_validation} summarises the most relevant cases. The full wavelength-dependent model is strongly preferred for the fiducial event, with $\Delta\ln Z\simeq230$ relative to the grey Arnett recovery, and remains preferred for the sparsely sampled high-$M_{\rm Ni}$ case, with $\Delta\ln Z\simeq16$. The recovered $M_{\rm Ni}$ values are close to the injected values, while the recovered $M_{\rm ej}$ lies below the injected value in both selected realisations. The wrong fixed distance scale shifts the recovered nickel mass upward by $\simeq15$-$16\%$, illustrating the sensitivity of the absolute nickel-mass scale to the adopted cosmology. 
These tests support three conclusions that are important for interpreting the full sample. First, distance assumptions directly affect the absolute nickel mass, with a shift of $\approx 15\%$ between a Planck and an SN cosmology. Second, wavelength-dependent opacity is needed to reproduce the multi-band structure, especially the red-band shoulder and secondary-maximum region, although sparse $i$-band data weakens the evidence.  In Fig.~\ref{fig:validation_corner}, we show the corner plot which illustrates the dominant degeneracy between $M_{\rm ej}$ and $f_{\rm Ni}$, while the light-curve panel shows that broad, degenerate regions of parameter space can still produce acceptable fits to the ZTF-like $gri$ data. Notably, the input value for burning efficiency is lower than the inferred max posterior but consistent with the true input at the $90\%$ credible interval (and similarly for $M_{\rm ej}$), showcasing how individual-event posterior estimates can easily lead to the wrong interpretation across a full large sample with broad posteriors.

\subsection{Pseudo-bolometric Arnett comparison}
\label{sec:pseudobolometric}
As an independent analysis of the ejecta properties compared to full multi-band model, we also construct pseudo-bolometric light curves from the empirical SALT2 fits supplied with ZTF DR2. In particular, for each SN, we evaluate the fitted-colour SALT2 SED on a rest-frame wavelength grid from $2000$--$9200$~\AA\ over phases $-10$ to $+45$ days relative to SALT2 maximum, deredden the SED for Milky Way extinction using the tabulated $E(B-V)_{\rm MW}$ values, integrate the SED over wavelength, and convert the resulting flux to luminosity using $H_0=73.7~{\rm km~s^{-1}~Mpc^{-1}}$. We assign a conservative $5\%$ luminosity uncertainty to the reconstructed pseudo-bolometric points.
We then fit these pseudo-bolometric light curves with the grey-opacity one-zone model in Sec.~\ref{sec:model_physics}, allowing $M_{\rm ej}$, $f_{\rm Ni}$, $v_{\rm ej}$, the optical opacity, gamma-ray opacity to vary. We use an identical prior on all parameters as our full analysis on the photometry. This analysis deliberately removes the wavelength-dependent SED modelling from the inference and asks what bolometric diffusion parameters are required to reproduce the luminosity scale and temporal width implied by the SALT2 reconstruction.
We stress that the nickel masses from this procedure should be interpreted as a lower limit. The reconstructed luminosity excludes flux outside the adopted wavelength range, depends on the SALT2 template and colour treatment, and does not correct separately for host extinction or UV/NIR missing-flux uncertainties. If radioactive decay is the only power source, any unaccounted luminosity would increase the inferred $M_{\rm Ni}$. The corresponding $M_{\rm ej}$ and $v_{\rm ej}$ estimates are diffusion-model parameters from the same bolometric fit, but their interpretation remains tied to the assumed opacity and Arnett model.
We use this pseudo-bolometric analysis in two ways below. First, it provides an independent comparison for the full multi-band physical parameters in Figs.~\ref{fig:mass_distributions} and~\ref{fig:mass_plane}. Second, because the pseudo-bolometric analysis gives systematically lower nickel masses while the individual full-model posteriors are broad, it motivates the hierarchical treatment in Sec.~\ref{sec:hierarchicalinference}, as population-level inference is the correct way to interpret the broad, model-dependent individual-event posteriors rather than relying solely on point estimates.

\section{Results}\label{sec:results}
In this section, we showcase our results from analysis on the entire sample of 2,205 events. In particular, we will first present results in relation to the physical parameter distributions, followed by correlations with SALT parameters and host-galaxy properties. Finally, we highlight some results from analysis on a redshift-limited sub-sample ($z < 0.06$) where the DR2 sample is believed to be relatively complete (at least for the SALT2 model). We primarily present the results here and leave the discussion to Sec.~\ref{sec:discussion}. In the Appendix, we present some additional analyses where we fit the light curve of SN~2011fe. This reproduces results from different analyses and exploring the impact of gamma-ray opacities on our results. Where relevant, we compare the full multi-band model to the pseudo-bolometric Arnett analysis described in Sec.~\ref{sec:pseudobolometric}.

\subsection{Physical Parameter Distributions}
\label{sec:physics}
\begin{figure*}
\centering
\includegraphics[width=0.95\textwidth]{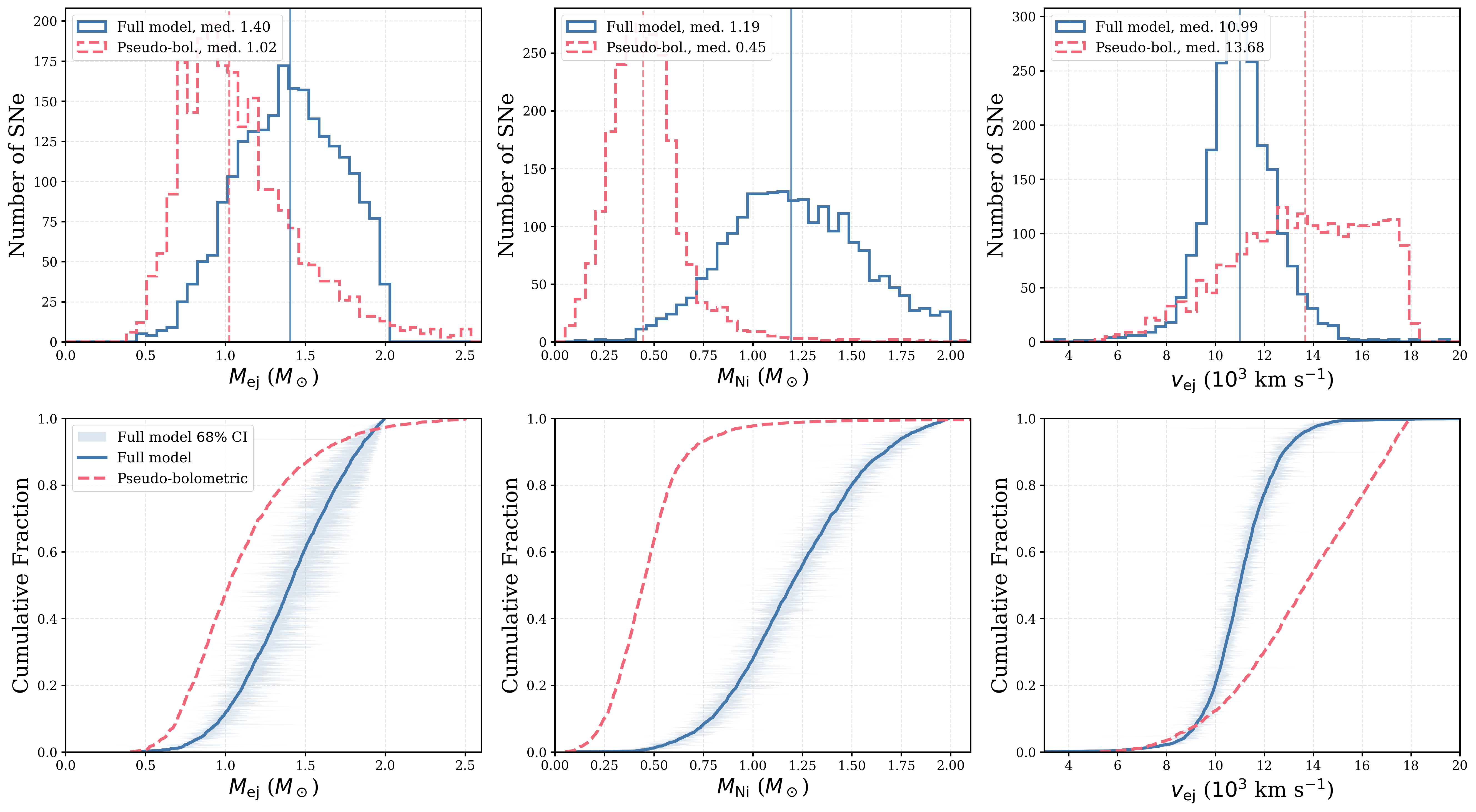}
\caption{Top row: Distributions of ejecta mass, nickel mass, and ejecta velocity for 2,205 SNe~Ia. Bottom row: Cumulative distributions showing the uncertainty bands ($68\%$ credible interval). Each panel compares the baseline full multi-band wavelength-dependent model with fixed $\kappa_\gamma$ to the pseudo-bolometric Arnett fits. The full model peaks near $M_{\rm ej}\simeq1.40~M_\odot$ and $M_{\rm Ni}\simeq1.19~M_\odot$, while the pseudo-bolometric analysis gives lower median values of $M_{\rm ej}\simeq1.02~M_\odot$ and $M_{\rm Ni}\simeq0.45~M_\odot$ and a median velocity of $1.37\times10^4~{\rm km~s^{-1}}$. The CDF uncertainty bands show the full-model posterior uncertainty; the pseudo-bolometric distributions are maximum-likelihood summaries. We stress that the histograms hide the true uncertainty represented by the individual posteriors, which is reflected partially by the coloured region in the CDF below.}
\label{fig:mass_distributions}
\end{figure*}

We first focus on the physical parameters that can be extracted directly from the light curves themselves. In Fig.~\ref{fig:mass_distributions}, we show histograms and cumulative distribution functions (CDF) for ejecta mass, nickel mass, and ejecta velocity of the full sample. Each panel compares the parameters inferred from the baseline full multi-band wavelength-dependent model with fixed $\kappa_\gamma=0.03~{\rm cm^2~g^{-1}}$ to those inferred from the pseudo-bolometric Arnett fits. The cumulative distribution function (CDFs) also incorporate the $68\%$ credible region from the posterior which is highlighted by the shaded region for the full multi-band model; the pseudo-bolometric Arnett distributions are maximum-likelihood summaries and do not include posterior uncertainty bands. The ejecta mass distribution from the full model has a median of $M_{\rm ej}=1.40~M_{\odot}$ with a standard deviation $0.32~M_{\odot}$, while the pseudo-bolometric Arnett fit gives a lower median $M_{\rm ej}=1.02~M_\odot$ with a broad distribution. 
The distribution itself spans 0.9--1.9 M$_\odot$ (5th--95th percentiles), and the percentages of the selected population that lie within the three relevant physical regimes are: 28.2\% sub-$M_{\rm Ch}$ ($M_{\rm ej} < 1.2$ M$_\odot$), 32.9\% near-$M_{\rm Ch}$ ($1.2 < M_{\rm ej} < 1.5$ M$_\odot$), and 38.9\% super-$M_{\rm Ch}$ ($M_{\rm ej} > 1.5$ M$_\odot$). 
For the pseudo-bolometric Arnett fits, the corresponding fractions are $69.0\%$ sub-$M_{\rm Ch}$, $17.5\%$ near-$M_{\rm Ch}$, and $13.5\%$ super-$M_{\rm Ch}$. We stress that the histograms hide the true uncertainty represented by the individual posteriors, and caution against drawing strong conclusions on the population from these point estimates on broad posteriors. The difference between the full-model and pseudo-bolometric summaries demonstrates the level at which SED, opacity, distance, and bolometric-correction assumptions can shift point-estimate population summaries.

The nickel mass distribution for the selected population has a mean $M_{\rm Ni}= 1.21~M_{\odot}$ with a standard deviation of $0.3$ M$_\odot$, higher than the canonical $\sim$0.6~M$_\odot$~\citep[e.g.,][]{Stritzinger2006, Scalzo2014}. However, this is consistent with constraints from nebular-phase spectra~\citep{Childress2015}, especially when considering potential systematic uncertainties in both modelling approaches. 
As we will discuss later, this high estimate is driven by selection biases as our sample excludes sub-luminous SNe~Ia (91bg-like, Iax) that would populate the low-$M_{\rm Ni}$ regime, and our sample is not volume complete, so under-luminous events (with less nickel) are more likely to be missed. However, some differences could be due to a modelling systematic or a product of marginalisation over flat, wide priors, such as the velocity. In the Appendix, we explore the impact of changing the gamma-ray opacity, which shifts this median down to $\approx 1$ M$_{\odot}$. The pseudo-bolometric Arnett analysis gives a much lower median nickel mass, $M_{\rm Ni}=0.45~M_\odot$, and a median nickel fraction $f_{\rm Ni}=0.43$. This estimate is based on a finite-wavelength pseudo-bolometric reconstruction and excludes the host-galaxy extinction and therefore serves as a conservative lower-limit comparison rather than a replacement for the full multi-band model. The offset between this scale and the full multi-band point estimates should therefore be read as evidence that the inferred nickel masses are sensitive to SED, extinction, distance, and opacity assumptions, rather than as a direct contradiction between two complete physical measurements. Moreover, combined with Fig.~\ref{fig:validation_corner}, it exemplifies how the broad posterior on the nickel fraction can skew posterior summaries higher than they otherwise would be.
The ejecta velocity distribution has a mean velocity $v_{\rm ej} = 11,060$ and standard deviation $1,596$ km s$^{-1}$, consistent with spectroscopic Si~II measurements~\citep[e.g.,][]{Foley2011, Mandel2014, Siebert2020, Dettman2021}. The pseudo-bolometric fits prefer a broader and somewhat higher velocity distribution, with median $v_{\rm ej}=1.37\times10^4~{\rm km~s^{-1}}$, reflecting the remaining diffusion-time degeneracy between $M_{\rm ej}$, $v_{\rm ej}$, and opacity when fitting bolometric luminosities alone.
The distributions can be summarised by single-population distributions, but this does not demonstrate that the underlying population is homogeneous or exclude substructure such as normal- and high-velocity groups, as also seen in ZTF DR2 and other Ia samples~\citep{Dettman2021, Burgaz2025_spec_diversity}. A hierarchical analysis with a mixture-model is necessary to test the existence of such subpopulations robustly. We note that given the width of the posteriors such constraints would be prior-driven.

\begin{figure*}
\centering
\includegraphics[width=0.95\textwidth]{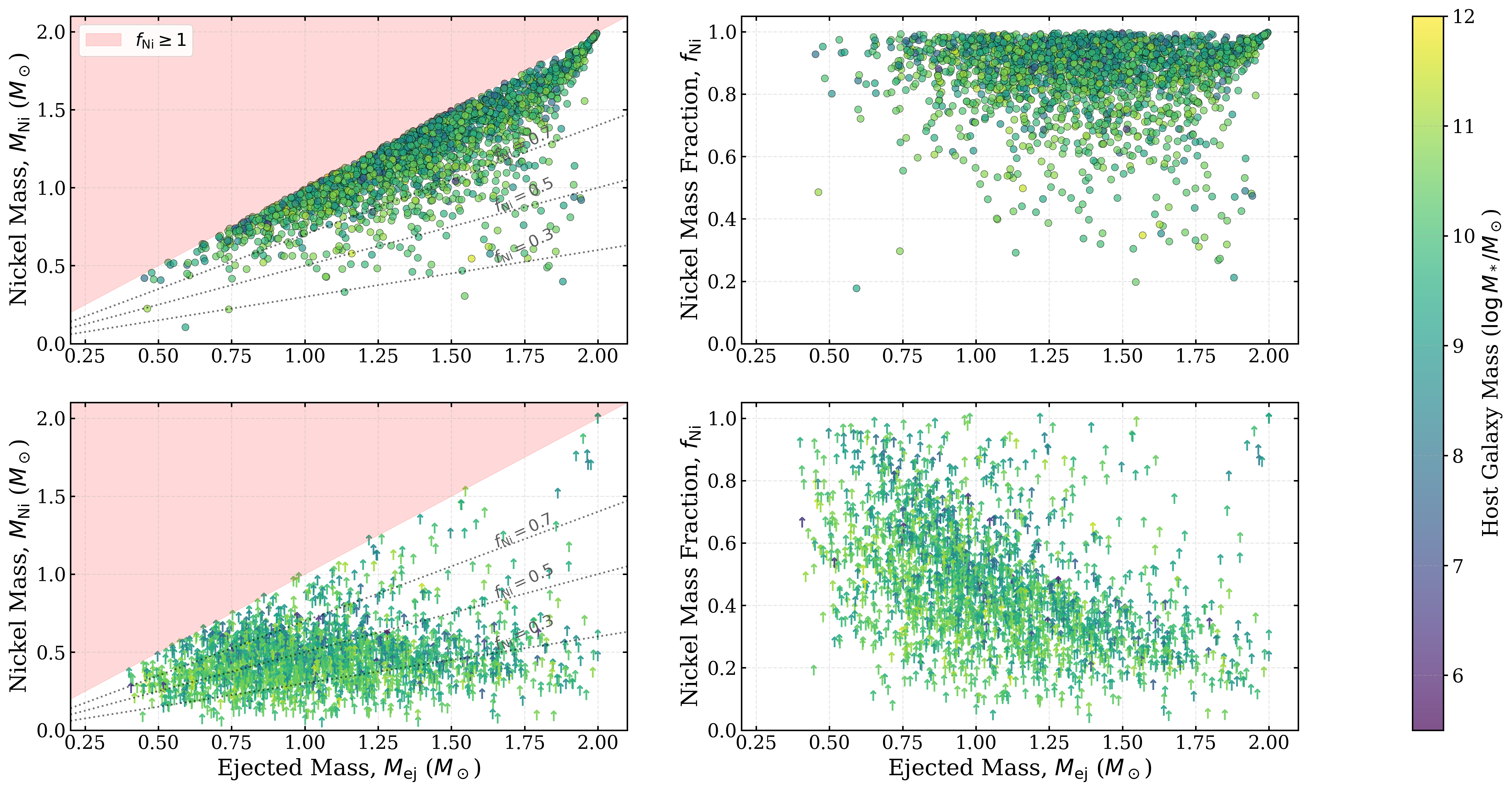}
\caption{Left: The SN~Ia mass plane ($M_{\rm Ni}$ vs $M_{\rm ej}$), coloured by host galaxy mass. All events respect the physical prior constraint $M_{\rm Ni} < M_{\rm ej}$ (forbidden region shaded red). Gray dotted diagonal lines show constant burning efficiency. Right: Burning efficiency $f_{\rm Ni} = M_{\rm Ni}/M_{\rm ej}$ versus ejecta mass. The top row shows the baseline full multi-band wavelength-dependent model with fixed $\kappa_\gamma$. The bottom row shows the pseudo-bolometric Arnett fits, with the nickel masses plotted as upward arrows to indicate their lower-limit interpretation. Both panels share the host mass colour bar with no strong indication of correlation with host-mass.}
\label{fig:mass_plane}
\end{figure*}

In Fig.~\ref{fig:mass_plane} we show the $M_{\rm Ni}$--$M_{\rm ej}$ plane coloured by host galaxy mass. The top row shows the baseline full multi-band model with fixed $\kappa_\gamma$, while the bottom row repeats the same diagnostic using the pseudo-bolometric Arnett fits. As enforced by the prior on the model parameters, all SNe lie below the forbidden $f_{\rm Ni} = 1$ line (dashed red). However, most of our sample is strongly clustered around the limit, suggesting that the cosmological sample are highly efficient burners. In the right panel, we show the burning efficiency versus ejecta mass, revealing a high mean $f_{\rm Ni} = 0.88 \pm 0.15$, which is broadly constant across all masses. Burning efficiency alone is not a unique channel diagnostic: near-$M_{\rm Ch}$ delayed-detonation models and sub-$M_{\rm Ch}$ detonation models span broad and overlapping $^{56}$Ni fractions depending on progenitor mass, central density, metallicity, ignition geometry, and radiative-transfer treatment~\citep[e.g.,][]{Blondin2022,Fink2010,Sim2010,Shen2018}. However, combined with other parameters such as ejecta mass and velocity it can help distinguish between different channels. The pseudo-bolometric lower-limit comparison shifts most events to lower $M_{\rm Ni}$ and lower $f_{\rm Ni}$, with only $\sim7.4\%$ of events above $f_{\rm Ni}=0.8$, showing that the high burning fraction point estimates from the photometric analysis are not driven by the implied absolute bolometric luminosity of the data. We note that the pseudo-bolometric one-zone model fits still include events with $f_{\rm Ni} \to 1$, so the tension is not removed entirely. As these measurements are conservative lower limits it points towards a modelling systematic at a bolometric level. In particular, the residual high-efficiency tail may indicate remaining model misspecification in the one-zone diffusion treatment such as nickel mixing and ejecta stratification, or a more realistic temperature and opacity structure.

\begin{figure*}
\centering
\includegraphics[width=0.95\textwidth]{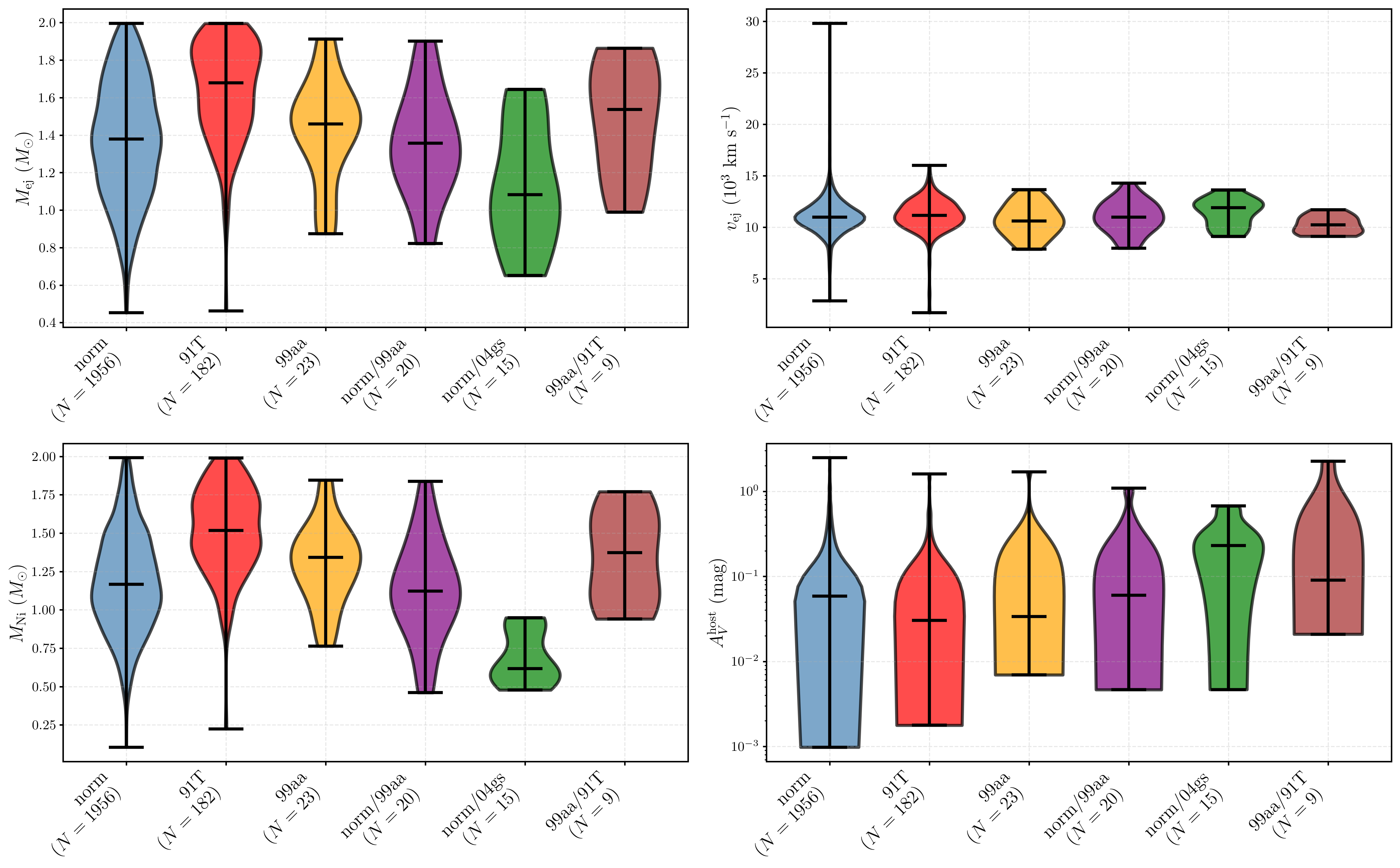}
\caption{Violin plots of the physical properties by spectroscopic subtype. Horizontal lines mark median values.}
\label{fig:subtypes}
\end{figure*}

In Fig.~\ref{fig:subtypes} we compare physical properties across spectroscopic subtypes with violin plots showing the full distributions across our sample, and with horizontal lines marking medians. SN~1991T-like events (red, 8.3\% of the sample) are systematically distinct from normal SNe~Ia (blue) and typically have a higher ejecta mass ($M_{\rm ej} = 1.64$ vs $1.38M_\odot$), slightly higher velocities ($v_{\rm ej} \sim 11,200$ vs 11,000 km s$^{-1}$), and $30\%$ more nickel. SN~1999aa-like events (orange) show intermediate properties. The host extinction distributions are similar across subtypes. Notably, $84\%$ of SN~1991T-like SNe have $M_{\rm ej} > 1.4$~M$_\odot$, suggesting they may be super-$M_{\rm Ch}$ candidates and possibly correspond to a different progenitor channel.

\subsection{Correlations with SALT2 Parameters}
\label{sec:saltcorr}
\begin{figure}
\centering
\includegraphics[width=0.48\textwidth]{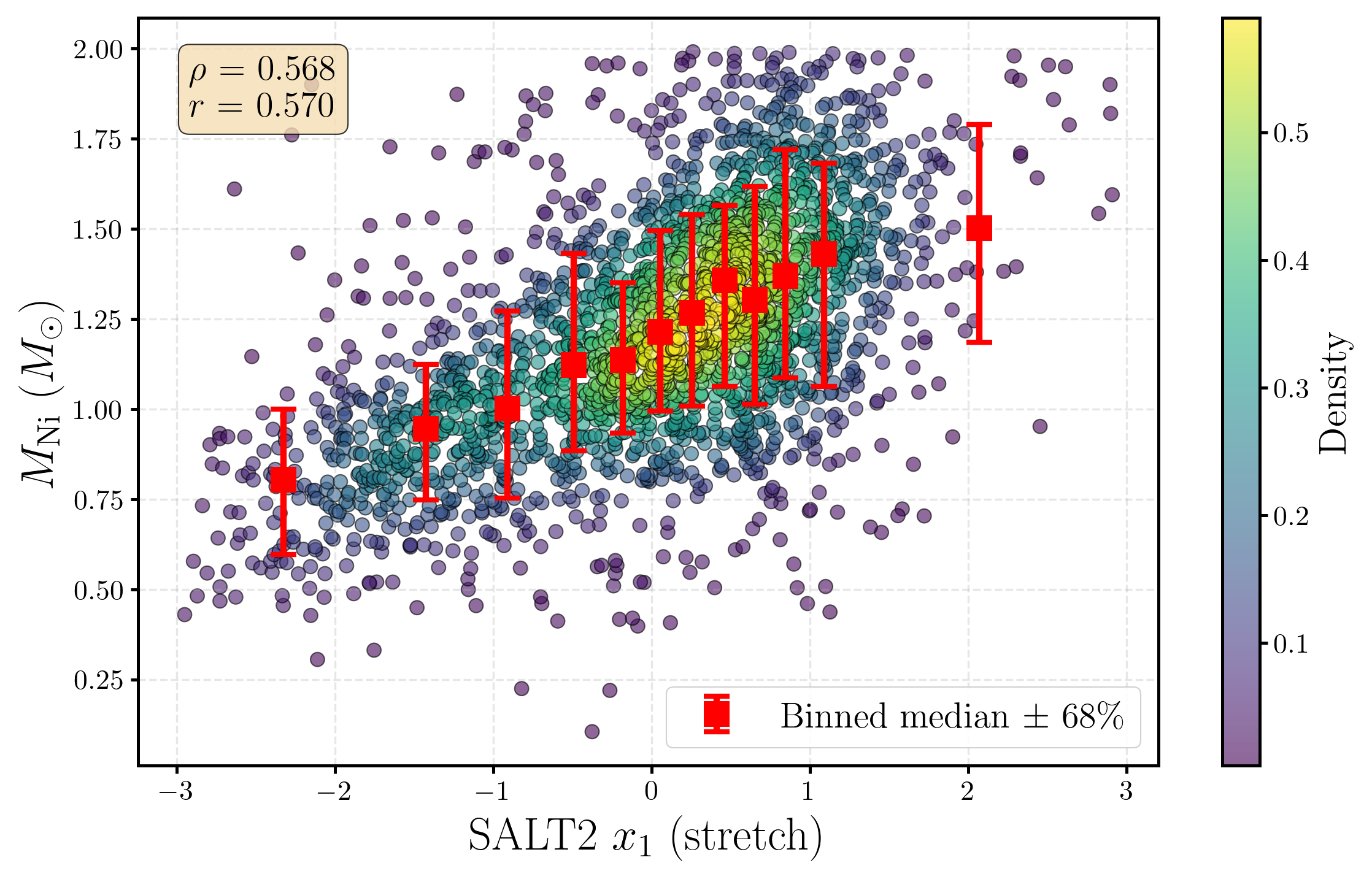}
\caption{Nickel mass versus SALT2 stretch ($x_1$). The strong correlation ($\rho = 0.57$) confirms that brighter, slower-declining SNe synthesize more $^{56}$Ni. Points are coloured by density of events, with red squares showing binned medians with $68\%$ credible intervals.}
\label{fig:phillips}
\end{figure}

We now explore how the derived physical parameters correlate with the empirical SALT model parameters.
Unless otherwise stated, the quoted correlation coefficients in this section are calculated using posterior medians for each SN and therefore do not fully propagate individual-event posterior uncertainties. We use the posterior widths and the hierarchical analysis in Sec.~\ref{sec:hierarchicalinference} to assess population-level uncertainty.
We start by exploring the empirical Phillips relation i.e., brighter supernovae have longer lasting peaks.
In Fig.~\ref{fig:phillips} we show the nickel mass for all our supernovae against their corresponding SALT2 stretch, which highlights a strong positive correlation. To further visualise the correlation, we also plot a binned estimate of the nickel mass at different SALT2 $x_{1}$ bins alongside the $68\%$ credible interval for that bin, clearly highlighting the positive correlation.
This shows that slower-declining SNe ($x_1 > 0$) synthesize systematically more $^{56}$Ni and therefore are more luminous. In the left-hand side we also plot the Pearson and spearman scores for the two parameters, which again reinforce the positive correlation.

\begin{figure}
\centering
\includegraphics[width=0.48\textwidth]{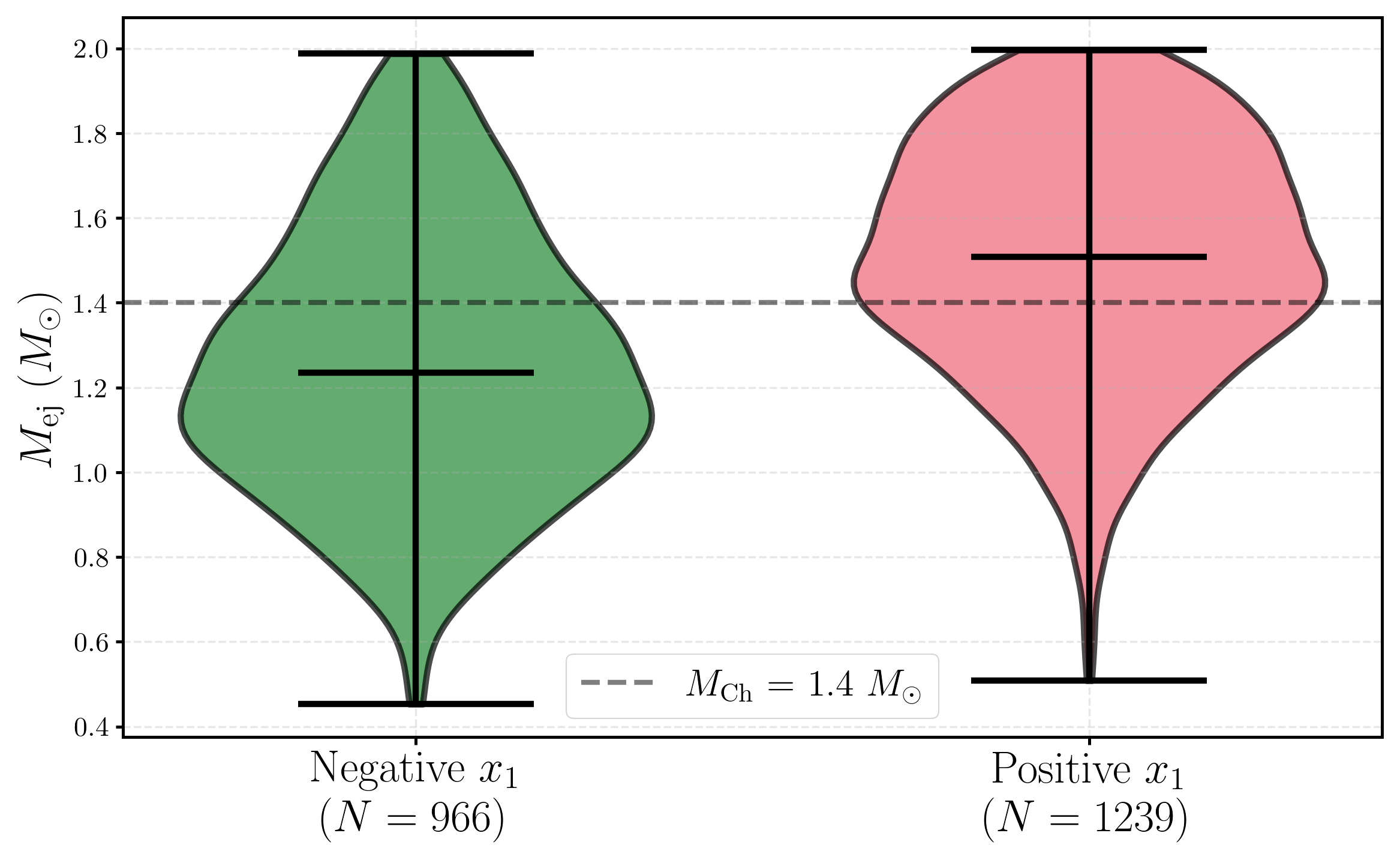}
\caption{Ejecta mass distributions split by SALT2 stretch. Fast decliners ($x_1 < 0$) have lower median $M_{\rm ej}$ than slow decliners ($x_1 \geq 0$), indicating a higher sub-$M_{\rm Ch}$ fraction among fast decliners. Violin plots show the full distributions; horizontal lines mark the median of each distribution.}
\label{fig:mej_stretch}
\end{figure}

In Fig.~\ref{fig:mej_stretch} we show ejecta mass distributions split by SALT2 stretch. This highlights that fast decliners ($x_1 < 0$, green) have systematically lower $M_{\rm ej}$ than slow decliners ($x_1 \geq 0$, red), with medians of $1.23$ vs. $1.51~M_\odot$. This shows that stretch is a composite parameter sensitive to both peak luminosity (via $M_{\rm Ni}$) and diffusion time (via $M_{\rm ej}$). 
The Chandrasekhar mass $1.4$~M$_{\odot}$ (dashed line) lies between the median values for the two distributions.

\begin{figure*}
\centering
\includegraphics[width=0.95\textwidth]{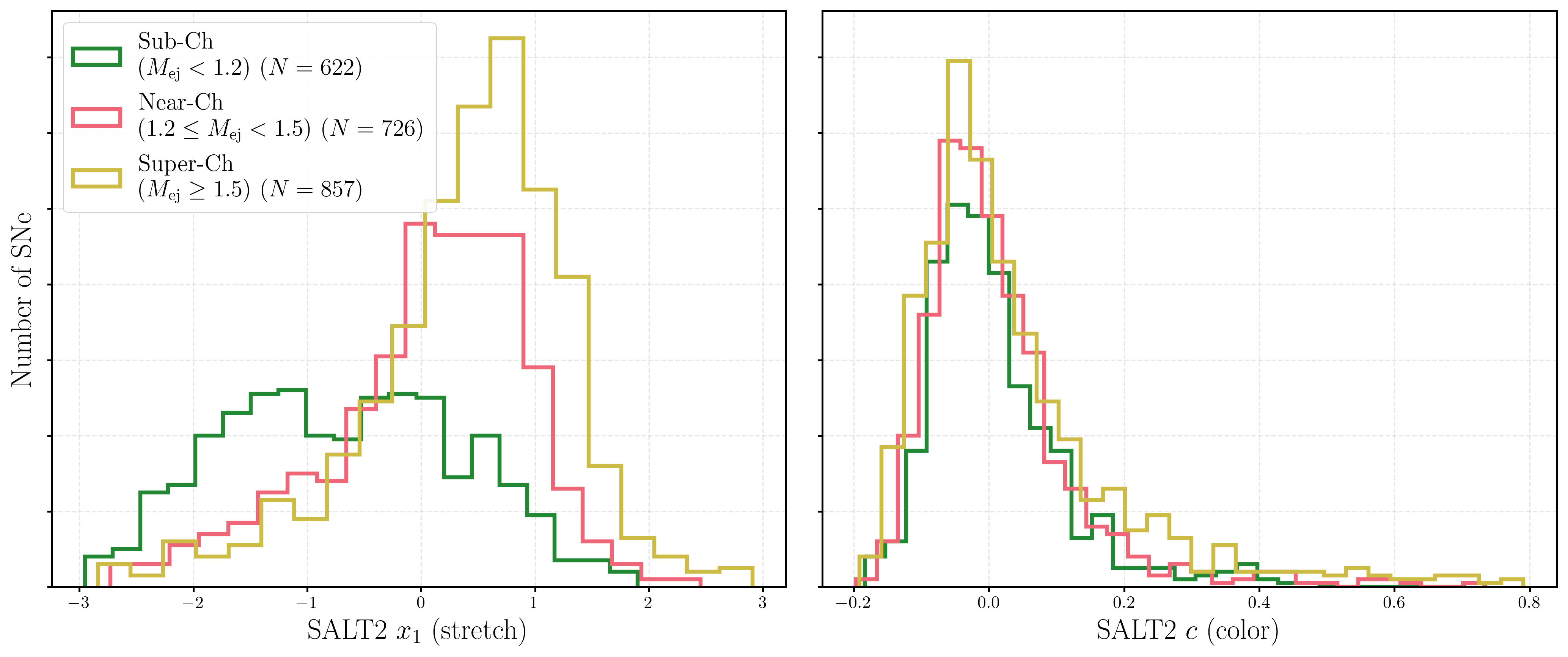}
\caption{SALT2 parameters binned by ejecta mass. Left: Stretch ($x_1$) distributions shift systematically toward higher values for more massive explosions, confirming the physical connection between $M_{\rm ej}$, $M_{\rm Ni}$, and light curve width. Right: Colour ($c$) distributions are largely independent of ejecta mass, indicating $c$ is affected by dust and intrinsic SED-level effects, and does not map cleanly onto a single explosion parameter such as ejecta mass.}
\label{fig:salt2_by_mej}
\end{figure*}

As our sample size is substantial, we can also explore substructure in different bins on physical parameters. In Fig.~\ref{fig:salt2_by_mej} we show the SALT2 parameter distributions binned by ejecta mass, to explore whether different bins (which may map to different progenitor channels) show any different features in these parameters. The left panel reveals that stretch ($x_1$) distributions shift progressively towards higher values as $M_{\rm ej}$ increases with sub-$M_{\rm Ch}$ events (green) peak at $x_1 \sim -1$, near-$M_{\rm Ch}$ (red) at $x_1 \sim 0$, and super-$M_{\rm Ch}$ (yellow) at $x_1 \sim +1$. This is not unexpected, as larger ejecta mass means longer diffusion times and therefore larger stretch. However, it is notable that none of the distributions follow the simple unit-Normal expectation often used as an empirical population prior for $x_1$. This is consistent with previous work showing that the ZTF DR2 $x_1$ distribution is not well described by a single unit Gaussian and may be bimodal~\citep{Ginolin2025}. 

The right panel of Fig.~\ref{fig:salt2_by_mej} shows colour ($c$) distributions are largely independent of $M_{\rm ej}$, with all three mass bins exhibiting similar spreads centred near $c \approx 0$. This is somewhat surprising given that $c$ is expected to trace a combination of intrinsic colour and host-galaxy extinction. For example, it is expected that sub-Chandrasekhar helium-shell double detonations will produce SNe that feature redder optical colours at maximum light \citep[e.g.,][]{Polin2019}, in which case we may expect that sub-Chandrasekhar explosions have larger values of $c$, yet this is not observed. Furthermore, if sub-, near-, and super-Chandrasekhar explosions have different progenitor channels, each with different delay time distributions \citep[e.g.,][]{Eitner23},
then each of these sub-types is likely to occur within different environments, and yet there is no clear correlation between SALT2 $c$ and $M_\mathrm{ej}$ (a proxy for explosion mechanism). This reinforces that stretch is physically connected to explosion properties like the ejecta mass, while SALT2 colour does not have a direct mapping to the physical properties captured by our simplified model.

\begin{figure}
\centering
\includegraphics[width=0.48\textwidth]{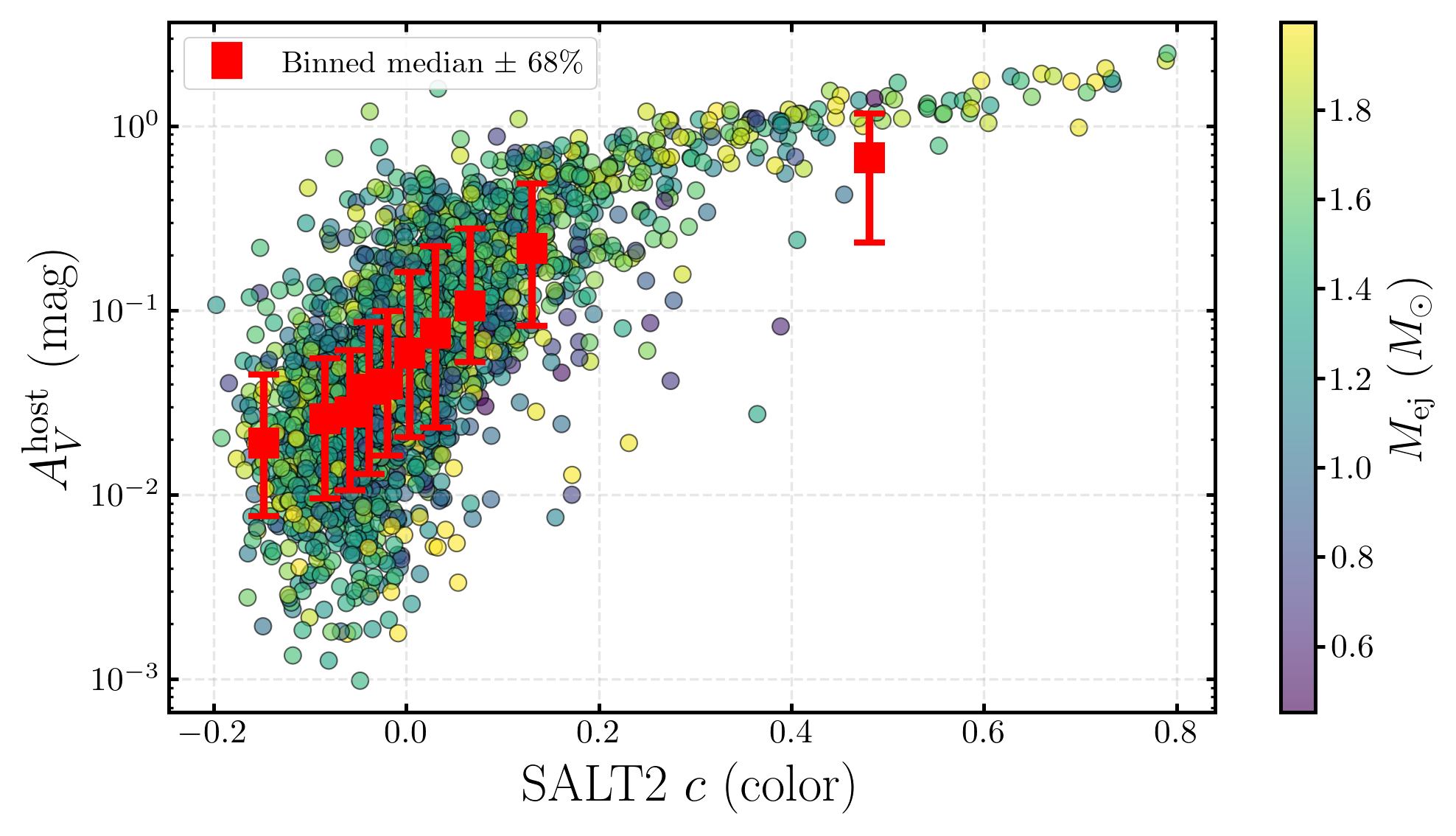}
\caption{Host extinction ($A_V^{\rm host}$) versus SALT2 colour ($c$). The strong correlation ($\rho = 0.83$) indicates dust contributes significantly to SALT2 colour, however the scatter at low $c \leq 0.2$, may also point to intrinsic colour variations (recombination, line blanketing) rather than purely extrinsic dust. The correlation is computed using $A_V^{\rm host}$, not $\log A_V^{\rm host}$. Points coloured by $M_{\rm ej}$ show no strong segregation i.e., the colour and dust extinction does not depend on the amount of ejecta.}
\label{fig:av_color}
\end{figure}

To further explore $c$, we show in Fig.~\ref{fig:av_color} how our inferred host extinction correlates with SALT2 colour, which indicates that dust reddening contributes significantly to $c$, with a large Spearman correlation score ($\rho = 0.83$). However, the scatter (at $c \lesssim 0.2$, the range containing the bulk of cosmology-quality SNe~Ia), suggests that $c$ may also capture some intrinsic colour variations rather than purely extrinsic host-galaxy dust. 
Points are coloured by ejecta mass, showing no strong mass-colour correlation (as also indicated by Fig.~\ref{fig:salt2_by_mej}).

\subsection{Host Galaxy Correlations}\label{sec:hostcorr}
We now explore correlations with the host galaxy. We do not perform any additional modelling of the host galaxy, but use values available in the DR2 release~\citep{Rigault2025}. We note that only 2151 events out of our selected population have host-galaxy constraints in the DR2 release. 

\begin{figure}
\centering
\includegraphics[width=0.48\textwidth]{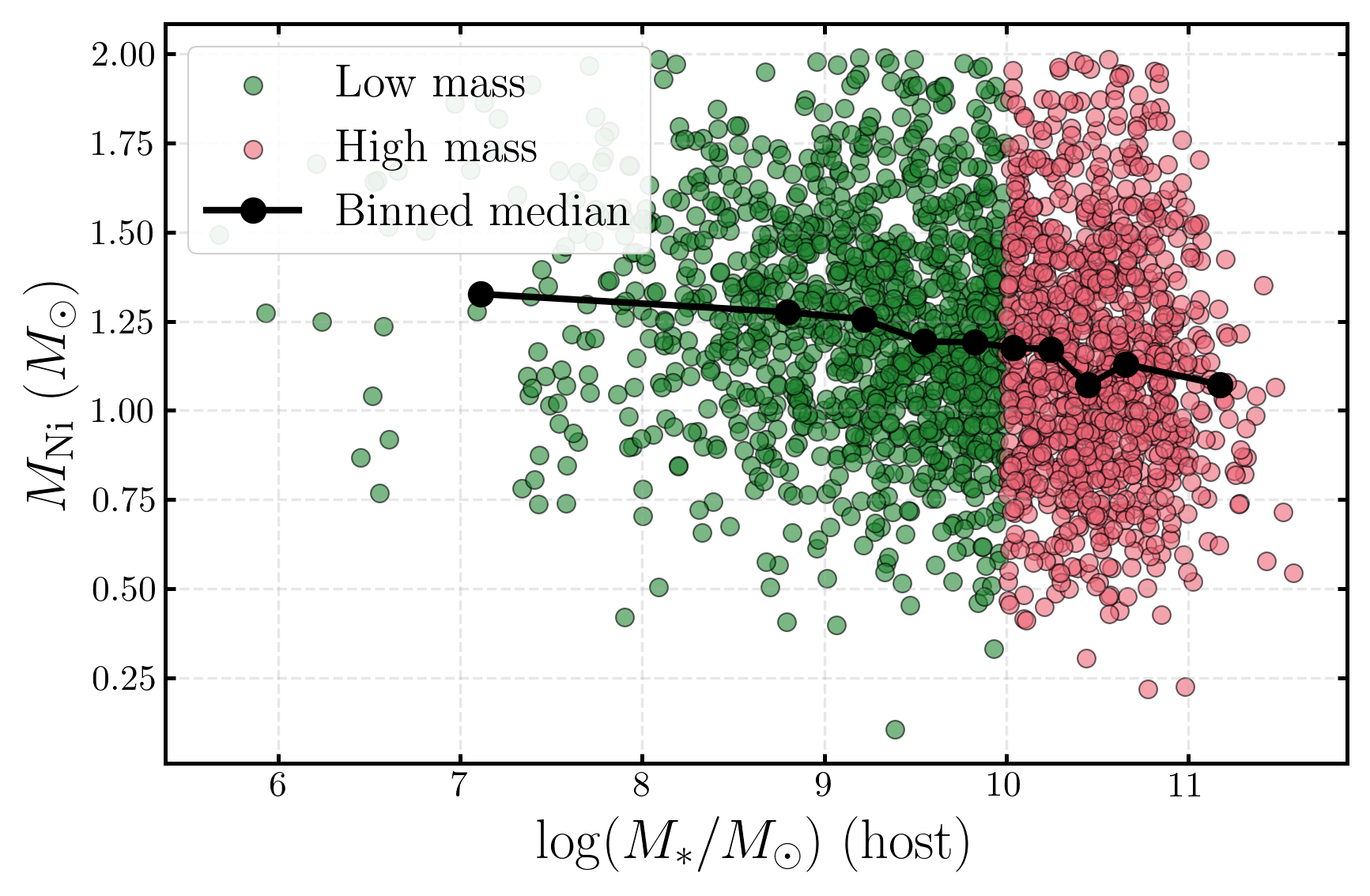}
\caption{Nickel mass versus host galaxy mass. The binned median (black line) shows declining $M_{\rm Ni}$ with increasing host mass. Low-mass host SNe (green) synthesize systematically more nickel than high-mass host SNe (red).}
\label{fig:mni_hostmass}
\end{figure}

In Fig.~\ref{fig:mni_hostmass} we show nickel mass versus host galaxy mass. The binned trend (black line) reveals declining $M_{\rm Ni}$ with increasing host mass, which could potentially provide a hint towards the origin of a mass step (which is associated with a step near $\log_{10}(M_*/M_\odot) \sim 10$). SNe in low-mass, star-forming hosts (green) synthesize more nickel than those in high-mass, passive hosts (red). 
The scatter is substantial, but there is a strong hint of a downward trend with increasing host galaxy stellar mass. This suggests that less massive, likely younger galaxies have a SN~Ia channel that is more efficient at producing nickel, potentially hinting at a metallicity dependence in nickel production \citep{Timmes2003, Howell2009, PadillaGonzalez2025}.
Given that the nickel mass strongly correlates with $x_1$, this plot is directly comparable to the known and expected trends between $x_1$ and host galaxy properties \citep[e.g.,][]{Sullivan2010}. 

\begin{figure*}
\centering
\includegraphics[width=0.95\textwidth]{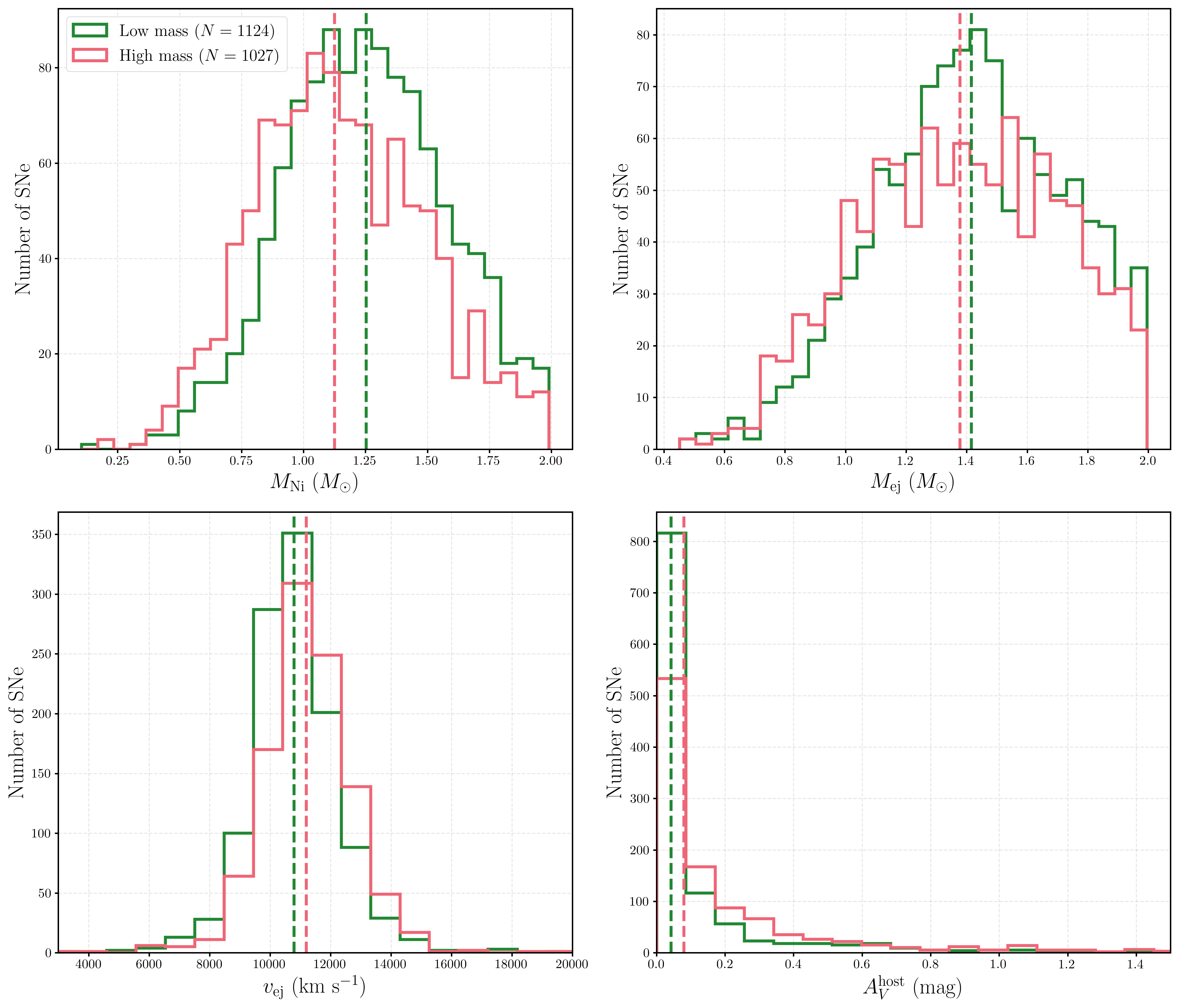}
\caption{The host-galaxy mass step in physical parameters. Step histograms compare low-mass (green, $\log_{10} (M_*/M_{\odot}) < 10$) versus high-mass (red, $\log_{10} (M_*/M_{\odot}) \geq 10$) hosts. The largest offset is in nickel mass ($\Delta M_{\rm Ni} = 0.13~M_\odot$), suggesting intrinsic progenitor differences between environments. Dashed lines mark medians. We note that only 2151 events out of our selected population have host-galaxy constraints in the DR2 release.}
\label{fig:host_step}
\end{figure*}

In Fig.~\ref{fig:host_step} we show the host galaxy mass step in physical parameters. In particular, all four panels compare low-mass ($\log_{10} (M_*/M_{\odot}) < 10$, green) versus high-mass ($\log_{10} (M_*/M_{\odot}) \geq 10$, red) hosts, while dashed lines mark the medians for each population. The most significant offset occurs in nickel mass: $\Delta M_{\rm Ni} = 0.13~M_\odot$, with SNe Ia in low-mass hosts producing $12\%$ more nickel. 
The host-extinction shows a broader tail in high-mass galaxies, while the ejecta velocity and ejecta mass show little differences.

\begin{figure}
\centering
\includegraphics[width=0.48\textwidth]{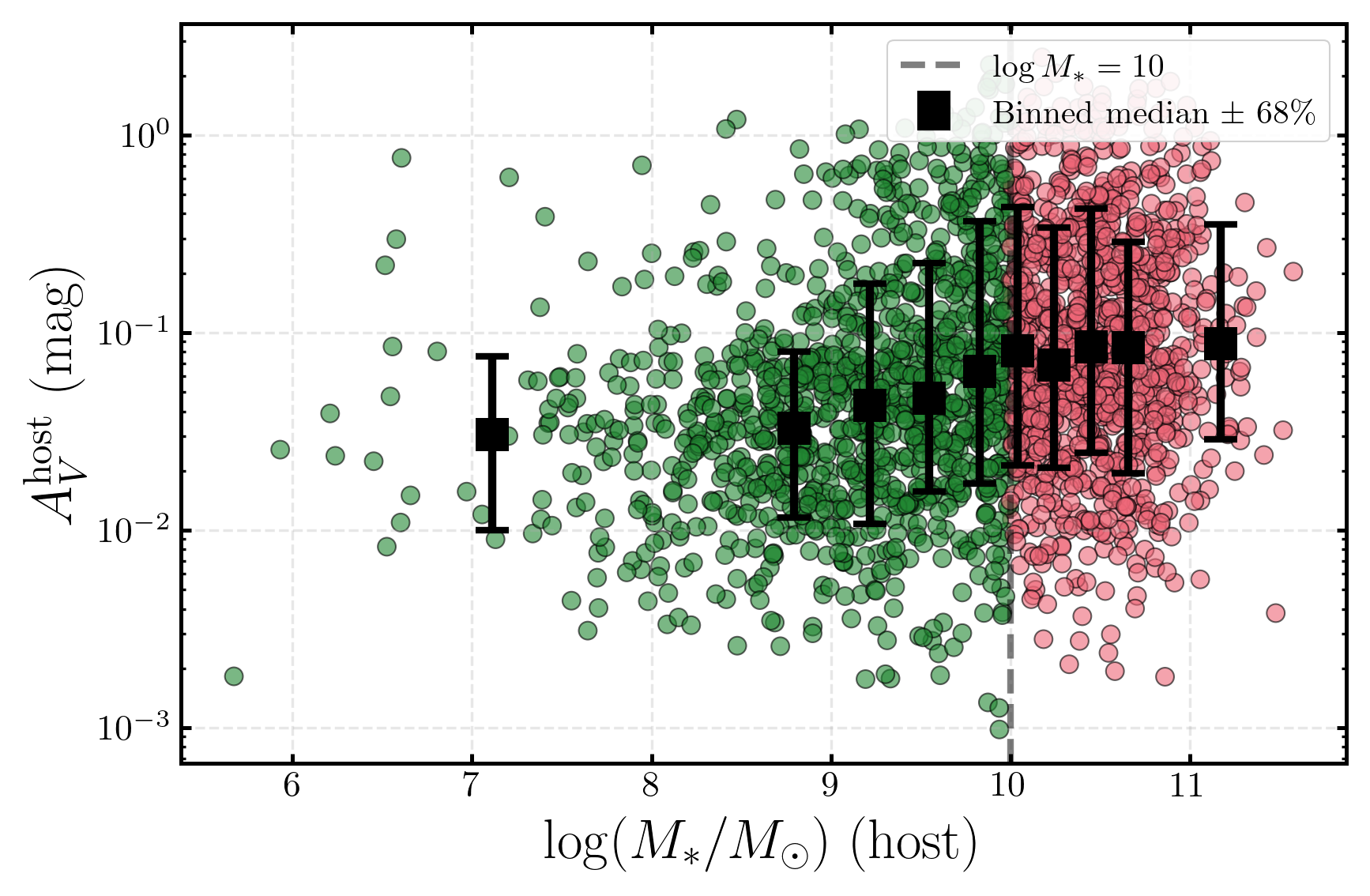}
\caption{Host extinction ($A_V^{\rm host}$) versus galaxy mass. Binned statistics (black squares with error bars) show increasing extinction with host mass. Massive galaxies ($\log_{10} (M_*/M_{\odot}) > 10$) have systematically higher dust content, although with a large scatter. The vertical dashed line marks $\log_{10} (M_*/M_{\odot}) = 10$.}
\label{fig:av_hostmass}
\end{figure}

We further investigate the host extinction vs host-mass relation in Fig.~\ref{fig:av_hostmass}, which shows our inferred host extinction versus galaxy mass for all SNe in our sample. The binned trend reveals increasing $A_V$ with host mass, with massive galaxies exhibiting $\sim$50\% higher median extinction (in AB magnitude) than low-mass galaxies. The transition again occurs near $\log_{10} (M_*/M_{\odot}) = 10$. Higher extinction in massive hosts may reflect higher metallicity and dust content. However, this requires more detailed studies of the hosts. 

\begin{figure*}
\centering
\includegraphics[width=0.95\textwidth]{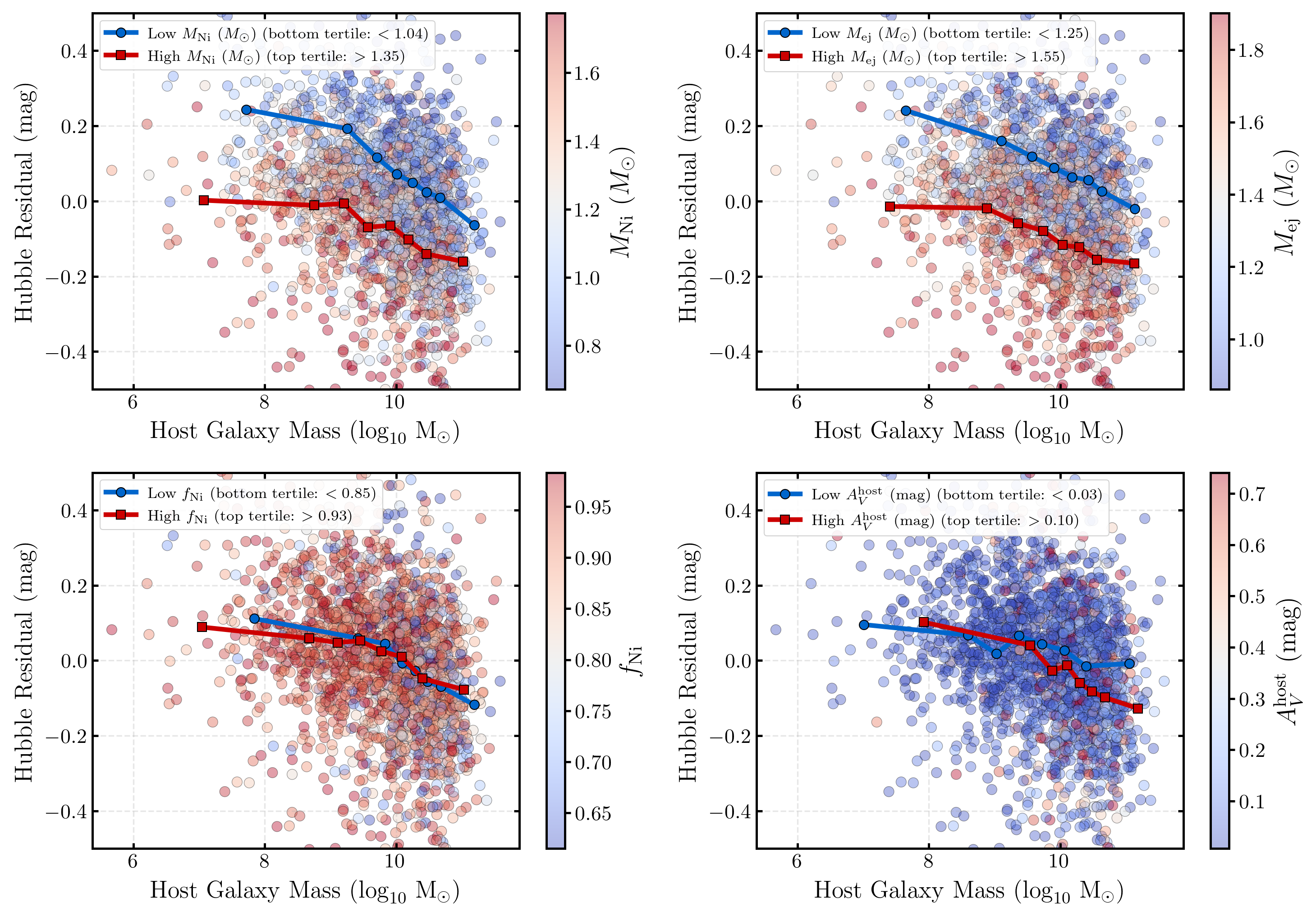}
\caption{Hubble residuals versus host galaxy mass, with panels coloured by nickel mass, ejecta mass, burning efficiency, and host extinction from top left to bottom right, respectively. Scatter plots show individual SNe coloured by the parameter value. Overlaid binned trends compare bottom tertile (blue circles) and top tertile (red squares) for each parameter. High-$M_{\rm Ni}$ SNe~Ia are systematically over-luminous, while low-$M_{\rm Ni}$ SNe are under-luminous, demonstrating incomplete SALT2 standardization.}
\label{fig:hubble}
\end{figure*}

We now investigate the link between standardised Hubble residuals and SN Ia parameters. The Hubble residuals have been corrected for SALT stretch and colour using the \texttt{standax} package presented in \cite{Ginolin2025a, Ginolin2025}, leading to $\alpha=0.161$ and $\beta=3.06$. As we aim to explore the dependence of the mass step on SN Ia properties, we do not correct for environmental dependencies of SN Ia magnitudes. 
In Fig.~\ref{fig:hubble} we examine the Hubble residuals as a function of host galaxy mass, with panels coloured by different physical parameters. Each panel shows scatter plots colour-coded by the indicated parameter (colour bar shown on the right), with binned trends for the bottom tertile (blue circles) and top tertile (red squares) overlaid. The top left panel shows residuals binned by $M_{\rm Ni}$. The top right panel shows residuals binned by $M_{\rm ej}$: similar trends emerge, with massive ejecta producing brighter-than-expected SNe. 

These plots reveal interesting features in the SN~Ia population. Of particular note are SNe~Ia with high nickel mass (red curve), which in low-mass galaxies are relatively well standardized, but systematically produce negative residuals (over-luminous) in high-mass galaxies. Meanwhile, low-$M_{\rm Ni}$ SNe (blue curve) produce positive residuals (under-luminous) across the host galaxy mass bins, strongly suggesting that the mass-step may be rooted in an incomplete standardization of the intrinsic luminosity--$M_{\rm Ni}$ relation. Differences in Hubble residual between high and low $M_{\rm Ni}$ and $M_{\rm ej}$ are relatively consistent across host galaxy stellar mass. 

The bottom left and bottom right panels show residuals binned by $f_{\rm Ni}$ and by $A_V^{\rm host}$, respectively. SNe Ia with different burning efficiencies in low-mass host galaxies show a clear difference in Hubble residual, yet are consistent and trend towards more negative for higher mass host galaxies. Conversely, SNe Ia with different host extinctions in low-mass hosts have consistent Hubble residuals, but as we move to higher host masses, the difference in average Hubble residual between high and low host-extinction bins increases, with the high host-extinction sample trending to lower Hubble residuals than the low extinction sample.  

\subsection{Redshift-limited sub-sample}\label{sec:volsubsample}
As discussed in Sec.~\ref{sec:sample}, our sample includes cosmology-quality SNe~Ia with a spectroscopic classification, but it does not include all different SN~Ia subtypes and is not volume complete. 
This bias is towards brighter events, which is evident in our higher than expected ${}^{56}$Ni estimate of the population, as well as the fact that we are missing the sub-luminous classes. We can mitigate this by performing full hierarchical inference and inverting the selection function, which we discuss further below. Here, instead, we briefly examine a redshift-limited subsample ($z < 0.06$, $N = 902$), where ZTF DR2 is believed to be relatively volume-complete for normal SNe~Ia~\citep{Amenouche2025}. However, we note this estimate is based on the recovery efficiency of SALT2 injections and unlikely to be the same for our model as SALT2 and a one-zone radioactive decay model do not have a one-to-one mapping. This low-redshift cut employed here minimizes Malmquist bias, as sub-luminous events are more likely to be detected at these distances, though our photometric quality and SALT2 parameter cuts still exclude several events.

In Fig.~\ref{fig:hyperpost}, we show histograms of the ejecta mass and nickel masses of all the events in this volume-limited sample. As expected, the values drop towards smaller values; in particular, our inferred ${}^{56}$Ni-mass distribution (i.e., the median of the posterior of all 902 events) has a mean of $1.02~M_{\odot}$ with a standard deviation of $0.3~M_{\odot}$. This is again more consistent with estimates from smaller, albeit (also biased) samples~\citep{Scalzo2014, Childress2015, Bora2024}, especially given the large uncertainty. We also see a decrease in the ejecta mass, which shifts to smaller values with a median of $1.2~M_{\odot}$. This decrease also translates to a drop in the burning efficiency, where our median ($\sim 0.8$) is less extreme than the higher efficiency indicated by the full sample ($\sim 0.9$). As we describe below, our results even within this lower redshift sample are susceptible to selection effects but with a more minimized impact. 

Our other results, such as the correlations with SALT2 and host-galaxy parameters show similar trends in the low-redshift subsample. Of particular note, the relationship seen in Fig.~\ref{fig:phillips} maintains a similar strength, demonstrating that the $M_{\rm Ni}$--light curve width correlation is not an artifact of our selection function. The host galaxy mass step also persists, again providing confidence that this is a true effect rather than an artefact of our selection function. We show a few additional plots for the low-redshift sample in the Appendix. 

The correlations seen in Fig.~\ref{fig:mni_hostmass} and Fig.~\ref{fig:av_hostmass} weaken significantly. However, the binning with host mass also further reduces the sample size, so we are partly limited by Poisson noise in many bins.
The Hubble residuals compared to host-mass (Fig.~\ref{fig:hubble}) shows similar trends, but with more scatter due to the smaller number of SNe at low galaxy masses. 
Future analyses incorporating hierarchical Bayesian inference and including galaxy properties with explicit selection functions may enable precise corrections of these biases to recover the true population-level distributions.

\subsection{Hierarchical inference on low-redshift sample}\label{sec:hierarchicalinference}
\begin{figure*}
\centering
\includegraphics[width=0.95\textwidth]{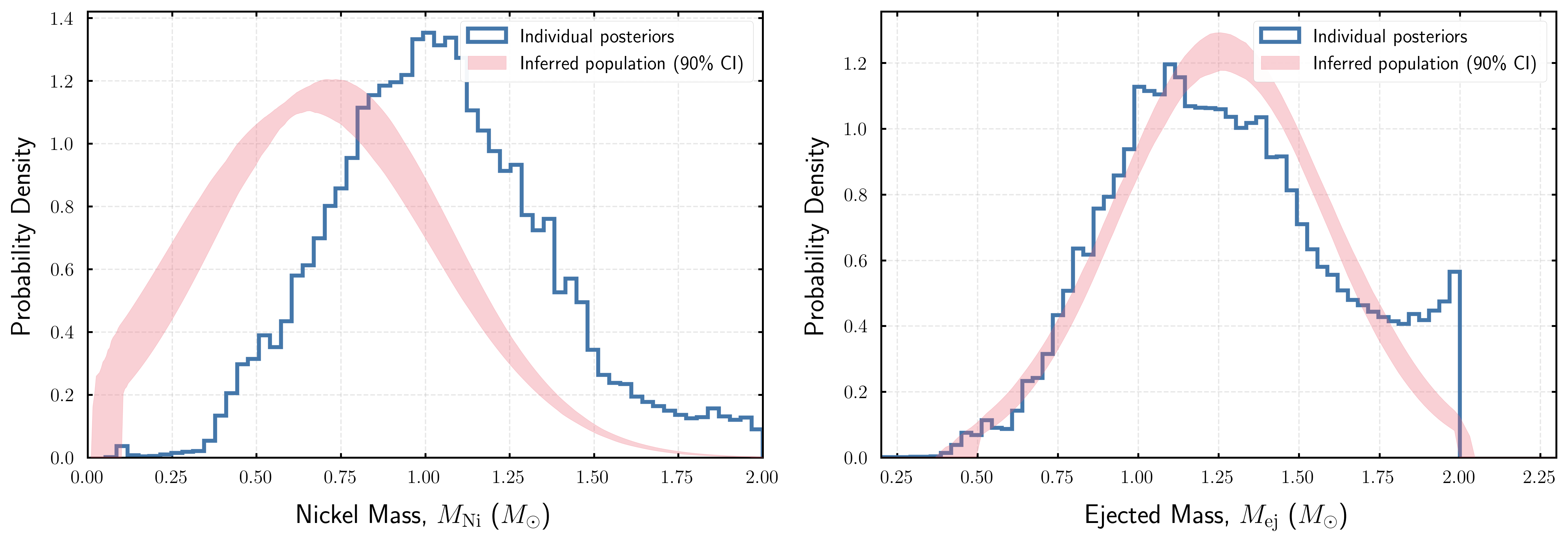}
\caption{Low-redshift ($z \leq 0.06$) sample of 902 SN Ia, with medians of the nickel mass (left) and ejecta mass (right) in blue, the pink curves indicate the $90\%$ credible interval from hierarchical modelling on this sample.}
\label{fig:hyperpost}
\end{figure*}

In this section, we perform hierarchical Bayesian modelling to infer the population properties of the physical parameters. Hierarchical Bayesian inference was first applied to SN Ia samples by \citet{Mandel09, Mandel2011} using empirical light curve models. 
With the aim to further understand the true distribution of SN Ia explosion properties, we perform hierarchical modelling on our redshift-limited sample ($z \leq 0.06$), consisting of 902 supernovae with
well-sampled light curves. We want to extract the population hyperparameters $\boldsymbol{\Lambda}$, in particular, the parameters describing the distribution of ${}^{56}$Ni and ejecta masses. The hyper-parameter posterior is given by~\citep{Mandel2019, Thrane2019},
\begin{equation}
p(\boldsymbol{\Lambda}|d_i, N_{\rm obs}) \propto \pi(\boldsymbol{\Lambda})  \mathcal{L}(d_i, N_{\rm obs}|\boldsymbol{\Lambda}).
\label{eq:hyperpost}
\end{equation}
Here $d_i$ represents the data (light curves) for the $N_{\rm obs} = 902$ observed events, $\pi(\boldsymbol{\Lambda})$ is the prior on our population parameters, and $\mathcal{L}$ is the likelihood for the population hyperparameters.

The population likelihood is related to the individual-event likelihoods via, 
\begin{equation}
\mathcal{L}(d_i, N_{\rm obs}|\boldsymbol{\Lambda}) = \prod_{i=1}^{N_{\rm obs}} \mathcal{L}(d_i|\boldsymbol{\Lambda}),
\end{equation}
where the single-event likelihood $\mathcal{L}(d_i|\boldsymbol{\Lambda})$ marginalises over the individual event parameters $\boldsymbol{\theta}_i$ such as the nickel and ejecta masses.
\begin{equation}
\mathcal{L}(d_i|\boldsymbol{\Lambda}) = \int \mathcal{L}(d_i|\boldsymbol{\theta}_i)  \pi(\boldsymbol{\theta}_i|\boldsymbol{\Lambda})  d\boldsymbol{\theta}_i.
\label{eq:single_event}
\end{equation}
Here $\mathcal{L}(d_i|\boldsymbol{\theta}_i)$ is the individual event likelihood (proportional to the posterior from our light curve fits), and $\pi(\boldsymbol{\theta}_i|\boldsymbol{\Lambda})$ is the
population model.

A common technique to reduce the computational load for hierarchical inference is to approximate the integral in Equation~\ref{eq:single_event} using posterior samples $\boldsymbol{\theta}_i^{(j)}$ from our individual fits \citep{Thrane2019},
\begin{equation}
\mathcal{L}(d_i|\boldsymbol{\Lambda}) \approx \frac{1}{N_{\rm samples}} \sum_{j=1}^{N_{\rm samples}} \frac{\pi(\boldsymbol{\theta}_i^{(j)}|\boldsymbol{\Lambda})}{\pi(\boldsymbol{\theta}_i^{(j)})},
\label{eq:importance_sampling}
\end{equation}
where $\pi(\boldsymbol{\theta}_i^{(j)})$ is the prior used in the individual fit. This importance sampling approximation allows us to ``reweight'' the posterior samples according to the population model, accounting for the difference between the individual-event priors and the population distribution.

As described above, this sample is plagued with selection biases that are non-trivial to invert. Specifically, the full pipeline for selection involves at a minimum: 1) Spectroscopic classification
as a SN Ia, which inherently introduces some selection based on brightness and matches to spectral templates that exclude SNe Ia with smaller nickel yields; 2) Good light curve coverage, which again implicitly favours brighter sources; 3) A ``good fit'' with the SALT2 model; and 4) A selection cut which excludes sub-luminous Type Ia such as 91bg-like SNe. 
The smaller redshift-limited sample minimises these biases, but the sample is still not free from selection effects for our model.

To account for selection effects, we must correct the likelihood to reflect the fact that our observed sample is not a fair draw from the underlying population. This is typically done through modification of the population likelihood, with a selection-term. 
\begin{equation}
\mathcal{L}_{\rm sel}(d_i|\boldsymbol{\Lambda}) = \frac{\mathcal{L}(d_i|\boldsymbol{\Lambda})}{p_{\rm det}(\boldsymbol{\Lambda})},
\label{eq:selection_correction}
\end{equation}
where $p_{\rm det}(\boldsymbol{\lambda})$ is the fraction of the population which surpasses the selection criteria, and can be expressed in terms of event-level parameters as,
\begin{equation}
p_{\rm det}(\boldsymbol{\Lambda}) = \int \pi(\boldsymbol{\theta}|\boldsymbol{\Lambda})  p_{\rm det}(\boldsymbol{\theta})  d\boldsymbol{\theta}.
\label{eq:beta}
\end{equation}
Here $p_{\rm det}(\boldsymbol{\theta})$ is the probability that an event with parameters $\boldsymbol{\theta}$ would be detected and included in our sample. As the predominant impact of our selection procedure in this low-redshift sample is on brightness, we model the selection bias through a detection probability function that incorporates both a hard lower threshold and brightness-dependent selection. We assign a relative detection probability
\begin{equation}
p_{\rm det}(M_{^{56}\rm Ni}) = \begin{cases}
0 & M_{^{56}\rm Ni} < M_{\rm threshold}
\\
\left(\frac{M_{^{56}\rm Ni}}{M_{\odot}}\right)^{\Gamma} & M_{^{56}\rm Ni} \geq M_{\rm threshold}
\end{cases}
\end{equation}
This functional form reflects two distinct selection effects: (1) a lower cutoff at $M_{\rm threshold}$, below which sub-luminous events (such as those classified as 91bg-like) are absent from the sample, and (2) a power-law brightness selection above the threshold, where $\Gamma$ parameterizes the strength of the brightness bias. For $\Gamma = 0$, the detection probability is flat above the threshold (no brightness bias), while $\Gamma > 0$ indicates preferential detection of more luminous events. 
This form is motivated by the empirical correlation between nickel mass and peak luminosity \citep[e.g.,][]{Arnett1982, Pinto2000}, and follows analogous selection functions used in gravitational-wave population analyses \citep{Abbott2023}. Both $M_{\rm threshold}$ and $\Gamma$ are treated as free parameters in our inference. 

We model the intrinsic population as independent truncated Gaussians for both $M_{^{56}\rm Ni}$ and $M_{\rm ej}$, i.e., the hyperprior is
\begin{equation}
\pi(\boldsymbol{\theta}|\boldsymbol{\Lambda}) = \pi(M_{^{56}\rm Ni}|\Lambda_{\rm Ni}) \times \pi(M_{\rm ej}|\Lambda_{\rm ej}),
\end{equation}
where $\pi(M_{^{56}\rm Ni}|\Lambda_{\rm Ni})$ is the truncated Gaussian distribution for nickel mass with hyperparameters $\mu_{\rm Ni}$, $\sigma_{\rm Ni}$, $M_{\rm Ni}^{\rm min}$, and $M_{\rm Ni}^{\rm      
max}$, with an analogous set of four parameters describing the ejecta mass distribution.

We adopt flat uninformative uniform priors on all 8 population hyperparameters, as well as the 2 selection function parameters $\Gamma \in [0, 1.5]$ and $M_{\rm threshold} \in [0.2, 0.4]~M_\odot$. 
We use \program{pymultinest} \citep{Feroz2009, Buchner2016} to sample the joint hyperparameter and selection function posterior, accounting for the volume-time (VT) correction through numerical integration over the population model weighted by the selection function (see \citealt{Mandel2019} for formalism). 
In Fig.~\ref{fig:hyperpost}, we show the credible intervals of the inferred ejecta and nickel mass population distributions alongside the individual event posteriors from our 902 SNe. This illustrates how correcting for the selection bias and marginalising over the uncertainty in the individual posteriors in our sample shifts both nickel and ejecta mass distributions to lower values, becoming much more consistent with theoretical estimates. This shift is also consistent with the pseudo-bolometric comparison in Sec.~\ref{sec:pseudobolometric}, which independently points to lower conservative nickel masses than the raw full-model medians. The agreement is not exact, because the pseudo-bolometric analysis has different assumptions and does not propagate all uncertainties, but it supports the broader conclusion that posterior width, selection effects, and modelling systematics must be handled at the population level. We stress that while the correction in our framework is via a selection term in Eq.~\ref{eq:selection_correction}, a similar correction could be put into the hierarchical model to correct for systematics in the modelling by assuming each posterior is a biased measurement of the true event properties.

The hierarchical inference recovers well-constrained population parameters. For the ${}^{56}$Ni distribution, we find a mean of $\mu_{\rm Ni} = 0.64 \pm 0.06~M_{\odot}$ with an intrinsic scatter of $\sigma_{\rm Ni} = 0.42 \pm 0.02~M_{\odot}$ where uncertainties represent the $68\%$ credible interval of the posterior, truncated between $M_{\rm Ni}^{\rm min} = 0.08~M_{\odot}$ and $M_{\rm Ni}^{\rm max} = 2~M_{\odot}$. 
For the ejecta mass distribution, we infer $\mu_{\rm ej} = 1.26 \pm 0.01~M_{\odot}$ with scatter $\sigma_{\rm ej} = 0.33 \pm 0.01~M_{\odot}$, truncated between $M_{\rm ej}^{\rm min} = 0.46 \pm 0.04~M_{\odot}$ and $M_{\rm ej}^{\rm max} = 2.00 \pm 0.01~M_{\odot}$. The mean ejecta mass is consistent with explosions near the Chandrasekhar mass, while the moderate scatter ($\sigma_{\rm ej}/\mu_{\rm ej} \approx 0.26$) suggests a somewhat narrower range in progenitor masses compared to nickel production. We find a threshold of $M_{\rm threshold} = 0.4 \pm 0.05 M_{\odot}$, suggesting our detection probability drops to $0$, around this mass, which likely reflects the absence of the sub-luminous Ia subtypes in our sample. 

With our hyperparameters, we can again re-examine the fraction of events in each ejecta mass bin. The inferred ejecta mass distribution reveals similar-size split between sub-Chandrasekhar ($M_{\rm ej} < 1.2~M_{\odot}$) and near-Chandrasekhar ($1.2 \leq M_{\rm ej} \leq 1.5~M_{\odot}$) mass explosions, with $43 \pm 2\%$ and $34 \pm 1\%$ of events in each category, respectively. $24 \pm 2\%$ of events are super-Chandrasekhar ($M_{\rm ej} > 1.5~M_{\odot}$), suggesting that the majority of SNe~Ia arise from progenitors with masses at or below the Chandrasekhar limit. 
The smooth, continuous distribution spanning these regimes, combined with the similar fractions of sub- and near-Chandrasekhar events, is well described by the single-population hierarchical model adopted here. However, because that model assumes a unimodal population, this result should not be interpreted as evidence against multiple channels. A mixture-population hierarchical model would be required to test directly whether the data prefer distinct sub-populations.  

\section{Discussion}\label{sec:discussion}
Our analysis of 2,205 SNe~Ia with a wavelength-dependent one-zone model represents the largest homogeneous extraction of physical parameters from SN~Ia light curves to date. The sample size provides unprecedented statistical power to address fundamental questions about SN~Ia progenitors, explosion mechanisms, and the physical basis for cosmological standardization. In this section, we interpret our findings presented in Sec.~\ref{sec:results} in the context of SN~Ia cosmology followed by discussions of the progenitor channels, theoretical explosion models, and ejecta properties. We then further discuss systematics, a comparison with previous studies, and future directions.

We reiterate that our sample is not complete, and that there are selection biases inherent in our sample. In particular, our cosmology-quality selection is biased toward intrinsically luminous, slowly declining SNe~Ia due to the requirement of high signal-to-noise, multi-band coverage. 
Normal events are retained with $\sim 80\%$ efficiency and SN~1991T-like explosions with $\sim 81\%$ efficiency and no SN~1991bg-like or SN~Iax events make it into the cosmological sample~\citep{Rigault2025}. 
This preference for luminous light curves shifts the stretch distribution towards higher $x_1$, and also explains the lack of significant low-ejecta-mass, nickel-poor explosions in our sample. Consequently, our measured ejecta-mass distribution likely overestimates the true population median, and the nickel-mass fraction $f_{\rm Ni} \sim 0.8$ reflects the efficiently burning subset of SNe~Ia rather than the full thermonuclear transient population. Our analysis on the local redshift population ($z \leq 0.06$) minimises this Malmquist bias, however our sample selection still excludes some under-luminous events. We further note that while the ZTF DR2 sample is considered complete to $z \leq 0.06$~\citep{Amenouche2025}, some events are excluded from the release for calibration purposes or fail to pass our other cuts, so even this sample is not free from selection biases, even for the SALT model. 

In Sec.~\ref{sec:hierarchicalinference}, we derived estimates of the ejecta and nickel mass distributions, aiming to correct for selection effects assuming a toy model for selection based on only the nickel mass. This captures the true selection function in the low-redshift sample well, however, for truly robust hierarchical modelling, we require a more complete treatment of $p_{\rm det}$. As the selection is mostly a Malmquist bias, the predominant impact is on the Nickel mass, and more weakly on the ejecta mass, therefore, our other results are likely robust. 
The selection-biases induced due to spectroscopic classification or forcing good SALT2 fits into the selection procedure are somewhat intrinsic to cosmology samples and precisely match the population used for Hubble diagram analyses. Therefore, in general, our results (apart from those derived with the hierarchical model) should be considered under the lens that they characterize the ``cosmologically useful'' population rather than the full diversity of SNe~Ia.
The pseudo-bolometric Arnett comparison strengthens this point. It shows that a conservative luminosity-based reconstruction can move the point-estimate distributions of $M_{\rm Ni}$, $M_{\rm ej}$, and $f_{\rm Ni}$ substantially relative to the full multi-band model, especially when only considering a maximum-posterior estimate. At the same time, the pseudo-bolometric approach is not free of systematics and relies on a finite-wavelength SALT2 reconstruction, conservative extinction and distance assumptions, and the same one-zone diffusion model. Neither point-estimate distribution should be treated as exact as the individual-event constraints are broad and model-dependent enough that population-level inference is critical. The hierarchical analysis provides one route to combine the individual posteriors and its inferred nickel-mass distribution lies between the conservative pseudo-bolometric lower-limit estimates and the raw full-model max posterior estimates.
The two synthetic tests provide an internal check of the inference pipeline for two selected parameter sets and cadences, rather than a population-level calibration or an external validation of the physical model. They demonstrate sensitivity to distance and model assumptions, but cannot determine whether the real-sample $M_{\rm Ni}$, $M_{\rm ej}$, or $f_{\rm Ni}$ estimates are biased, or in which direction. Independently, the substantially lower pseudo-bolometric nickel-mass scale shows that the high full-model $M_{\rm Ni}$ and $f_{\rm Ni}$ point estimates are sensitive to the adopted SED treatment. However, because those pseudo-bolometric values are incomplete lower-limit estimates with their own systematics, they do not define an unbiased alternative scale. All physical parameter values and correlations discussed below should therefore be understood as conditional on the one-zone SED and opacity prescription.
\subsection{Physical Basis for Cosmological Standardization}
\label{sec:standardization}
Our detection of a strong correlation between nickel mass and SALT2 stretch ($\rho = 0.57$, Fig.~\ref{fig:phillips}) provides a model-dependent physical interpretation for the brighter-slower relation: brighter, slower-declining SNe are inferred to synthesize more $^{56}$Ni. Within our one-zone modelling treatment this is directly due to such brighter SNe also having more ejecta mass, and therefore longer diffusion times. This is consistent with the long-standing theoretical expectation~\citep{Arnett1982,Phillips1993} but with strong statistical significance given our sample size. Here and below, the significance describes correlations among the model-inferred parameters; it does not include uncertainty from model misspecification. The lower nickel-mass scale obtained from the pseudo-bolometric reconstruction means that the absolute calibration of this relation, rather than the existence of a correlation within the full-model inference, is particularly sensitive to the SED prescription.
The correlation is not perfect ($\sigma_{M_{\rm Ni}} \sim 0.25~M_\odot$ at fixed $x_1$), indicating that light curve width depends on additional factors beyond nickel mass. This is expected theoretically, as the width in effect traces the diffusion time. Larger $M_{\rm ej}$ increases the diffusion time $t_{\rm diff}$, broadening light curves independently of $M_{\rm Ni}$, as demonstrated by correlations seen in past works~\citep[e.g.,][]{Scalzo2014, Sharon2025}. However, this may also point towards different effects, such as higher nickel masses also driving larger opacities~\citep{Wygoda2019}. Full radiative transfer modelling, with more sophisticated treatment of opacity values due to composition and ionization further alter both peak brightness and diffusion time, and asymmetric explosions produce orientation-dependent light curves~\citeg{Kasen2006_pol,Boos2021}.

The empirical SALT2 standardization corrects peak magnitudes using $x_1$ and $c$. Physically, the $x_1$ correction accounts for $M_{\rm Ni}$ variations as the primary effect plus $M_{\rm ej}$/opacity contributions as secondary effects, while the $c$ correction includes both intrinsic colour variations from recombination and line blanketing and extrinsic dust reddening. The success of SALT2 standardization in reducing scatter to $\sim 0.14$ mag reflects the fundamental physics that total radiated energy scales with nickel mass, which is captured by light curve shape and colour.

The SALT2 parameter distributions binned by ejecta mass (Fig.~\ref{fig:salt2_by_mej}) reveal an interesting asymmetry between stretch and colour. The stretch ($x_1$) distributions shift systematically toward higher values as $M_{\rm ej}$ increases, with sub-$M_{\rm Ch}$ events ($M_{\rm ej} < 1.2~M_\odot$) peaking at $x_1 \sim -1$, near-$M_{\rm Ch}$ events peaking at $x_1 \sim 0$, and super-$M_{\rm Ch}$ events ($M_{\rm ej} > 1.5~M_\odot$) peaking at $x_1 \sim +1$. This confirms the physical connection between ejecta mass and light curve width, as more massive explosions have longer diffusion times and broader light curves. In contrast, the colour ($c$) distributions are largely independent of $M_{\rm ej}$, with all three mass bins exhibiting similar spreads centred near $c \sim 0$. 
This demonstrates that while stretch is fundamentally tied to explosion physics through the $M_{\rm ej}$--$M_{\rm Ni}$--$t_{\rm diff}$ relationship, colour is influenced by both environmental effects, such as dust extinction, and intrinsic SED-level effects produced by the explosion physics and nucleosynthetic yields. 
The independence of $c$ from $M_{\rm ej}$ also suggests that the intrinsic colour component, driven by recombination and line opacity, does not strongly correlate with the total ejecta mass, consistent with it arising from secondary processes in the outer ejecta layers rather than bulk explosion properties, or largely dominated by extrinsic effects. 
This dichotomy between the physical origin of $x_1$ and $c$ explains why the brighter-slower relation has a clear interpretation in terms of nickel mass while the colour correction remains more empirical, and it underscores the challenge of disentangling intrinsic and extrinsic contributions to $c$ for precision cosmology. 

We also find a strong significant correlation between host extinction $A_V^{\rm host}$ and SALT2 colour ($\rho = 0.83$, Fig.~\ref{fig:av_color}), showing that dust reddening contributes significantly to $c$. However, the large scatter at $-0.3 < c < 0.3$ which includes much of the cosmology-quality SN~Ia colour range, may also point towards intrinsic effects at these values, with the scatter at each colour bin reflecting variations in recombination epoch, line blanketing, and ejecta composition. 
This has important implications for the interpretation of the SALT2 $c$ correction in cosmological standardization. Binning by ejecta mass shows no strong segregation in the $A_V$--$c$ plane, suggesting that the intrinsic component of $c$ is largely independent of primary explosion parameters such as $M_{\rm ej}$ and $M_{\rm Ni}$. This suggests that intrinsic colour arises from secondary physics such as ionization state, velocity gradients in the ejecta~\citep{Foley2011_vel}, and the detailed structure of the photosphere during recombination, rather than being a direct consequence of the explosion energy or nickel production. The practical consequence is that the $\beta$ parameter in the \citet{Tripp1998} standardization formula is not purely a dust law but rather a hybrid correction encompassing both reddening and intrinsic colour physics, which complicates physical interpretation and may introduce redshift-dependent systematics if the balance between dust and intrinsic contributions evolves cosmologically \citep[e.g][]{Mandel2017}.

Our finding that SNe in low-mass hosts are inferred to synthesize $12\%$ more nickel ($\Delta M_{\rm Ni} = 0.13~M_\odot$, Fig.~\ref{fig:host_step}) suggests a possible physical origin for the Hubble residual mass step. The most plausible interpretation is progenitor age and metallicity, which lead to different burning efficiencies. This would be good to explore in detail on a small sample with high-quality galaxy spectra and multi-wavelength supernova data. 
The magnitude of the physical offset ($\Delta M_{\rm Ni} = 0.13~M_\odot$) corresponds to $\Delta M_{\rm bol} \sim 0.3$ mag at peak luminosity. After SALT2 standardization, which partially corrects for this via the $x_1$--$M_{\rm Ni}$ correlation, a residual $\sim 0.06$ mag offset remains, which is consistent with the observed Hubble residual mass step across multiple samples~\citeg{Sullivan2010,Childress2013,Kelsey2021,Kelsey2023}.
This suggests that the mass step is not a systematic error in standardization but rather an intrinsic physical difference between SN~Ia populations in different environments, and in particular the difference manifests in their nickel production (effectively their brightness). Given the host-mass step is one of the largest systematic uncertainties in SN cosmology \citep{Vincenzi2024}, for example, potentially biasing dark energy equation-of-state measurements at the $\Delta w \sim 0.05$ level~\citep{Betoule2014}, a confident physical origin for this property would be a remarkable result.

The Hubble residual analysis (Fig.~\ref{fig:hubble}) further highlights a residual correlation between luminosity and the nickel mass inferred by the full model. High-$M_{\rm Ni}$ SNe remain systematically over-luminous even after $x_1$ and $c$ corrections, while low-$M_{\rm Ni}$ SNe remain under-luminous.
Within the adopted model, this suggests that the $x_1$ parameter, while strongly correlated with $M_{\rm Ni}$, does not capture the full luminosity-$M_{\rm Ni}$ relation, leaving residual model-inferred scatter that could further introduce cosmological systematics if the $M_{\rm Ni}$ distribution also evolves with redshift. The Hubble residual analysis also reveals additional structure in the standardization. SNe with high nickel mass are relatively well standardized in low-mass galaxies but systematically produce negative residuals (over-luminous) in high-mass galaxies. Meanwhile, low-$M_{\rm Ni}$ SNe produce positive residuals (under-luminous) across host galaxy mass bins. This again suggests that the mass step may be rooted in incomplete standardization of the intrinsic luminosity--$M_{\rm Ni}$ relation, where the SALT2 $x_1$ correction does not fully capture the environmental dependence of the nickel mass production. The pseudo-bolometric analysis does not provide the an independent extinction estimate and is conditioned on a SALT template, so it cannot establish whether this residual trend is preserved under its lower nickel-mass scale.
\subsection{Progenitor Channels and the Ejecta Mass Distribution}
\label{sec:progenitors}

A key result of our work is the ejecta mass distribution (Fig.~\ref{fig:mass_distributions}), which shows a peak at $M_{\rm ej} = 1.40~M_\odot$, coincident with estimates of the Chandrasekhar mass. This suggests, within our model assumptions, a substantial contribution from near-Chandrasekhar-mass white dwarf explosions to the cosmologically-selected SN~Ia population. Moreover, the distribution is broad and does not show obvious bimodality in these point-estimate summaries, although this alone cannot rule out normal- and high-velocity subgroups or multiple progenitor channels. The width of the distribution ($\sigma \sim 0.30~M_\odot$) likely reflects a combination of intrinsic diversity, measurement uncertainty, selection effects, and possible model systematics.

Within the near-$M_{\rm Ch}$ paradigm, our results are qualitatively compatible with delayed-detonation or deflagration-to-detonation transition (DDT) models for the near-$M_{\rm Ch}$ subset~\citep{Khokhlov1991,Hoeflich1996}. In these scenarios, a carbon-oxygen white dwarf accretes matter from a companion (single-degenerate channel) or otherwise approaches conditions for carbon ignition near $M_{\rm Ch}$. Thermonuclear runaway begins as a subsonic deflagration, which later transitions to a supersonic detonation that unbinds the star. The observed $M_{\rm ej}$ distribution is consistent with minimal mass loss during the explosion for events whose true ejecta masses are close to $M_{\rm Ch}$.
Pure deflagration models predict substantial ($\sim 0.2$--$0.5~M_\odot$) unburned ejecta remaining bound~\citep{Nomoto1984}, which would shift the peak below $M_{\rm Ch}$. The apparent concentration near $1.4~M_\odot$ is therefore compatible with efficient unbinding in successful near-$M_{\rm Ch}$ explosions, but should not be read as uniquely selecting DDT over other channels.

The absence of clear bimodality in the ejecta mass distribution is notable given theoretical expectations of distinct progenitor channels. Though see past work that supports the hypothesis of a single channel for the ZTF-BTS supernovae sample~\citep{Sharon2022}. While we define sub-$M_{\rm Ch}$ ($M_{\rm ej} < 1.2M_\odot$, $28.2\%$) and super-$M_{\rm Ch}$ ($M_{\rm ej} > 1.5M_\odot$, $38.9\%$) populations based on nominal mass thresholds, the distribution itself shows no clear separation between these regimes. This absence of clear separation should be interpreted cautiously, it could simply reflect a genuinely continuous population, but it could also be produced by broad individual posteriors, selection biases, or the restricted spectroscopic diversity of the cosmology-quality sample. Future work with more extensive hierarchical modelling of the population may help answer this question. As discussed, our sample excludes sub-luminous 91bg-like and SN Iax events that would populate the $M_{\rm ej} < 0.9~M_\odot$ region, potentially creating the appearance of a smooth tail rather than a distinct low-mass component. Including these missing populations might reveal bimodality, with the cosmological sample representing only the high-mass mode. Our modelling may also be smearing the true bimodality, as the individual posteriors themselves are broad.
A DDT interpretation is physically relevant primarily for near-$M_{\rm Ch}$ progenitors, while sub-$M_{\rm Ch}$ events are more naturally associated with detonations or double detonations of lower-mass white dwarfs, and super-$M_{\rm Ch}$ point estimates would require either mergers, rapidly rotating massive white dwarfs, or residual modelling bias. The pseudo-bolometric Arnett comparison reinforces this caution, where the inferred fractions change to $69.0\%$ sub-$M_{\rm Ch}$, $17.5\%$ near-$M_{\rm Ch}$, and $13.5\%$ super-$M_{\rm Ch}$. This demonstrates that the apparent high-mass tail in the full multi-band model is not a robust channel census by itself, and that ejecta-mass point estimates should be interpreted together with posterior width, selection effects, and model systematics.

Generically, the events with $M_{\rm ej} < 1.2~M_\odot$ are consistent with double-detonation explosions~\citep{Livne1990,Fink2010,Sim2010, Blondin2017}, in which a thick helium shell accreted onto a sub-$M_{\rm Ch}$ CO white dwarf detonates, triggering secondary detonation of the core. Notably, \citet{Dhawan2017} previously identified two classes of fast-declining SNe~Ia that may correspond to distinct sub-$M_{\rm Ch}$ progenitor masses.
However, the smooth transition from sub-$M_{\rm Ch}$ to near-$M_{\rm Ch}$ masses without a clear gap does not by itself identify a common explosion mechanism; it may also arise from broad posteriors, selection effects, or the exclusion of peculiar sub-luminous classes. We note that the \citet{Dhawan2017} sample was dominated by 91bg-like supernovae, which are excluded from our analysis.  
The density at which a deflagration transitions to a detonation can be treated as a free parameter in near-$M_{\rm Ch}$ DDT calculations, but sub-$M_{\rm Ch}$ detonations and merger-driven explosions need not undergo a deflagration phase. Thus, similar inferred burning efficiencies across this mass range should be interpreted cautiously and may reflect modelling systematics or selection rather than a single DDT-like physical regime.

The high-mass tail ($38.9\%$ with $M_{\rm ej} > 1.5~M_\odot$) is intriguing but requires cautious interpretation. Interestingly, \citet{Bravo2022} argue from nucleosynthesis constraints that Chandrasekhar-mass explosions may represent only a small fraction of SNe~Ia, contrary to our observed dominance, though this may reflect differences between the full population and the cosmology sample.
One possibility is white dwarf mergers, where head-on collisions of two CO white dwarfs can produce super-$M_{\rm Ch}$ explosions with $M_{\rm ej} \sim 1.6$--$2.0~M_\odot$~\citep{Pakmor2010,Pakmor2012}. These merger models do not correspond to DDT explosions, and therefore would represent a distinct physical channel if they contribute substantially to the high-mass tail. However, merger rates may be insufficient to explain a large fraction of SNe~Ia~\citep{Ruiter2013}. Another possibility is rapidly rotating white dwarfs, where centrifugal support can stabilize white dwarfs beyond $M_{\rm Ch}$~\citep{Yoon2005,Hachisu2012}, though these appear to require high angular momentum and may not be plausible for a large population. Moreover, rapidly rotating massive-WD explosion calculations do not currently provide a straightforward explanation for a large population of spectroscopically normal SNe~Ia with super-$M_{\rm Ch}$ ejecta masses; for example, \citet{Fink2018} found that detonation models of rapidly differentially rotating WDs are very luminous but do not match known SN~Ia subclasses. The pseudo-bolometric comparison reduces the super-$M_{\rm Ch}$ point-estimate fraction to $13.5\%$ for $M_{\rm ej}>1.5~M_\odot$ ($9.8\%$ for $M_{\rm ej}>1.6~M_\odot$), suggesting that at least part of the full-model high-mass tail is likely produced by modelling assumptions rather than a literal population of super-$M_{\rm Ch}$ normal SNe~Ia.

Notably, we find that the SN~1991T-like population ($8.3\%$ of our sample) is physically distinct, with systematically higher $M_{\rm ej}$ (median $1.64~M_\odot$) and $M_{\rm Ni}$ ($1.38~M_\odot$, $33\%$ more than normals). This is further reinforced by the finding that $84\%$ of 91T-like SNe have $M_{\rm ej} > 1.4~M_\odot$, which suggests that 91T-like SNe may preferentially arise from super-$M_{\rm Ch}$ progenitors or represent the high-mass tail of the near-$M_{\rm Ch}$ distribution. Alternative interpretations include more energetic CO+CO mergers producing both high $M_{\rm ej}$ and efficient burning~\citep{Pakmor2012}, model-dependent near-$M_{\rm Ch}$ delayed detonations with high nickel yields~\citep{Hoeflich1996}, or different C/O ratios in the white dwarf increasing both explosion energy and nickel yield~\citep{Timmes2003}. The rarity of 91T-like SNe ($8.3\%$) compared to normals ($88.7\%$) suggests their progenitor channel has a lower volumetric rate or requires special conditions such as specific mass ratios in mergers or unusual accretion histories. However, our crude estimate does not incorporate the luminosity function, see e.g., \citet{Dimitriadis2025}.

Our derived ejecta masses depend on model assumptions which could influence our results. In particular, we assume spherical symmetry, but real SNe~Ia can be intrinsically asymmetric~\citeg{Cikota2019}. Multi-dimensional simulations show that DDT models produce $\sim 20\%$ variations in observables with viewing angle~\citeg{Kasen2006_pol}, while double detonations show even larger effects with ``polar'' versus ``equatorial'' viewing producing $\Delta M \sim 0.5$ mag~\citep{Townsley2019}. Our spherical model captures orientation-averaged properties, but individual SNe may induce viewing-angle-dependent biases of $\sim 0.1$--$0.2~M_\odot$ in $M_{\rm ej}$. Additionally, we assume one zone of $^{56}$Ni, and a simplified diffusion model, more detailed numerical and semi-analytical modelling shows this simplified diffusion under-predicts the bolometric luminosity over time~\citep{Pinto2000, Kasen2006}, this could shift our inferred $M_{\rm ej}$ downward~\citep{Khatami2019}. If systematic biases shift the distribution downward by $\sim 0.2M_\odot$, the super-$M_{\rm Ch}$ fraction would decrease to values more consistent with theoretical expectations. As we explore in the Appendix, different assumptions on gamma-ray opacity also shift the inferred median substantially. Detailed comparison with multi-dimensional radiative transfer simulations and more modelling is required to quantify these systematics.

Of further note are the high burning efficiencies (and indirectly high $^{56}$Ni) of our sample (see Fig.~\ref{fig:mass_plane}), which are on the higher end of theoretical expectations for many channels and modelling~\citep{Scalzo2014, Magee2024}. However, as indicated by our volume-limited analyses, this is influenced by Malmquist bias, and the volume-limited sample has both the burning efficiency $f_{\rm Ni} \approx 0.8$ and $^{56}M_{\rm Ni}\approx 1.0M_{\odot}$, which are less extreme than the full-sample point estimates but still high relative to many published model grids~\citep{Ruiter2025}. 
The pseudo-bolometric Arnett analysis provides an additional, independent reference scale for this tension. The reconstructed pseudo-bolometric luminosity is incomplete, so the resulting $M_{\rm Ni}$ values should be treated as lower limits; nevertheless, they show that the observed luminosity scale does not require most events to sit near $f_{\rm Ni}=1$. The gap between the full multi-band point estimates and the pseudo-bolometric lower-limit estimates is therefore an illustration, rather than a calibrated measure, of the sensitivity associated with SED redistribution, opacity modelling, extinction, and distance assumptions. Importantly, the pseudo-bolometric fits still include a tail of events with high $f_{\rm Ni}$, so the remaining tension is likely not solely an opacity-law artefact. It may instead reflect limitations common to both analyses, especially the one-zone assumption. In realistic SN~Ia ejecta, nickel mixing, compositional stratification, and a depth-dependent temperature and ionisation structure can alter the relation between peak luminosity, light-curve width, gamma-ray trapping, and the inferred ratio $M_{\rm Ni}/M_{\rm ej}$. These effects could reduce the apparent need for extreme nickel fractions in some objects without requiring those events to be nearly pure iron-group ejecta.

To quantify these population fractions while accounting for measurement uncertainties and selection effects, we performed hierarchical Bayesian inference on the nearby ($z \leq 0.06$) subsample. This analysis is in effect an attempt to understand the true SN~Ia population rather than just the selected cosmological sample. We model the population-level distributions of $M_{\rm ej}$ and $M_{\rm Ni}$ as truncated Gaussians with an approximate brightness-dependent selection function. 
The hierarchical analysis recovers well-constrained population parameters: for ejecta mass, we infer $\mu_{\rm ej} = 1.26 \pm 0.02~M_\odot$ with intrinsic scatter $\sigma_{\rm ej} = 0.33 \pm 0.01~M_\odot$, while for nickel mass we find $\mu_{\rm Ni} = 0.64
\pm 0.06~M_\odot$ with scatter $\sigma_{\rm Ni} = 0.42 \pm 0.02~M_\odot$. 
These constraints provide a model-dependent population summary with a substantial near-Chandrasekhar contribution. Critically, the inferred population fractions are $43 \pm 2\%$ sub-$M_{\rm Ch}$ ($M_{\rm ej} < 1.2~M_\odot$), $34 \pm 1\%$ near-$M_{\rm Ch}$ ($1.2 \leq M_{\rm ej} \leq 1.5~M_\odot$), and $24 \pm 2\%$ super-$M_{\rm Ch}$ ($M_{\rm ej} > 1.5~M_\odot$), significantly revising our initial naive estimates.
The similar-size split between sub- and near-$M_{\rm Ch}$ explosions, combined with the smooth continuous distribution ($\sigma/\mu \sim 0.5$),\textemdash{}although this is enforced by the hyperprior\textemdash{} is a useful single-population summary, but should not be interpreted as a test excluding multiple explosion mechanisms. In particular, a DDT interpretation is physically most relevant to near-$M_{\rm Ch}$ progenitors, while sub-$M_{\rm Ch}$ and merger scenarios can represent distinct channels. 
This merits further investigations with more flexible hyperprior and by considering the burning efficiencies, which are predictions of these channels. 
Simply considering the two distributions, the hierarchical model implies a burning efficiency distribution with mean $f_{\rm Ni} = 0.52 \pm 0.30$, indicating that on average approximately half of the ejecta mass is synthesized into ${}^{56}$Ni. The broad distribution reflects substantial diversity in burning conditions, with $\sim 23\%$ of events achieving high burning efficiencies ($f_{\rm Ni} > 0.8$) and $\sim 13\%$ approaching complete burning($f_{\rm Ni} \approx 1$). These values overlap the broad range spanned by published near-$M_{\rm Ch}$ delayed-detonation and sub-$M_{\rm Ch}$ detonation grids, but the burning fraction alone is not a unique channel diagnostic.
The reduced super-$M_{\rm Ch}$ fraction ($24\%$ versus the naive $\sim 39\%$) after accounting for selection effects and measurement uncertainties alleviates tension with theoretical merger rates, although exotic super-$M_{\rm Ch}$ channels may still contribute at the $\sim 20\%$ level. However, this estimate could be affected by a different selection model with a more explicit dependence on $M_{\rm ej}$.
\subsection{Systematic Uncertainties and Model Limitations}
\label{sec:systematics}
Our wavelength-dependent Arnett model relies on several simplifications that may introduce systematic biases. We assume homologous expansion with $v(r,t) = r/t$, which is generally a good approximation after the early hydrodynamic phase of many SN~Ia explosion calculations, including DDT models that are evolved to homologous expansion. The larger concern for our analysis is therefore not the assumption of late-time homology itself, but the one-zone density, opacity, and SED approximations used to map the bolometric power into multi-band photometry. 
Departures from homology are therefore unlikely to be the dominant systematic for the normal SNe~Ia in this sample, but deviations from the one-zone approximation will likely matter for individual events. 
Our wavelength-dependent opacity formula is also phenomenological and not derived from detailed radiation transport. Our model is also a simple one-zone model, which implicitly will fail to capture the full dynamics that could be revealed by a multi-zone model or full radiation hydrodynamics simulation. Such detailed modelling, including radiation hydrodynamics or full non-LTE radiative transfer treatment~\citep{Dessart2014}, would be more accurate but computationally prohibitive for 2,205 SNe. However, modified approaches, such as those using surrogate models~\citep{Magee2024, Sarin2025}, may provide a potential path forward. The key question is whether our parameterization introduces systematic biases affecting the population mean versus random scatter increasing uncertainty. Comparison with detailed models and a smaller subsample of events with more extensive multi-wavelength coverage would help quantify these effects.

The injection-and-recovery results in Sec.~\ref{sec:ztfvalidation} show that visually good fits can coexist with broad parameter degeneracies and recovery offsets. In the two selected realisations, $M_{\rm Ni}$ is recovered close to its injected value and $M_{\rm ej}$ below it, while changing the assumed distance scale shifts $M_{\rm Ni}$ by approximately $15\%$. These examples expose sensitivity to cadence, distance, and model choice, but are limited to establish the size of any systematic bias. Consequently, fit quality and statistical posterior width alone do not establish the external accuracy of the event-level masses, burning fractions, population distributions, or correlations. Further work on a curated sample and with numerical simulations could provide an answer here.

Photometric systematics also contribute uncertainty. We add $0.05$ mag systematic uncertainty in quadrature to account for ZTF photometric calibration uncertainties~\citep{Rigault2025}. Residual calibration errors could introduce colour-dependent biases affecting $A_V$ and $\kappa_{\rm opt}$ or magnitude zero-point offsets biasing $M_{\rm Ni}$ systematically. Our SED also ignores any strong spectral features and effectively only models a continuum, so errors in the SED shape due to line blanketing or strong spectral features will propagate to our inferred parameters. Our inferred host-extinction assumes the ~\citet{Fitzpatrick1999} extinction law with $R_V = 3.1$, which may not be reflective of extinction in all galaxy types.
The pseudo-bolometric analysis has a different set of systematics. It avoids the phenomenological wavelength-dependent opacity prescription, but it inherits the assumptions of the SALT2 SED template, the adopted wavelength range, the distance scale, and the treatment of dust and missing UV/NIR flux. These effects mostly make the pseudo-bolometric luminosity incomplete, so the inferred nickel masses are best treated as lower-limit-like estimates. The method is therefore not a replacement for the full multi-band analysis; rather, it brackets the scale of plausible systematic shifts and provides an external check on whether the high nickel fractions from the full model are robust.

Another source of uncertainty is the treatment of gamma-ray thermalisation. We rely on methods commonly used in semi-analytical modelling of a wide variety of supernovae. However, this is still somewhat susceptible to the choice of gamma-ray opacity, $\kappa_{\gamma}$ in Eq.~\ref{eq:gamma-rayleakage}, which we fix to $0.03$ cm$^2$ g$^{-1}$, while this value instead covers a larger range and is also time-dependent~\citep{Guttman2024}. In the Appendix, we show the impact of marginalising over a physics-motivated prior on the gamma-ray opacity. Notably, marginalising over this prior shifts our inferred $M_{\rm Ni}$ distribution to smaller values, with a median of $1.05~M_{\odot}$ (cf. $1.19~M_{\odot}$), but individual event-level posteriors remain consistent with the estimate for a fixed $\kappa_{\gamma}$. Our hierarchical modelling analysis does not change significantly based on what gamma-ray opacity we assume, as we are dominated by selection bias and that the individual posteriors are already significantly broad. 
\subsection{Comparison with Previous Studies}
\label{sec:comparison}

Our sample of 2,205 SNe~Ia with physical parameter measurements is roughly $100$ times larger than previous lightcurve modelling efforts. \citet{Scalzo2014} applied hierarchical Bayesian inference to 337 SNe~Ia with $z < 0.7$ using an empirical relation between ejected mass and light-curve width, finding that $25$--$50\%$ of normal SNe~Ia are inconsistent with Chandrasekhar-mass explosions and that the distribution displays a long tail towards sub-$M_{\rm Ch}$ masses but cuts off sharply above $1.4$ M$_\odot$. Their conclusion that super-Chandrasekhar-mass explosions comprise no more than $1\%$ of spectroscopically normal SNe~Ia is notably lower than our super-$M_{\rm Ch}$ fraction of $38.9\%$, though as discussed in Sec.~\ref{sec:progenitors}, systematic biases in our spherical model may inflate this fraction. The qualitative agreement on sub-$M_{\rm Ch}$ contributions and the sharp cutoff near $M_{\rm Ch}$ is reassuring despite methodological differences.
Recently, \citet{Bora2024} performed physical modelling of 28 SNe~Ia using multiband photometry and optical spectroscopy, finding that the majority show ejecta masses below the Chandrasekhar limit with mean $M_{\rm ej} \approx 1.1 \pm 0.3$ M$_\odot$, which is lower than our median of $1.40$ M$_\odot$ in the cosmological sample. This difference likely reflects our sample selection that excludes sub-luminous 91bg-like events, which would preferentially populate the low-$M_{\rm ej}$ regime. Their identification of a subset with higher ejecta masses between $1.2$--$1.5$ M$_\odot$ consistent with near-Chandrasekhar explosions agrees well with our near-$M_{\rm Ch}$ peak. 
Our hierarchical modelling results are generally more consistent with these previous studies. Moreover, the agreement within measurement uncertainties and methodological differences across studies spanning different sample sizes, redshift ranges, and fitting techniques strengthens confidence in our results. In particular, that the near-$M_{\rm Ch}$ peak is a robust feature of the cosmological SN~Ia population.

Our median $v_{\rm ej} \sim 11,000$ km s$^{-1}$ agrees well with Si~II $\lambda 6355$ velocities at maximum light from \citet{Foley2011}, who measured a mean of $v_{\rm Si~II} \sim 10,500$ km s$^{-1}$ for their sample, validating our model's velocity parameter as a physical quantity. The consistency between our model-derived bulk ejecta velocity and spectroscopic line velocities is non-trivial, as the two measurements probe different aspects of the explosion: our $v_{\rm ej}$ represents the characteristic velocity of the entire ejecta assuming homologous expansion, while Si~II velocities trace specific layers in the ejecta where silicon is only partially ionized. The agreement suggests that these layers are representative of the bulk flow and that our estimates are reasonable.

\subsection{Implications for Future Surveys}
\label{sec:future_surveys}
The key implication of our analysis and modelling is a demonstration that physical parameters can be extracted from light curves and such models can describe the light curves at a comparable level to the empirical SALT2 model (at least in the limited bands and cases we currently explored). 
The relatively small CPU cost of an analysis on 2,205 SNe ($\approx 10$ minutes per supernova) can be further improved with the use of GPUs, which strongly supports the idea that population-level physical modelling is feasible for next-generation samples with $N \sim 10^5$ SNe. 
This analysis also opens the door to simultaneous hierarchical modelling to probe progenitor and explosion population properties, in contrast to the currently more practical approach of ``recycling'' single-event posteriors that is ubiquitous in gravitational-wave astronomy~\citep{Abbott2023}.

Our results can also help inform the analysis of next-generation SN~Ia samples in upcoming surveys including Vera Rubin Observatory, the \textit{Nancy Grace Roman Space Telescope}, and Euclid. 
As highlighted by Fig.~\ref{fig:fit_quality}, the inclusion of wavelength-dependent recombination effects is vital for physical modelling of multi-band light curves. It also highlights that more detailed modelling in that direction and therefore observations in rest-frame $i$-band and redder observations are particularly valuable, especially at phases where the red-band shoulder or secondary maximum is expected, for understanding SN~Ia physics and the true root of the SALT2 colour parameter. 
Across our sample, we find a recombination epoch with mean $t_{\rm recomb}= 32$ days with an average width $\sigma_{\rm recomb} = 6$ days, which indicates that observations between $25$--$45$ days rest-frame are crucial for constraining wavelength-dependent opacity. 
However, this modelling is also complex, especially from first principles, so observations could instead be prioritised to create robust bolometric corrections for SNe~Ia across a range of subtypes. 
An alternative but complementary approach may be to have dedicated small surveys or follow-up campaigns with the explicit aim of obtaining multi-wavelength coverage on a small subsample of SNe~Ia to robustly extract the bolometric light curves, which are easier to model. One of our key results is that the host mass-step may be rooted in nickel production, which is likely tied to the age and metallicity of the local environment. This strongly suggests the need to further spectroscopically explore the host galaxy and local environments of a large sample of SNe~Ia. 
\subsection{Future Directions}
\label{sec:future}
Our analysis extracts physical parameters for individual SNe but treats each as independent. A logical next step would be to extend the hierarchical Bayesian modelling in Sec.~\ref{sec:hierarchicalinference}, to more accurately capture the selection procedure and infer more population-level parameters. 
Fitting this hierarchical model simultaneously to all 2,205 SNe while fully incorporating a selection-function by modelling the detection efficiency as a function of $M_{\rm ej}$, $M_{\rm Ni}$, and redshift, would produce strong constraints on population fractions of sub-$M_{\rm Ch}$, near-$M_{\rm Ch}$, and super-$M_{\rm Ch}$ channels, measure intrinsic scatter in physical parameters at fixed observables, detect correlations between parameters while marginalizing over measurement uncertainties. They would also enable more detailed tests of correlations with host-galaxy properties and explore the redshift evolution of different parameters to test for cosmic time dependence of progenitor properties. 
This approach has been successfully applied to gravitational-wave events to infer black hole and neutron star population properties~\citeg{Abbott2023} and would yield robust constraints on progenitor demographics for SNe~Ia.

Beyond hierarchical inference, a natural extension of this work is to develop a physically-motivated distance estimator that directly uses the extracted explosion parameters rather than empirical light curve features. The strong correlation between nickel mass and luminosity suggests that a standardization of the form $M = M_0 + \alpha M_{\rm ej} + \beta A_{\rm v, host} + \Delta f(\theta)$, could provide distances with comparable precision to SALT2, with the advantage that the coefficients $\alpha$ and $\beta$ would have direct physical interpretations. Here, the ejecta mass term captures the primary luminosity driver (total radioactive energy release, through its strong link to the nickel mass), while also capturing at some level diffusion time effects that modulate the light curve width independently of peak brightness. Additional model parameters could also be included to further standardize the SNe from first principle physics.
Such a physical standardization would be advantageous for several reasons. First, it would enable direct modelling of systematic evolution: as galaxy demographics and progenitor metallicity distributions change with redshift, we could predict how the $M_{\rm Ni}$ and $M_{\rm ej}$ distributions evolve using stellar population synthesis and delay time distribution models, rather than empirically calibrating the redshift dependence of $x_1$ and $c$. 
Second, it would allow propagation of theoretical uncertainties from explosion physics (e.g., opacity variations, asymmetries) directly into cosmological error budgets. Third, unlike SALT2 which requires training on a low-redshift sample to define the ``normal'' SN~Ia template, a physical model trained on local SNe could in principle extrapolate to high redshift with quantifiable systematic uncertainties rooted in explosion physics rather than empirical extrapolation.

Another key extension is directly towards the modelling, which would be to investigate the extended model's performance on a smaller subsample of SNe~Ia with extensive multi-wavelength coverage~\citep{Hicken2009, Burns2014}, and especially including the near-IR bands where secondary maxima are more pronounced would provide stronger constraints on wavelength-dependent opacity. Here, we could compare our analysis to more physical models or compare the systematics associated with multi-wavelength vs bolometric light curve modelling. A similar analysis could also investigate the consistency between our light curve fitting estimates compared to estimates derived from e.g., nebular-phase observations at $t > 100$ days where forbidden-line emission directly traces nickel mass and density structure~\citep{Mazzali2007}. Our current analysis also only uses photometry, but adding spectroscopic constraints from velocity measurements of Si~II and Ca~II lines that constrain $v_{\rm ej}$ independently, line ratios of Fe~II/Fe~III that constrain temperature and ionization during recombination, and nebular spectra where [Fe~II], [Fe~III], and [Co~III] line fluxes directly measure $M_{\rm Ni}$ and density~\citep{Childress2015} would substantially reduce parameter degeneracies.

\section{Conclusions}\label{sec:conclusions}

We have presented the largest homogeneous extraction of physical parameters from SN~Ia light curves to date, analysing 2,205 SNe~Ia from the Zwicky Transient Facility Data Release 2~\citep{Rigault2025} with a wavelength-dependent one-zone radioactive decay model. By incorporating a phenomenologically-motivated extension to capture Fe~III $\rightarrow$ Fe~II recombination physics in a one-zone radioactive diffusion model, we successfully model multi-band ($gri$) photometry and extract fundamental explosion parameters including ejecta mass ($M_{\rm ej}$), nickel mass ($M_{\rm Ni}$), and ejecta velocity ($v_{\rm ej}$). Our Bayesian analysis provides statistical posterior uncertainties for the full sample, while the validation tests reveal additional model sensitivity and degeneracy. The independent pseudo-bolometric comparison further shows that the absolute nickel-mass scale is sensitive to the SED treatment, although its incomplete wavelength coverage makes it a lower-limit comparison rather than an unbiased replacement. Our key findings are:

\begin{itemize}
\item Injection-and-recovery tests using realistic ZTF observing logs recover the injected nickel scale but bias the ejecta mass low for both tested cadences. A change from an SN-calibrated to a Planck-like distance scale shifts the inferred nickel mass by approximately $15\%$, and a grey Arnett recovery changes the inferred parameters and fit evidence. These tests show that acceptable light-curve fits do not guarantee unbiased individual-event parameters and motivate interpreting the sample through full posteriors, systematic comparisons, and hierarchical inference. 

\item We detect a highly significant correlation between model-inferred nickel mass and SALT2 stretch ($\rho = 0.57$), suggesting that the empirical brighter-slower relation is consistent with a physical relationship in which brighter, slower-declining SNe produce more $^{56}$Ni. The median nickel mass for the cosmological sample is $M_{\rm Ni}= 1.19 \pm 0.33~M_\odot$, higher than the canonical $\sim 0.6$ M$_\odot$ often quoted, but consistent with estimates from smaller samples and reflecting selection bias toward efficiently-burning events in cosmology samples. This median estimate is also strongly affected by the shape and broadness of the individual posteriors. Analysis of a restricted sample less affected by Malmquist bias, or correcting for selection biases, reduces this to a more consistent value. The much lower pseudo-bolometric estimate demonstrates that the absolute nickel-mass scale, and therefore the physical interpretation of this correlation, remains conditional on the SED, opacity, distance, and one-zone assumptions; neither comparison alone identifies the unbiased $M_{\rm Ni}$ distribution.

\item SNe~Ia in low-mass hosts ($\log_{10}(M_*/M_\odot) < 10$) are inferred to synthesize $12\%$ more nickel than those in high-mass hosts ($\Delta M_{\rm Ni} = 0.13~M_\odot$ on average). This suggests that the well-known Hubble residual mass step could arise from intrinsic physical differences between progenitor populations in different environments, likely reflecting progenitor age and metallicity effects. After SALT2 standardization partially corrects for this via the $M_{\rm Ni}$--$x_1$ correlation, a residual $\sim 0.06$ mag offset remains, consistent with observations. The Hubble residual (Fig.~\ref{fig:hubble}) shows a strong dependence on the nickel mass, strongly hinting that the standardisation process fails to properly account for the nickel production in different galaxies. This analysis suggests that the mass-step could be removed if standardisation was based on physically-derived parameters. 

\item Fast-declining SNe ($x_1 < 0$) have systematically lower ejecta masses ($M_{\rm ej} = 1.23~M_\odot$ versus $1.51~M_\odot$ for slow decliners, $\Delta = 0.28$ M$_\odot$), demonstrating that SALT2 stretch directly tracks the ejecta mass (as expected due to diffusion times) and indirectly tracks the progenitor channel, providing evidence for a higher sub-$M_{\rm Ch}$ fraction among fast decliners.

\item Considering our cosmologically-selected population, the ejecta mass distribution exhibits a strong peak at $M_{\rm ej} = 1.40~M_\odot$, consistent with the Chandrasekhar mass. This suggests a substantial near-Chandrasekhar contribution within the assumptions of our model. The distribution is broad and does not show clear bimodality in the point-estimate summaries, but this does not rule out multiple channels, velocity subgroups, or sub-populations blurred by posterior width and selection effects. The peak contains $42.7\%$ of the sample, with symmetric wings extending to sub-$M_{\rm Ch}$ ($28.2\%$ with $M_{\rm ej} < 1.2M_\odot$) and super-$M_{\rm Ch}$ ($29.1\%$ with $M_{\rm ej} > 1.6M_\odot$) regimes. However, the absence of clear separation between these regimes indicates that nominal mass thresholds may not correspond to fundamentally distinct progenitor channels.

\item SN~1991T-like events ($8.3\%$ of the sample) are physically distinct from normal SNe~Ia, with systematically higher ejecta masses ($M_{\rm ej} = 1.64$ M$_\odot$ versus $1.38$ M$_\odot$ for normals) and $30\%$ more nickel production. Notably, $84\%$ of 91T-like SNe have $M_{\rm ej} > 1.4M_\odot$, though whether this indicates super-$M_{\rm Ch}$ progenitors or simply represents the high-mass tail of the near-$M_{\rm Ch}$ distribution with exceptionally efficient burning remains unclear. We note that this is based on our photometric model which maybe systematically biased.

\item The opacity strength distribution has a mean $s=1.2$, which indicates that wavelength-dependent opacity variations during Fe~II recombination at an average time of $t_{\rm recomb}= 32 \pm 6$ days are a generic feature of SN~Ia light curves, demonstrating the necessity of extensions beyond the simple gray-opacity one-zone modelling to capture multi-band observations which probe the redder bands.

\item Summarising all posteriors solely by their median, the nickel mass fraction has a mean $0.9 \pm 0.1$ across all ejecta masses with no significant dependence on $M_{\rm ej}$. This uniformly high burning efficiency is likely partially driven by selection bias as our cosmology sample excludes sub-luminous, inefficiently-burning SNe~Ia such as SN~1991bg-like events with $f_{\rm Ni} \sim 0.2$--$0.4$ and SN~Iax with $f_{\rm Ni} < 0.3$. Our general results therefore characterize the standardizable population used for distance measurements rather than the full diversity of thermonuclear transients. However, the high inferred efficiencies should also be treated as a likely sign of model dependence in the individual-event point estimates, especially because both the full multi-band model and the pseudo-bolometric Arnett comparison contain events with large inferred $f_{\rm Ni}$, that are unlikely to be correct.

\item As an independent comparison, we reconstruct pseudo-bolometric light curves from the fitted-colour SALT2 SEDs, correct for Milky Way extinction, adopt $H_0=73.7~{\rm km~s^{-1}~Mpc^{-1}}$, and fit them with a grey Arnett model. This analysis gives lower median values, $M_{\rm Ni}\simeq0.45~M_\odot$, $M_{\rm ej}\simeq1.02~M_\odot$, and $f_{\rm Ni}\simeq0.43$. The pseudo-bolometric luminosity is incomplete and depends on template, distance, and extinction assumptions, so these nickel masses are lower-limit-like estimates. The comparison demonstrates that the high nickel fractions from the full multi-band model are sensitive to SED and opacity assumptions, while the remaining high-$f_{\rm Ni}$ tail in the pseudo-bolometric fits shows that simplified one-zone bolometric modelling also has systematics. This motivates interpreting individual-event point estimates through the broader posterior and hierarchical population analysis. Accordingly, the pseudo-bolometric and full-model values should be treated as complementary, model-dependent constraints rather than as a lower and upper measurement of a known systematic offset.

\item Hierarchical modelling on the redshift-limited sample ($z \leq 0.06$) to uncover the properties of the true Ia population, by accounting for measurement and posterior uncertainties and selection effects. With this analysis, we find that the Ia population can be well described by Gaussian distributions in ejecta and nickel mass with population-level parameters $\mu_{\rm ej} = 1.26 \pm 0.01~M_\odot$ with intrinsic scatter $\sigma_{\rm ej} = 0.33 \pm 0.01~M_\odot$ for ejecta mass, and $\mu_{\rm Ni} = 0.64 \pm 0.06~M_\odot$ with scatter $\sigma_{\rm Ni} = 0.42 \pm 0.02~M_\odot$ for nickel mass. The inferred population fractions are $43 \pm 2\%$ sub-$M_{\rm Ch}$ ($M_{\rm ej} < 1.2~M_\odot$), $34 \pm 1\%$ near-$M_{\rm Ch}$ ($1.2 \leq M_{\rm ej} \leq 1.5~M_\odot$), and $24 \pm 2\%$ super-$M_{\rm Ch}$ ($M_{\rm ej} > 1.5~M_\odot$), significantly revising estimates without selection and revealing a similar-size split between sub- and near-Chandrasekhar explosions. The reduced super-$M_{\rm Ch}$ fraction alleviates tension with theoretical merger rates, while the continuous smooth distribution with $\sigma/\mu \sim 0.5$ should be interpreted as a consequence of the adopted single-population hierarchical model rather than evidence against multiple channels. Moreover, the inferred burning efficiencies move away from the extreme $f_{\rm Ni}\simeq1$ boundary and overlap the broad range covered by published explosion-model grids, although burning efficiency alone does not identify a unique channel. This marginalises the individual statistical posteriors and the adopted selection model, but does not remove a systematic common to the light-curve model.

\end{itemize}

Our results have direct implications for cosmological applications and SN~Ia astrophysics. The strong correlations between our physical parameters and empirical SALT2 parameters suggests that standardization has a robust physical basis, with the brighter-slower relation fundamentally being a mass-luminosity relation. We note that this is predicated by the assumptions built into a one-zone model. However, the host galaxy mass step demonstrates that progenitor populations evolve with cosmic environment. As galaxy demographics change with redshift, the mix of young versus old and metal-poor versus metal-rich progenitors will shift, potentially introducing systematic errors in distance measurements. The likely coexistence of different explosion mechanisms in the standardizable sample means that if channel fractions evolve with redshift, this could introduce redshift-dependent systematics. Understanding and correcting for these selection effects through hierarchical Bayesian modelling of population evolution is crucial for precision cosmology.

As discussed above, our treatment of the selection-function is not rooted in a proper end-to-end analysis of a ZTF pipeline with a one-zone model. Furthermore, we only explore the distributions of ejecta and nickel masses and also assume they are independent. The logical next step is to further extend this analysis to fully account for selection and modelling the population-level distributions of multiple parameters across the full sample. This approach would lead to strong inferences into the intrinsic population fractions of different progenitor channels, redshift evolution of progenitor demographics, environmental dependencies, and intrinsic scatter in physical parameters at fixed observables. 
Such an analysis would provide comprehensive census of SN~Ia progenitor populations and enable direct comparison with theoretical population synthesis models. Furthermore, extending our wavelength-dependent model to near-IR bands where secondary maxima are more pronounced would provide tighter constraints on recombination physics, while nebular-phase observations where forbidden-line emission directly traces nickel mass would further validate our photometric measurements. Incorporating spectroscopic information would substantially reduce parameter degeneracies, and detailed comparison with 3D radiative transfer simulations would quantify systematic biases from our simplifying assumptions.

Our methodology is readily applicable to upcoming large SN~Ia samples from the Vera Rubin Observatory Legacy Survey of Space and Time with $N \sim 10^6$ SNe, the \textit{Nancy Grace Roman Space Telescope} with $N \sim 10^4$ high-redshift SNe. Physical parameter extraction for these samples will enable precision tests of progenitor evolution with redshift, measurement of the delay time distribution from cosmic star formation history, identification of rare explosion channels, and physical corrections to cosmological distance moduli that reduce systematic uncertainties in dark energy constraints.

By combining the statistical power of a 2,205 SN sample with physically-motivated modelling of wavelength-dependent opacity effects, we have compiled a comprehensive characterization of SN~Ia explosion physics. In our sample, near-Chandrasekhar ejecta masses are prominent in the standardizable population, but the broad distribution and model assumptions prevent us from ruling out distinct sub-populations or multiple channels.
We find evidence that variations in nickel synthesis, progenitor age, and metallicity contribute to the empirical brighter-slower relation and host-galaxy mass step, but note that this interpretation is model-dependent. 
Selection biases and accounting for measurement and modelling uncertainties are critical for interpreting physical parameter distributions, with the uniformly high burning efficiency likely reflecting cosmology sample selection, modelling systematics, and measurement and modelling uncertainties rather than suggestive of the true diversity of thermonuclear explosions. The two synthetic tests make this limitation explicit for the selected realisations: good fits coexist with broad degeneracies and sensitivity to distance and opacity assumptions. Together with the pseudo-bolometric nickel-mass lower limits, they show that the inferred mass scale is model-dependent. Neither analysis determines the direction or magnitude of bias across the real sample, so the reported masses, burning fractions, and correlations should be read as model-dependent measurements.
Wavelength-dependent opacity effects during Fe~II recombination are a generic feature necessitating multi-band observations, and hierarchical Bayesian inference of population properties is the essential next step for robust progenitor demographics. 
As next-generation surveys push toward samples of $10^5$--$10^6$ SNe~Ia, the physical modelling framework developed here will be valuable for translating observed light curves into progenitor population properties and may provide another standardisation procedure to reach precision constraints on cosmological parameters with well-understood systematic uncertainties rooted in explosion physics.
\section*{Acknowledgments}
We thank the anonymous referee whose comments have significantly improved the manuscript. We also thank Doron Kushnir for helpful comments and suggestions for further analysis. N. Sarin acknowledges support from the Kavli Foundation. LK acknowledges support for an Early Career Fellowship from the Leverhulme Trust through grant ECF-2024-054 and the Isaac Newton Trust through grant 24.08(w). C. M. B. O. acknowledges support from the Royal Society (grant Nos. DHF-R1-221175 and DHF-ERE-221005). E.E.H.\ is supported by a Gates Cambridge Scholarship (\#OPP1144). M. Ginolin, M. Grayling, A. Do, and K.S. Mandel are supported by the European Union’s Horizon 2020 research and innovation programme under ERC Grant
Agreement No. 101002652 (BayeSN; PI K. Mandel). A.A.M. is supported by Cottrell Scholar Award \#CS-CSA-2025-059 from the Research Corporation for Science Advancement and DoE award \#DE-SC0025599.

\section*{Data Availability}
All data are available from ZTF DR2 release~\citep{Rigault2025}. Median parameters with errors will be made available at \url{https://github.com/nikhil-sarin/sn_release} upon publication. Full posteriors and scripts available upon request. All analysis was performed using the open-source \program{Redback} v1.13.1 publicly available at \url{https://github.com/nikhil-sarin/redback}. This work made use of {\sc Redback} \citep{Sarin2024}, {\sc bilby} \citep{Ashton2019}, {\tt PyMultiNest} \citep{Buchner2016}, {\tt numpy} \citep{Harris2020}, {\tt scipy} \citep{Virtanen2020}, {\tt matplotlib} \citep{Hunter2007}, {\tt astropy} \citep{AstropyCollaboration2022}.

\bibliographystyle{mnras}
\bibliography{paper}

\appendix

\section{Validation with SN2011fe}
\label{app:validation}

To validate our modelling approach, we fit SN~2011fe, one of the best-studied Type Ia supernova to date with extensive multi-wavelength photometry and spectroscopy~\citep[e.g.,][]{Nugent2011, Pereira2013}. This nearby event ($z=0.000804$, $D_{L} \approx 6.4$ Mpc in M101) provides an ideal benchmark for testing our physically-motivated semi-analytic model against well-constrained observations.

\begin{figure}
\centering
\includegraphics[width=0.48\textwidth]{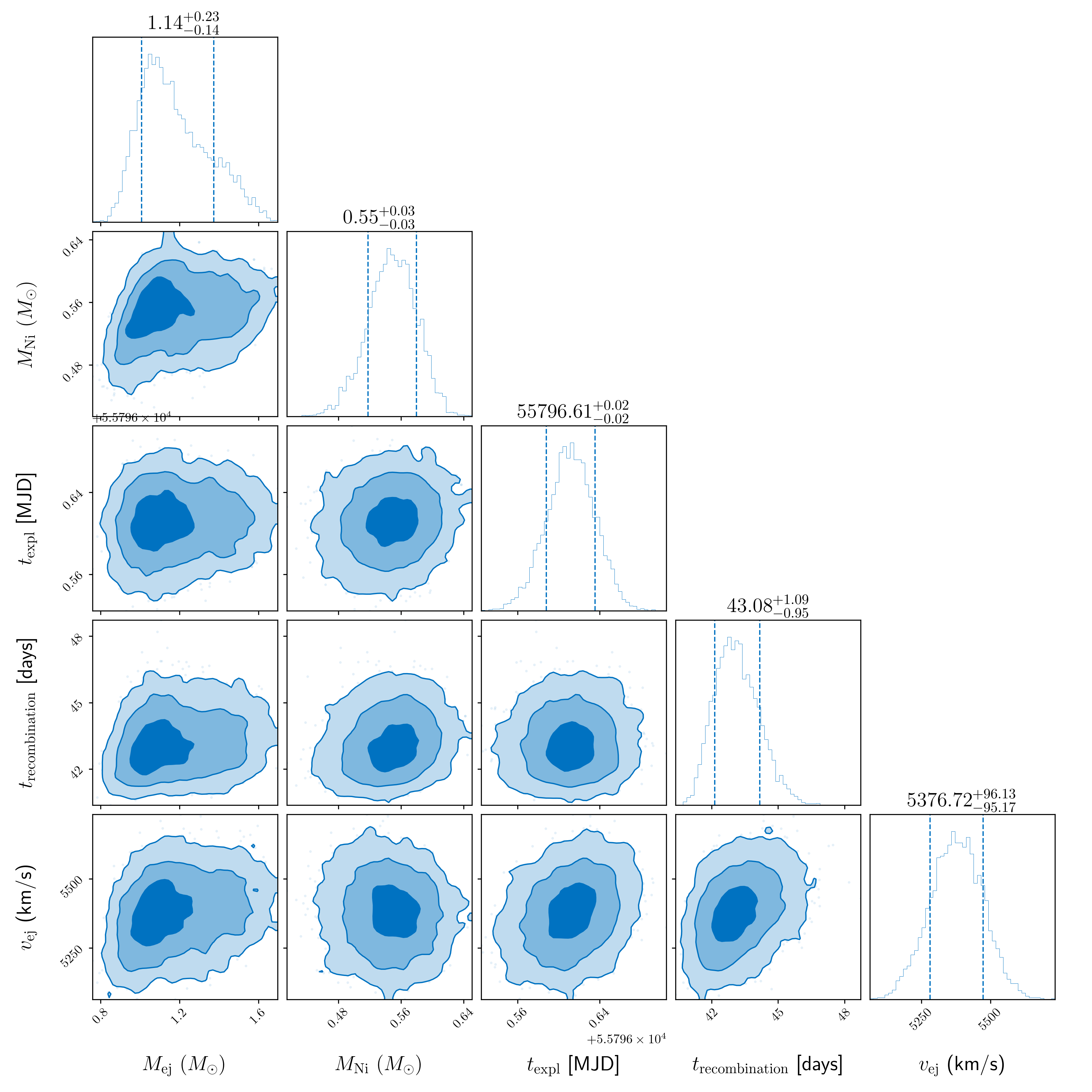}
\caption{Posterior on key parameters from our analysis on SN~2011fe}
\label{fig:sn2011fe_corner}
\end{figure}

\begin{figure}
\centering
\includegraphics[width=0.48\textwidth]{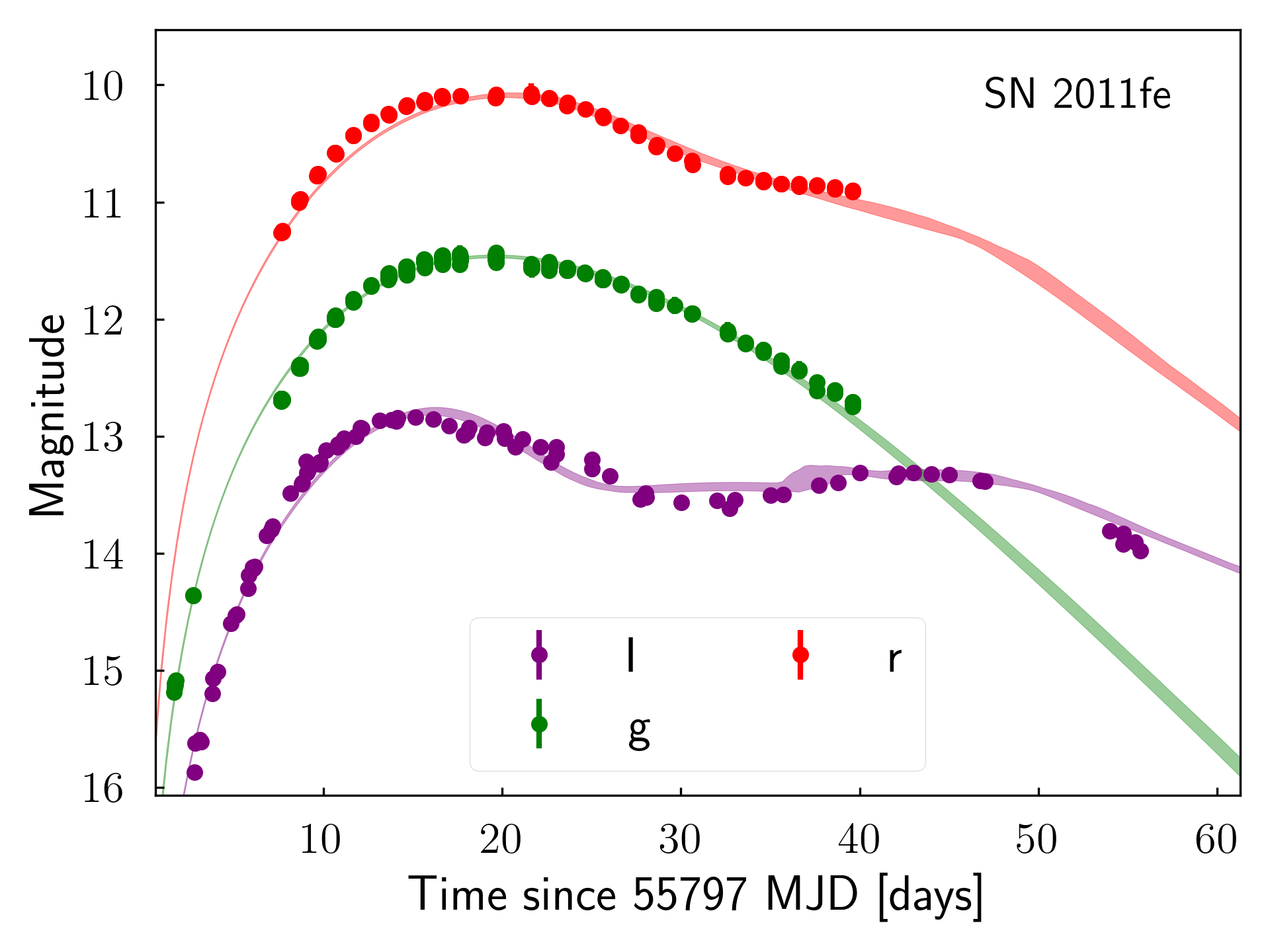}
\caption{Multi-band light curve fit for SN~2011fe using our extended model. The model (solid lines) captures the rise, peak, and decline across the g, r, and I-band filters.}
\label{fig:sn2011fe_lc}
\end{figure}

We fit the g, r, and I-band photometry of SN~2011fe using our extended model (as applied throughout this work). The inferred physical parameters are: ejecta mass $M_{\rm ej} = 1.14^{+0.23}_{-0.14}~M_\odot$, nickel mass $M_{\rm Ni} = 0.55\pm 0.03~M_\odot$, ejecta velocity $v_{\rm ej} = 5380\pm100$ km s$^{-1}$. The burning efficiency is $f_{\rm Ni} = 0.48^{+0.06}_{-0.08}$, with all uncertainties quoted at $68\%$ credible interval. We note that posteriors for SN~2011fe are narrower than the posteriors for almost all of our ZTF sample due to the higher quality and number of data available. Our posterior and fitted lightcurves are shown in Fig.~\ref{fig:sn2011fe_corner} and Fig.~\ref{fig:sn2011fe_lc}, respectively.

Our constraints and fits provide important context for interpreting our full sample results. The ejecta mass $M_{\rm ej} = 1.14^{+0.23}_{-0.14}$ M$_\odot$ is lower than the Chandrasekhar mass, consistent with estimates from previous studies~\citep{Nugent2011, Pereira2013}. 
The nickel mass $M_{\rm Ni} = 0.55$ M$_\odot$ is consistent with literature values from late-time spectroscopy and bolometric light curves~\citep[e.g.,][]{Pereira2013}, which found $M_{\rm Ni} \approx 0.4-0.6$ M$_\odot$. 
The burning efficiency $f_{\rm Ni} \approx 0.5$ places SN~2011fe in the regime predicted for Chandrasekhar-mass delayed-detonation models.
The relatively tight constraint on nickel mass (uncertainty $\sim 6\%$) despite broader ejecta mass uncertainty ($\sim 9\%$) reflects that the light curve peak is primarily sensitive to $M_{\rm Ni}$, while the diffusion timescale depends on both $M_{\rm ej}$ and $v_{\rm ej}$, leading to partial degeneracy. This could also be one of factors at play for our velocity estimate, which is inconsistent with theoretical expectations and spectroscopic constraints or it could point towards a different systematic within the model.

This validation confirms that our framework produces physically reasonable results for a well-studied event and a model that captures the data well, capturing the bump in i-band and the small shoulder in r-band. Given most ZTF DR2 events have worse data than SN~2011fe, this analysis also provides a good upper limit for the precision on key quantities in our analysis at an individual-object level.

\section{Impact of gamma-ray opacity}
Here we explore the impact of marginalising over a prior on the gamma-ray opacity. In particular, instead of fixing $\kappa_{\gamma} = 0.03~\rm{cm}^2\rm{g}^{-1}$, we use a physically-motivated uniform prior from $0.02-0.06~\rm{cm}^2\rm{g}^{-1}$, motivated by~\citet{Guttman2024} and marginalise over this prior. 

\begin{figure*}
\centering
\includegraphics[width=0.95\textwidth]{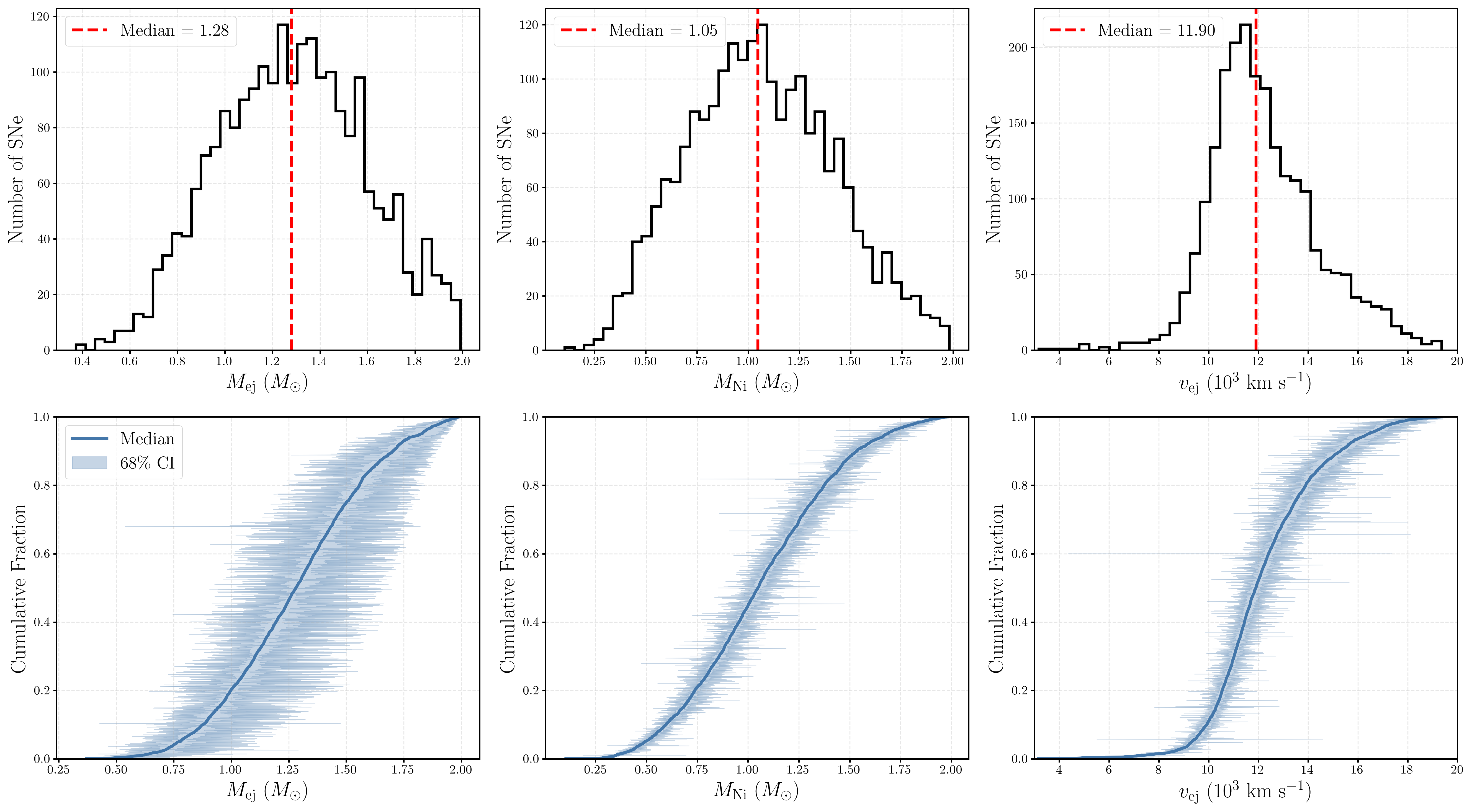}
\caption{Same as Fig.~\ref{fig:mass_distributions} but for analyses where we marginalise over a gamma-ray opacity prior.}
\label{fig:mass_distributions_opacity}
\end{figure*}

\begin{figure*}
\centering
\includegraphics[width=0.95\textwidth]{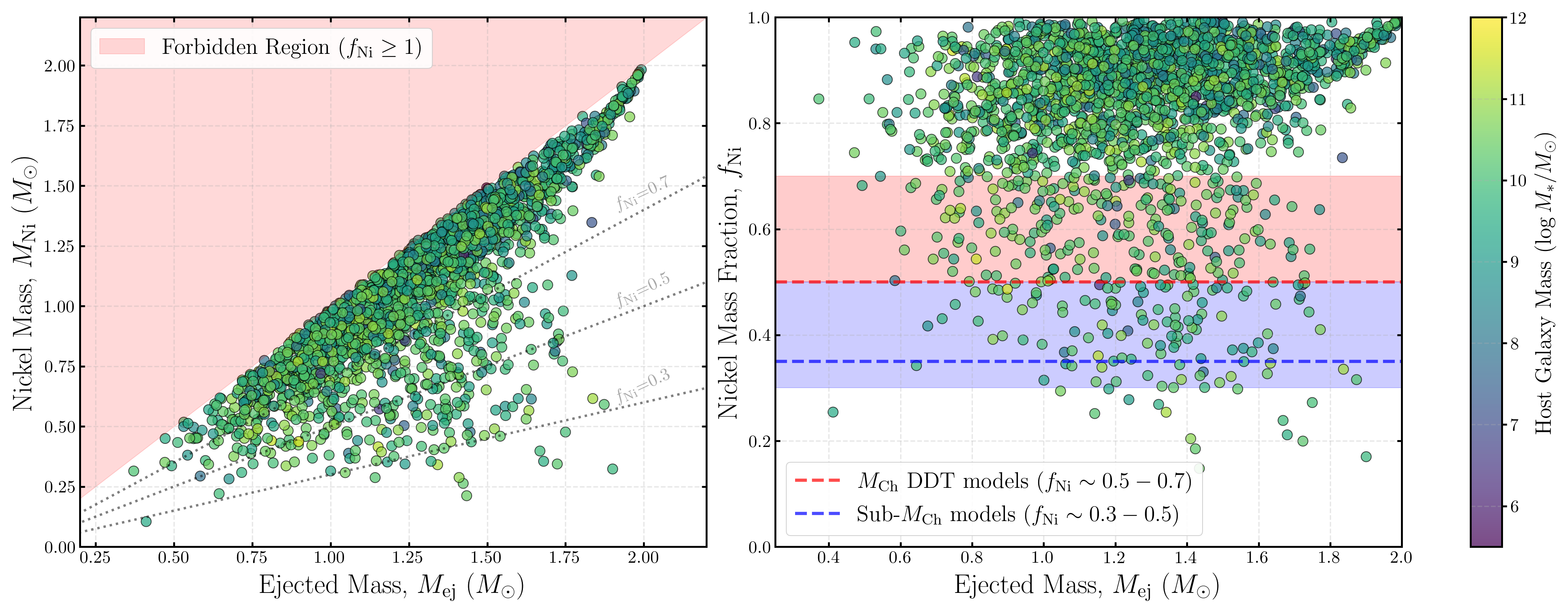}
\caption{Same as Fig.~\ref{fig:mass_plane} but for analyses where we marginalise over a gamma-ray opacity prior.}
\label{fig:mass_plane_opacity}
\end{figure*}

The most dominant impact of this change is on the burning efficiencies and in turn the nickel mass. This is as expected, as a larger gamma-ray opacity will lead to higher trapping, implying more energy remains in the system instead of escaping in gamma-rays. In Fig.~\ref{fig:mass_distributions_opacity}, we show the different inferred distributions for ejecta mass and nickel mass, which both shift to smaller values compared to our analysis with a fixed opacity, while velocity shifts to a larger value. In Fig.~\ref{fig:mass_plane_opacity}, we show the impact on burning efficiencies where marginalising over this prior shifts these efficiencies to lower values, moving the summary values away from the extreme $f_{\rm Ni}\simeq1$ boundary, although channel identification still requires comparison with specific explosion-model grids rather than fixed burning-fraction bands. 

We note that the shift in properties such as the nickel mass is largely only visible in the summary distributions, each individual single-event posterior still overlaps with the original analysis with a fixed gamma-ray opacity. In other words, the data is not informative about the gamma-ray opacity but marginalising over this prior has the general effect of shifting the broad estimates to require lower nickel masses. 

\section{Additional plots from Redshift-limited sample}
Here we show additional plots from our analysis on the redshift-limited sub-sample. 

\begin{figure*}
\centering
\includegraphics[width=0.95\textwidth]{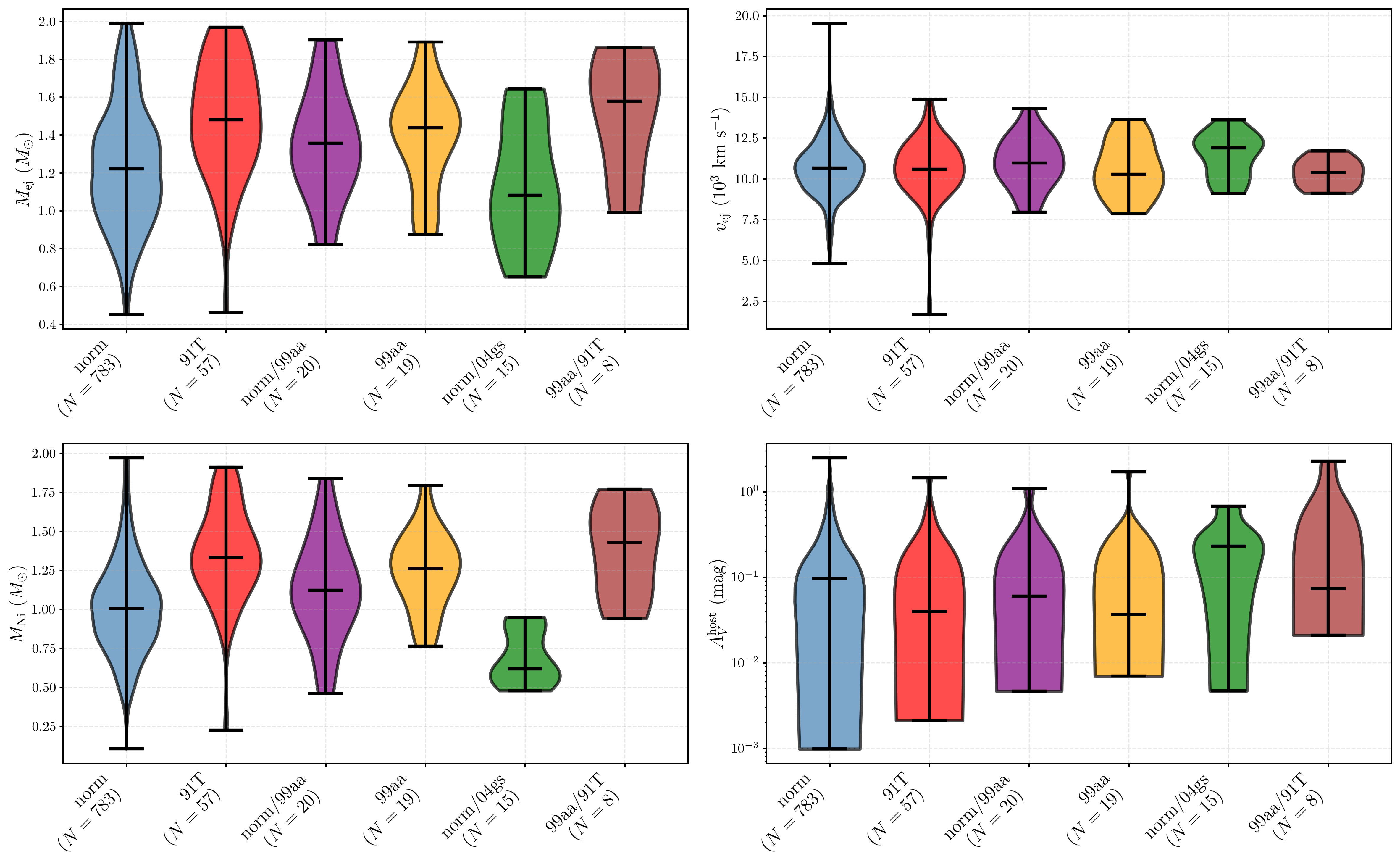}
\caption{Violin plots of the physical properties by spectroscopic subtype for the redshift-limited sample.}
\label{fig:subsample-subtypes}
\end{figure*}

\begin{figure}
\centering
\includegraphics[width=0.48\textwidth]{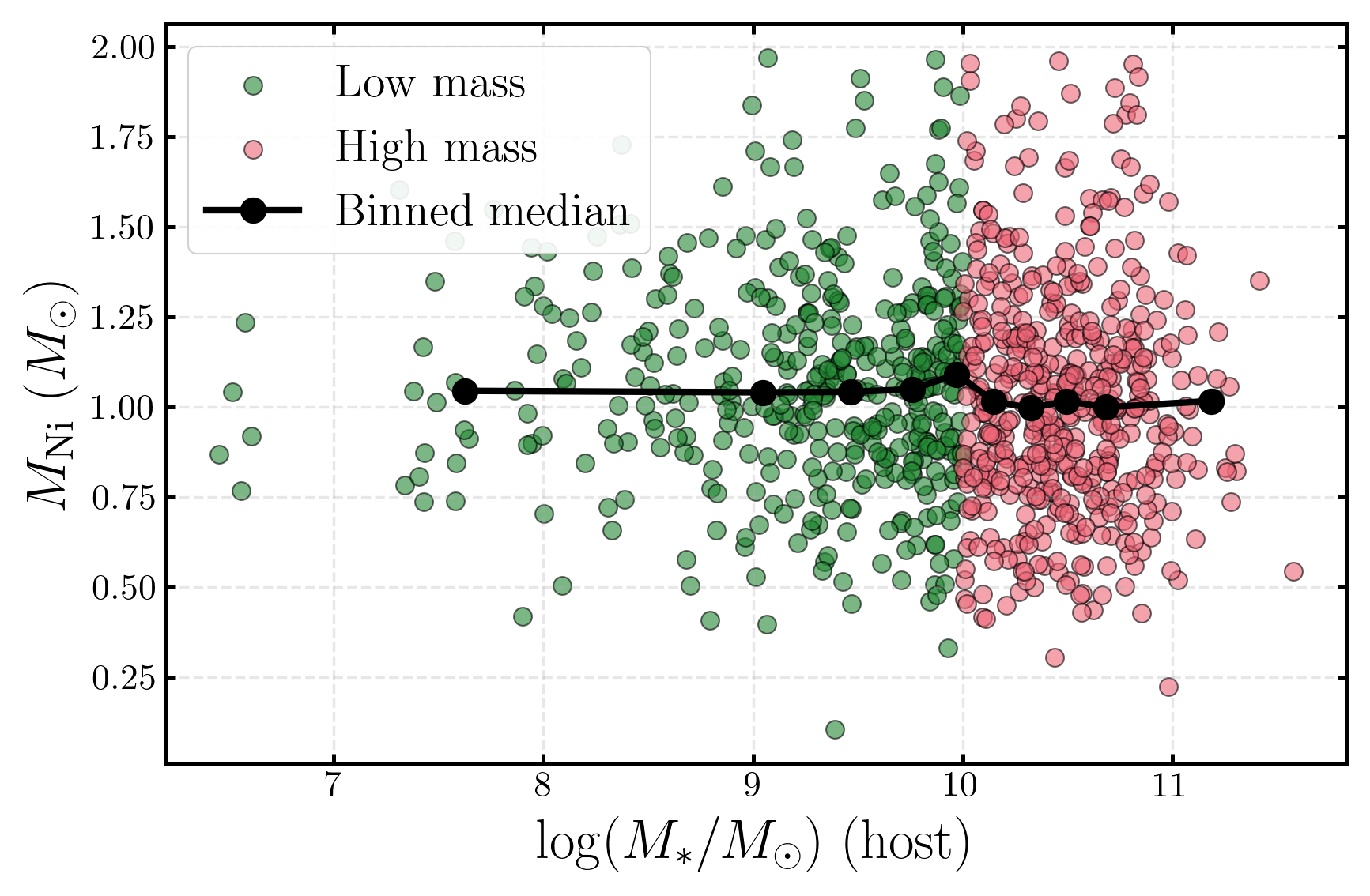}
\caption{Nickel mass versus host galaxy mass for the redshift-limited sample}
\label{fig:subsample-nickelvshost}
\end{figure}

\begin{figure}
\centering
\includegraphics[width=0.48\textwidth]{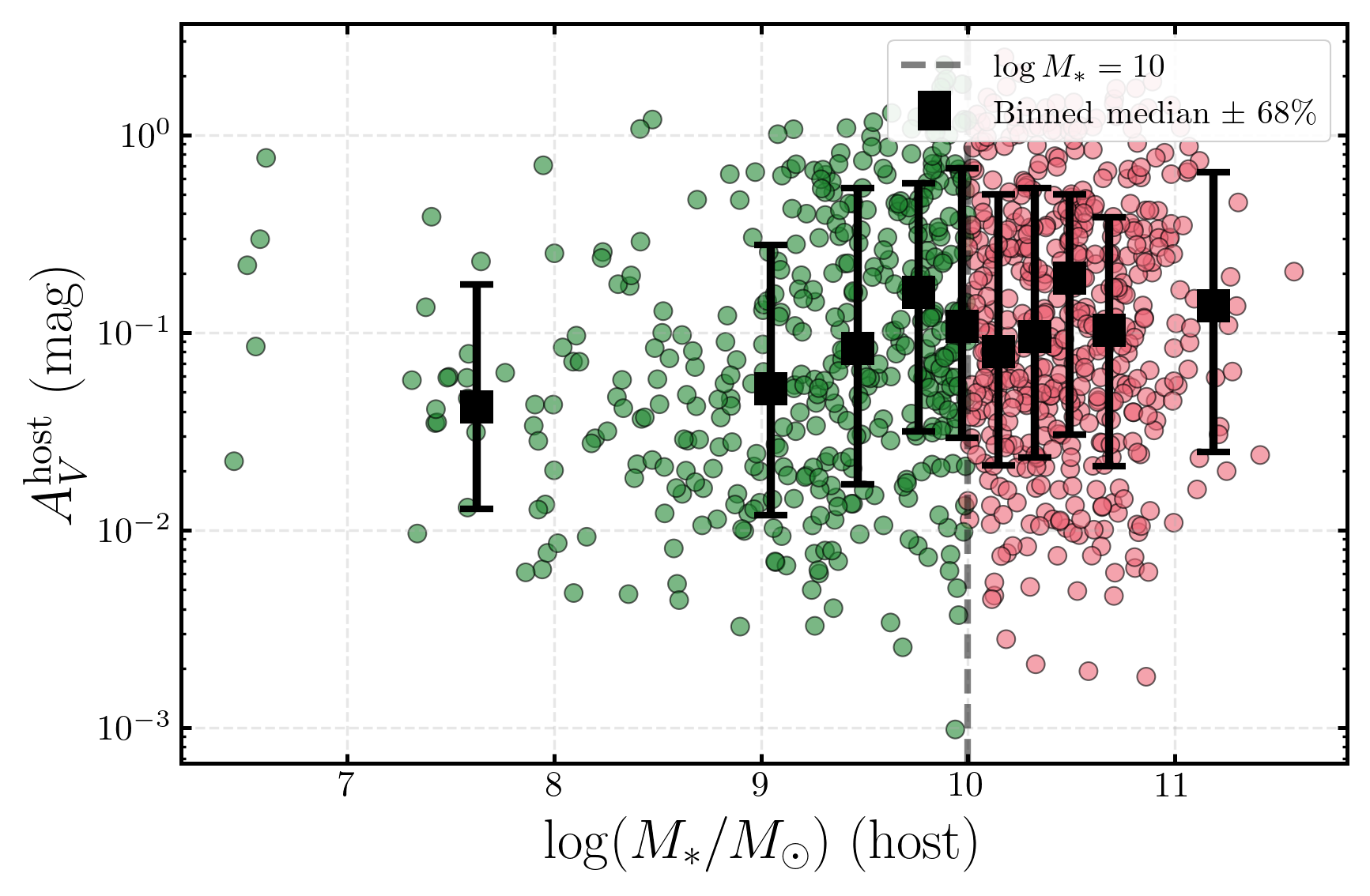}
\caption{Host-extinction versus host galaxy mass for the redshift-limited sample.}
\label{fig:subsample-hostextinctionvshost}
\end{figure}

\bsp	
\label{lastpage}
\end{document}